\documentclass[12pt]{article}
\usepackage{graphicx,epsfig,pslatex}  
\usepackage{picins}                   
\usepackage{subfigure}
\usepackage{fancybox,fancyhdr}       
\usepackage{hanging}
\usepackage{amsmath,amstext,amsfonts,amssymb}
\usepackage{amsthm}   
\usepackage{bm}       %
\usepackage{array}    
\usepackage{amscd,xy} 
\usepackage{exscale,relsize}
\usepackage{array}
\usepackage{color}
\usepackage{paralist}
\usepackage{indentfirst}
\usepackage{url}
\usepackage{fullpage}
\usepackage{times}

\newcommand{\black}{\color{black}}
\title{
\Large\textbf{
Modeling of Time-Structured Multi-Turn Injection\\
into Fermilab's Main Injector\\
(\large Micro-Bunch Injection with Uncontrolled Longitudinal Painting\normalsize)
}
}
\author{Phil~S.~Yoon\thanks{~E-mail:~phil.s.yoon@hotmail.com},
        ~David~E.~Johnson, and~Weiren~Chou\\ \\  
        Fermi National Accelerator Laboratory, Batavia, IL 60510, USA
  \thanks{~Work supported by Fermilab Research Alliance (FRA),~LLC 
          under contract No.~DE-AC02-07-CH11359
          with the United States Department of Energy}}
\date{Februrary~2008}
\begin{document}
\maketitle
   \begin{abstract}
       This article presents the modeling of
       time-structured multi-turn injection for 
       an upgraded version of Fermilab's Main Injector 
       with the 8-GeV Superconducting RF proton driver,
       or the International Linear Collider (ILC)-style linac, or the Project-X linac. 
       The Radio-Frequency (RF) mismatch between a linac 
       and the Main Injector will induce 
       uncontrolled longitudinal painting in RF-phase direction.
       Four scenarios have been explored 
       with different choices of RF parameters
       of a single RF system and a double RF system 
       in the presence of longitudinal space charge. 
       It is found from the studies of micro-bunch injection
       with the aid of ESME (2003) simulations
       that a dual RF system with an optimized choice of RF parameters 
       enables overcoming the space-charge limits set by beam intensity 
       during the multi-turn injection process.
       A double RF system with a harmonic ratio ($R_{H} = H_{2}/H_{1}$) 
       of 2.0 and a voltage ratio ($R_{V} = V_{2}/V_{1}$) of 0.5 
       are most favored to reduce both longitudinal and transverse 
       effects of space charge residing in the Main Injector.
       \black
   \end{abstract}   
%
\section{\label{sec:pd-microbunch}Micro-Bunch Injection
into the Main Injector from\\ a Superconducting RF Linac}
%
%
\par 
Following the method of time-structured multi-turn injection 
from the 400-MeV linac to Fermilab's Booster~\cite{yoon:thesis},
we have further explored the scheme of micro-bunch injection for 
applications to Fermilab's Main Injector, 
(from a future Superconducting RF (SCRF) linac 
to an upgraded Main Injector (MI-2))
\footnote{
We will use the Main Injector (MI) and an upgraded version of 
the Main Injector (MI-2) interchangeably throughout this article.
}
considering parasitic longitudinal painting.
The future SCRF linac referred here can be either 
the 8-GeV SCRF linac Proton Driver~\cite{foster-jam,foster:pac05,foster:pd-tdr},
or the Project-X linac~\cite{projectx}.
\black
	\subsection{\label{subsec:overview-mi}
	Overview of the Main Injector}
%
The Main Injector (MI) is a ring with a circumference of about 3.3 (km).
The central role of the MI is to connect to the Tevatron, the Booster, 
the Anti-Proton source, switchyard, and the Recycler Ring 
via a number of beam transport lines within the Fermilab accelerator complex.
The MI accelerates and decelerates particle beams with energy ranging from 8 (GeV) and 150 (GeV),
depending on the operation mode. The harmonic number of the MI is 588
and the harmonic RF at injection is 52.8114 (MHz)\footnote{
For the sake of brevity and convenience, 
\textbf{53 MHz} is referred to as MI RF hereafter.}.
\black
	\subsection{\label{subsec:overview-pdriver}
        Overview of the 8-GeV Superconducting RF Linac}
%
         An 8-GeV SCRF linac has been proposed as a single-stage 
         $H^{-}$ injector into the Main Injector as a replacement for 
         the aging 400-MeV Linac and the 8-GeV Booster.
         This new 8-GeV SCRF linac would be the highest-energy $H^{-}$ 
         multi-turn injection system in the world. 
         Fermilab has been carrying out design studies
         \cite{foster:pd-tdr,hminus:workshop} of the SCRF linac and 
         injection systems\cite{multi-turn:pac07} over the last several years.
         The linac design\cite{ostroumov} utilizes a warm-temperature 
         325-MHz RFQ and rebunching cavities to bunch the beam at 325 (MHz).
         At $\beta = 0.89$ (about $E_{kin} = 1.1~(GeV)$),
         the RF of the Superconducting (SC) cavities is 1.3 (GHz). 
         The ultimate bunch structure required for injection into the MI
         will be formed by a 325-MHz fast chopper system\cite{madrak}.
         The fast chopper system will be required to remove 
         individual 325-MHz bunches or bunch trains
         for matching to the MI RF structure and providing 
         a beam-abort notch: two out of every six micro-bunches\footnote{
         The \textbf{micro-bunch} is referred to 
         as a 325-MHz bunch hereafter.} are to be removed.
         \black
         \subsection{\label{subsec:mi-time-structure}
         Time Structure of the Main Injector}
\begin{figure}[t!h!]\centering
     \includegraphics[scale=0.40]{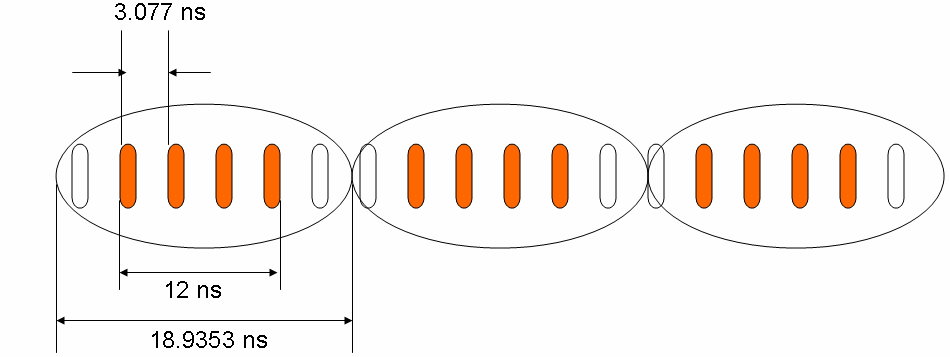}
     \caption{\label{fig:3rf-bucket}
     Three consecutive 53-MHz RF buckets used for the Main Injector.
     \black}
\end{figure}

\begin{figure}[h!]\centering
     \includegraphics[scale=0.17]{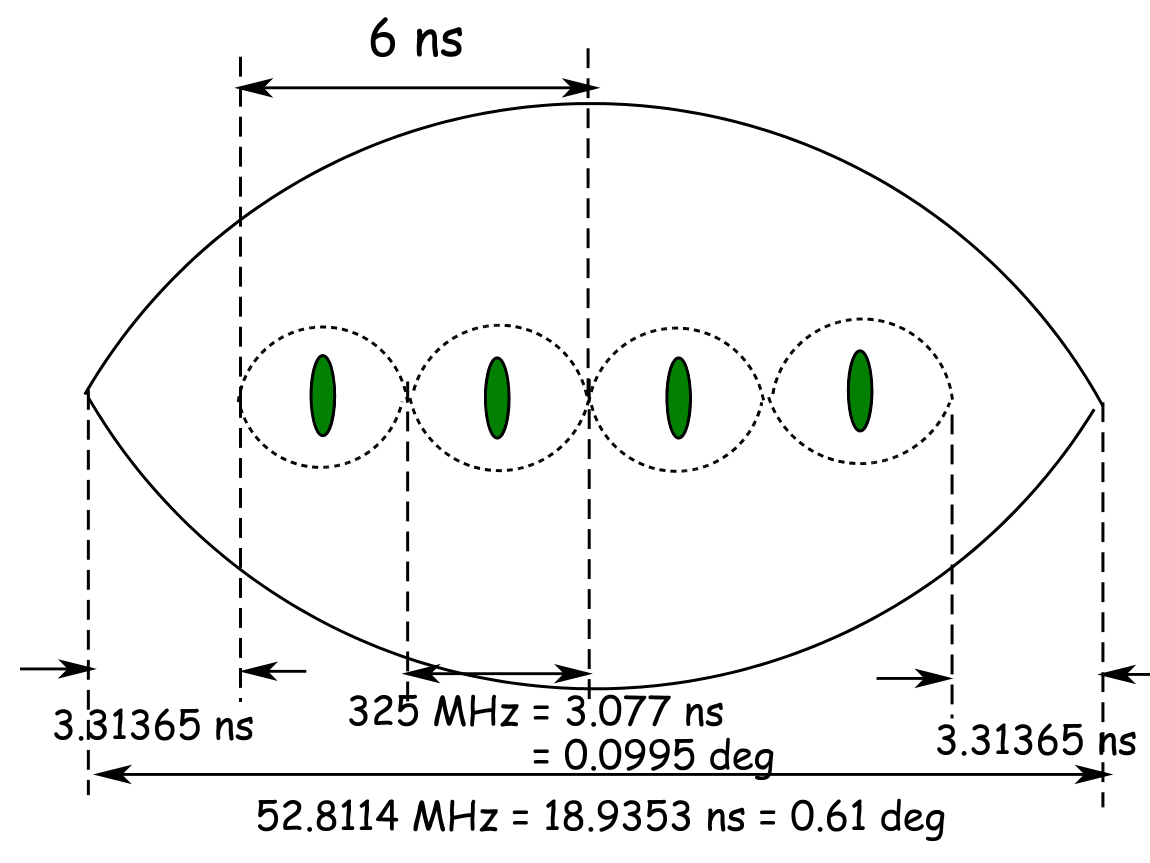}
     \caption{\label{fig:rf-bucket-fmi}Time structure of the single RF system used for the Main Injector}
\end{figure}

\begin{figure}[ht!]\centering
     \includegraphics[scale=0.40]{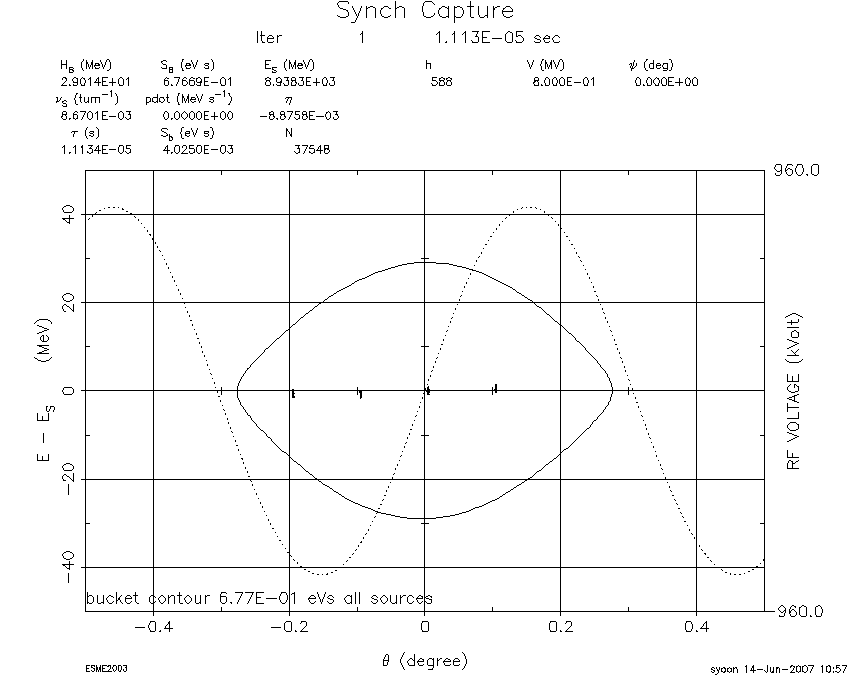}
     \caption{\label{fig:singlerf-waveform}
     Single RF system:~the injection of 4 micro-bunches
     at the $1^{st}$ turn with an RF waveform drawn 
     in the background\black}
\end{figure}

%
In the injection model 
the fast chopper system located at 
the front end of the 8-GeV SCRF linac
produces a train of four micro-bunches
for being injected into the MI.
Illustrated in Figure~\ref{fig:3rf-bucket} is 
a schematic for one MI RF bucket populated 
with an initial train of four micro-bunches.
The two chopped micro-bunches are represented by 
two consecutive empty 325-MHz RF buckets.
A train of four 325-MHz micro-bunches are 
synchronously injected into a standing 53-MHz bucket. 
The length of the two chopped micro-bunches is equivalent to 6 ns.
More details of the time structure of each MI RF bucket 
after the first synchronous injection from a SCRF linac 
are illustrated in Figure~\ref{fig:rf-bucket-fmi}.
A total beam notch per MI RF bucket is about 6 ns, 
which corresponds to two 325-MHz RF buckets, whereas
the principal RF harmonic of 52.8 (MHz) and 
the sub-harmonic of 325 (MHz).
Illustrated in Figure~\ref{fig:singlerf-waveform} is 
a phase-space ($\Delta E,\theta$) plot containing 
the very first train of four micro-bunches with 
an RF-voltage waveform drawn in the background.
\black
%
The following is a list of longitudinal parameters 
that can be seen in the header of each ESME phase-space plot 
presented in this article:
\par
\begin{center}
\fbox{\begin{minipage}{0.90\textwidth}
{\it iter} (number of turns), $H_{B}$ (bucket height), 
$S_{B}$ (bucket area), $S_{b}$ (bunch area),\\ 
$V$ (RF voltage), $E_{s}$ (synchronous energy), 
$\nu_{s}$ (synchrotron tune), pdot ($dp/dt$),\\
$\eta$ (slip factor), $\tau_{s}$ (revolution period),
h (harmonic number), $\Psi$ (synchronous phase), 
and\\ N (number of macro-particles)
\end{minipage}
}\end{center}
\par
\begin{figure}[t!]\centering
     \includegraphics[width=3.7214in,height=2.3in]
     {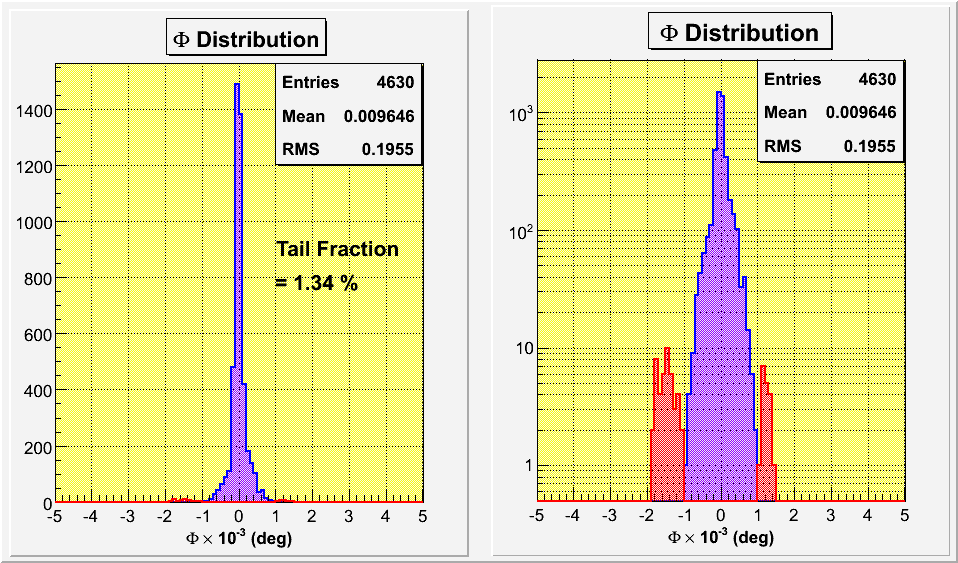}
     \caption{\label{fig:microbunch-phase-density}
     Zero-centered charge density with tail portion; 
     (left) on a linear scale (right) on a logarithmic scale}
\end{figure}
\begin{figure}[b!]\centering
     \includegraphics[width=3.72in,height=2.3in]
     {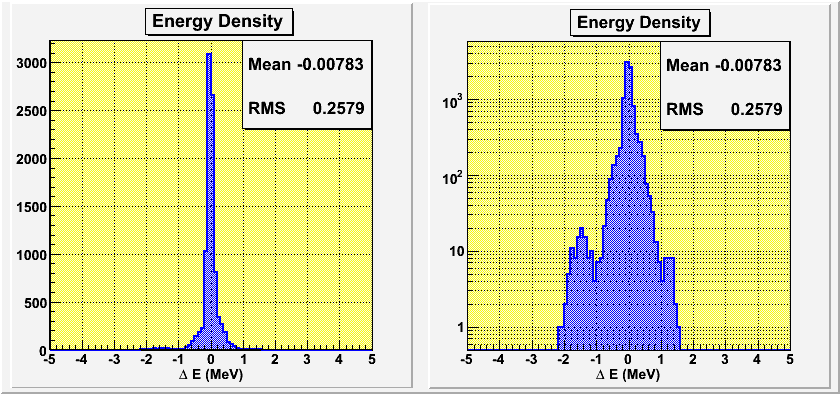}
     \caption{\label{fig:microbunch-energy-density}
      Zero-centered energy density;
     (left) on a linear scale (right) on a logarithmic scale}
\end{figure}

The horizontal axis represents $\Delta\theta$ ($= \theta - \theta_{s}$)
in units of degrees, and 
the vertical axis on the left is $\Delta E$ ($= E - E_{s}$)
in units of $MeV$, and the vertical axis on the right is $V_{rf}$ 
in units of $kV$ for the RF waveform.
In addition to the phase-space plot, 
the azimuthal-density\footnote{
The \textbf{azimuthal density} is referred to as 
the charge density, or the azimuthal profile of macro-particles.} 
and energy-density distributions\footnote{
The \textbf{energy-density distribution} is referred to as
the profile of macro-particles in energy ($\Delta E$) direction.} 
of one of four micro-bunches are plotted, using output data 
generated by ESME simulations as shown in Figures~\ref{fig:microbunch-phase-density} 
and~\ref{fig:microbunch-energy-density}.
\black
\newpage\clearpage
The root-mean-square (RMS) width of one micro-bunch is 
1.96 $\times 10^{-4}$ (deg), which corresponds to 6.05 (ps).
The area in red, indicating the tail portion of a micro-bunch, is about 1.34$\%$.
To make the tail portion of a micro-bunch stand out,
the density distribution is plotted on a logarithmic scale.
The RMS value of \textit{initial} energy spread is about 0.26 (MeV)
per each micro-bunch. Regarding the MI to be a 11.1338-$\mu$s, 
or 360-degree ring, we used the following conversion factor:
\begin{equation}\label{fig:ns2deg}
   0.032335~(deg/ns)
\end{equation}
\black
\newpage\clearpage
The principal MI RF is 53 (MHz) and the SCRF linac bunching frequency is 325 (MHz).
From the MI RF, we can obtain an integer harmonic number of 588.
The width of an MI RF bucket is 18.935 (ns), into which 
trains of micro-bunches are injected repeatedly during the first 270 turns, 
while filling in one MI RF bucket in the region of $\pm$6 (ns) around its center.
The beam notch that is kept free of beams is 3.3 (ns) long, 
on the basis of an earlier injection study performed with
a \textit{long bunch}\cite{jm:talk}. 
In the following sections, we will explore and discuss 
several different scenarios of micro-bunch injection, 
based upon the RF parameters used in the current MI operation.
\black 
	\subsection{\label{subsec:scenarios}
        Micro-Bunch Injection Scenarios for the Main Injector}
%
%
The goals for modeling the time-structured multi-turn injection
into the MI with a SCRF linac are three-fold:
\begin{compactenum}[(1)]\itemsep 0.1cm
   \item To find the optimized RF system of Fermilab Main Injector 
          with the following aspects:
      \begin{compactenum}[(a)]\itemsep 0.1cm
        \item Efficient RF capture methods with minimum particle losses
        \item Main Injector RF system with adequate RF parameters
        \item Minimizing space-charge effects in all degrees of freedom
        \item Optimizing total capture time and total injection time
      \end{compactenum}
   \item To design a fast beam-chopper system
         to make nearly loss-free injection attainable
   \item To investigate any limits on the intensity upgrades
         from the current Main Injector
\end{compactenum}
\black
\begin{table}[b!]\centering
         \caption {Main Injector Parameters for ESME Simulations}
         \label{tab:mi-param}
         \vspace{\belowcaptionskip}
         \begin{tabular}{@{}l r r r @{}}                          \hline 
         Mean Radius                               & 528.297  (m) \\
         Beam Momentum (at injection)              & 8.889 ($GeV/c$) \\
         $\gamma_{tr}^{2}$ (transition gamma)      & 466.53572   \\
         Principal Harmonic No.                    & 588         \\
         $\Phi_{s}$ (synchronous phase)            & 0.0 (deg)   \\
         effective beam radius (at injection)      & 0.0050 (m)  \\
         effective beam-pipe radius (at injection) & 0.051 (m)   \\
         No. of Space-Charge Bins                  & 64          \\
         No. of FFT Bins                           & 64          \\
         Micro-Bunch Intensity                     & 2.65 $\times 10^{8}$  \\
         Total Intensity (after injection is complete)
                                                   & 1.54 $\times 10^{14}$ \\
                                                                 \hline
      \end{tabular}
\end{table}
%
\newpage
With the above aims in mind, 
we modeled four different scenarios of synchronous injection 
over 270 injection turns, but with different RF harmonic systems 
and RF parameters.
Listed in Table~\ref{tab:mi-param} are
the key parameters of the Main Injector
that were employed in ESME simulations 
for all scenarios presented in this article.
\black
%
\section{Scenario I}
\par\smallskip
\fbox{\begin{Bitemize}[t]\itemsep 0.01mm
    \item synchronous injection of a train of 
          4 micro-bunches into a standing RF bucket
    \item single RF Harmonic (53 MHz)
    \item $V_{rf}$ = 800 (kV) (fixed RF voltage with no ramping)
    \item \textbf{270 injection turns} ($\sim$ 3 ms) 
          $\rightarrow$ 2,700 turns ($\sim$ 30 ms)
    \item longitudinal space charge included
    \item \textit{uncontrolled} longitudinal painting in RF-phase ($\theta$) direction
\end{Bitemize}}
\bigskip
\par
We begin with a single RF system of 53 (MHz) 
and the fixed RF voltage ($V_{rf}$) of 800 (kV).
Referring to Figure~\ref{fig:singlerf-waveform},
a set of four blobs captured in an RF bucket
represent the first train of micro-bunches 
that are synchronously injected into 
a standing RF bucket.
The dotted sinusoidal curve for RF voltage waveform
in the background of the phase-space figure indicates
the single RF harmonic and the amplitude of RF voltage.
\par
\begin{figure}[h!]\centering
     \subfigure[three trains of micro-bunches in an MI RF bucket]{\includegraphics[scale=0.28]
     {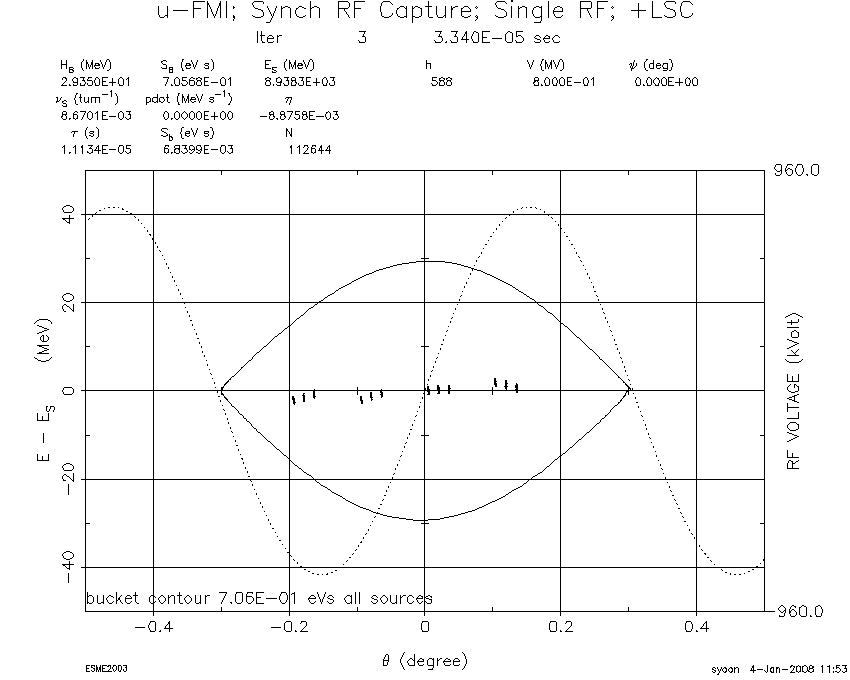}} 
     \subfigure[charge density distribution]{\includegraphics[scale=0.28]
     {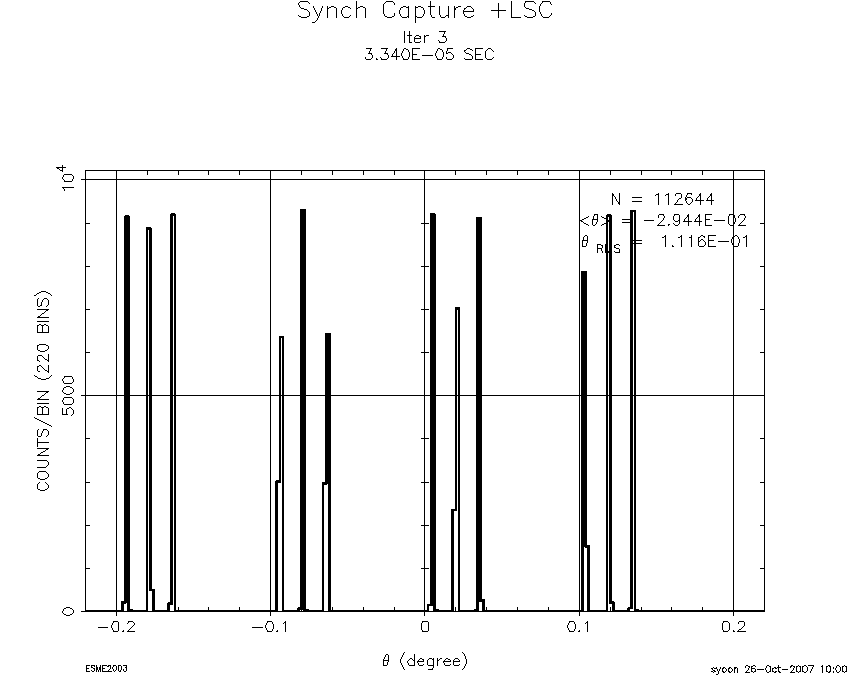}}
     \caption{\label{fig:singlerf-noramp-3t}[\textbf{Scenario I}]
              Single RF bucket and discrete charge density after 3 injection turns \black}
\end{figure}
\begin{figure}[h!]
     \centering
     \subfigure[270 turns]{\includegraphics[scale=0.35]{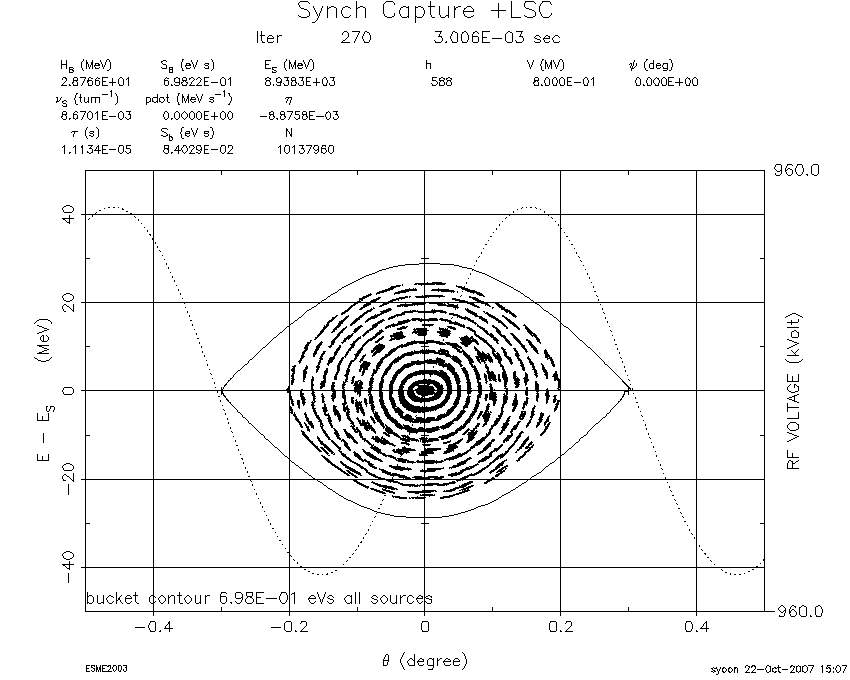}}
     \subfigure[2,700 turns]{\includegraphics[scale=0.35]{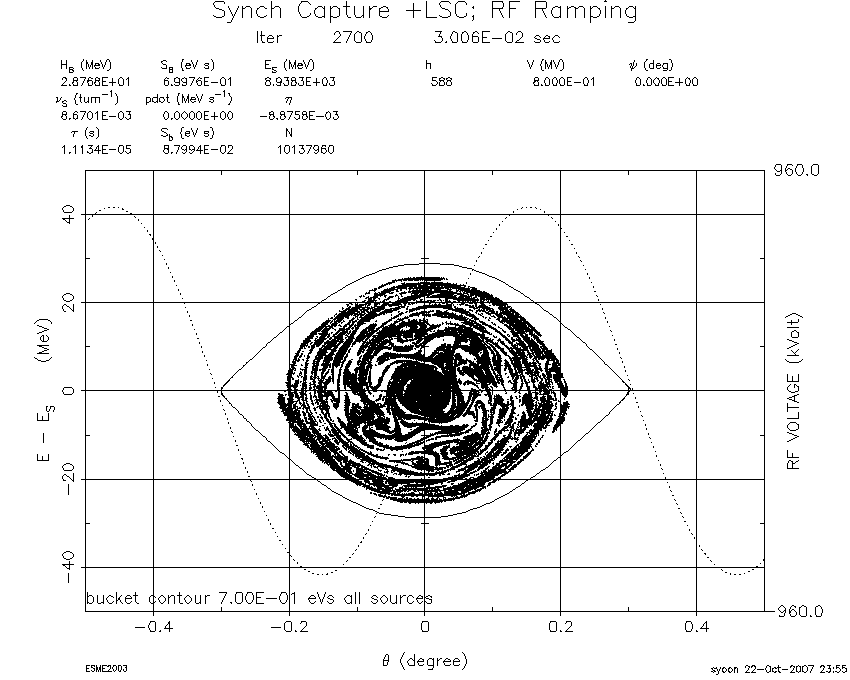}} 
     \caption{[\textbf{Scenario I}]~
              Synchronous injection of micro-bunches
              with a single RF system (a) after 270 turns
              (b) after 2,700 turns;
              note that \textbf{+LSC} in each figure title indicates
              \emph{inclusive of Longitudinal Space Charge}.\black
              }
     \label{fig:lps-singlerf-noramp-270t}
\end{figure}
\begin{figure}[hbt!]
   \begin{center}
   \subfigure[270 turns]{\includegraphics[scale=0.35]{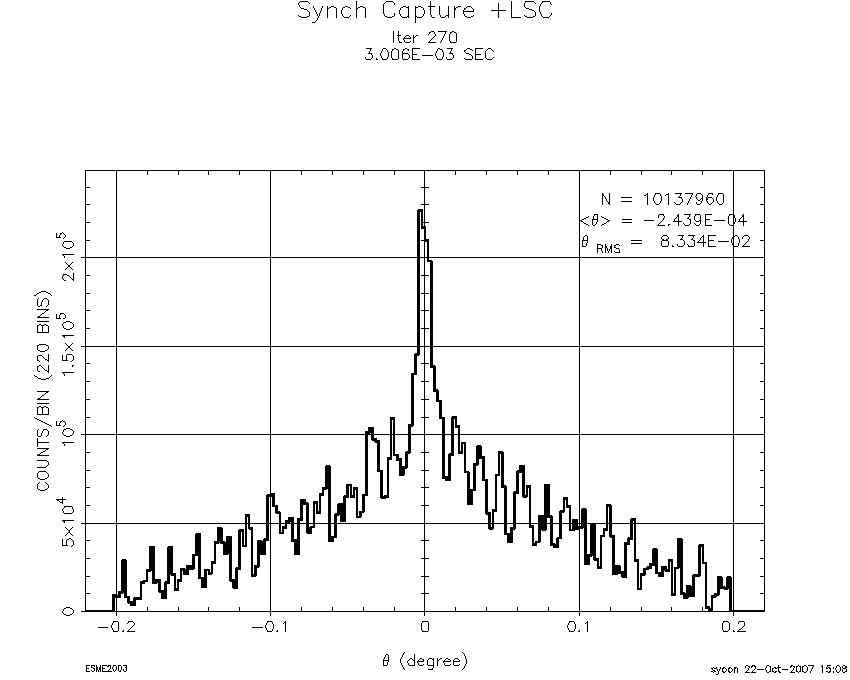}}
   \subfigure[2,700 turns]{\includegraphics[scale=0.35]{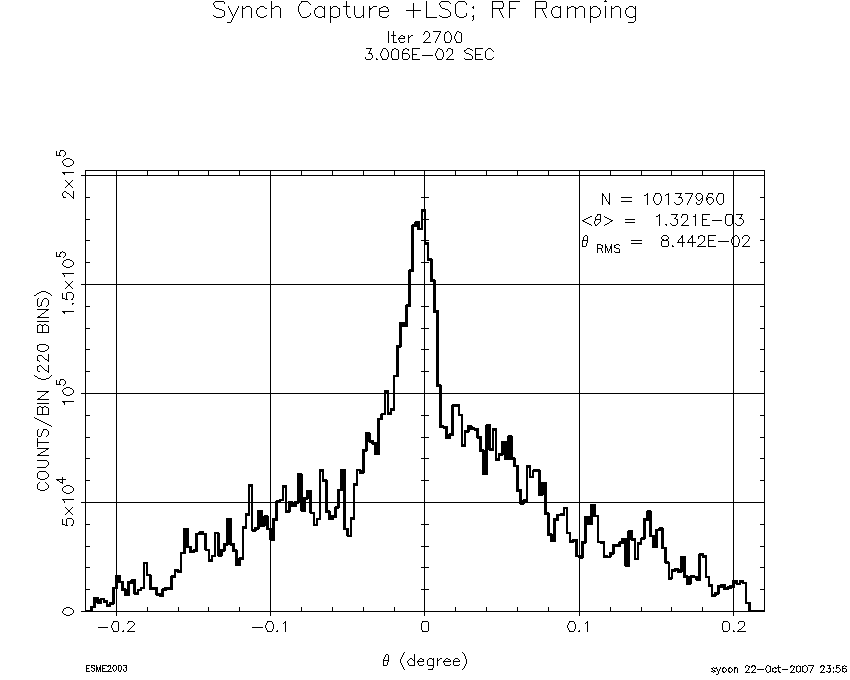}} 
   \caption{\label{fig:theta-singlerf-noramp-270t-2700t}
             \textbf{[Scenario~I]}
            ~Distribution of charge density
            with a single RF harmonic at the 270$^{th}$ turn 
            and at the 2,700$^{th}$ turn.\black
            }
   \end{center}
\end{figure}
\begin{figure}[h!t!]\centering
   \subfigure[Turn-by-turn evolution of longitudinal emittance
              with a single RF harmonic over 270 turns\label{fig:epsilon-l-singlerf-noramp-270t}]{
              \includegraphics[scale=0.25]{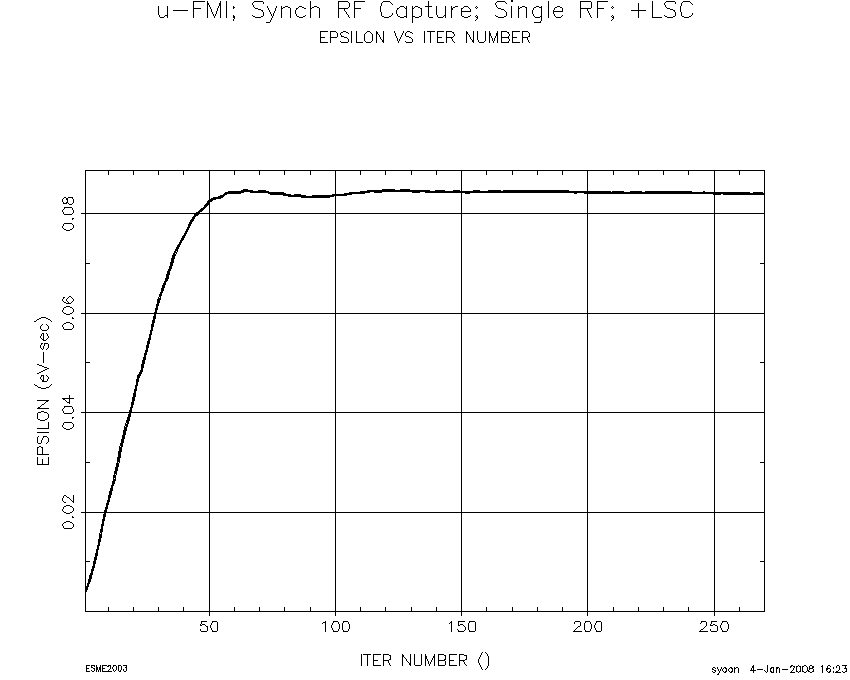}
              }
   \subfigure[Evolution of the number of injected macro-particles
              with a single RF harmonic over 270 turn]{
              \includegraphics[scale=0.25]{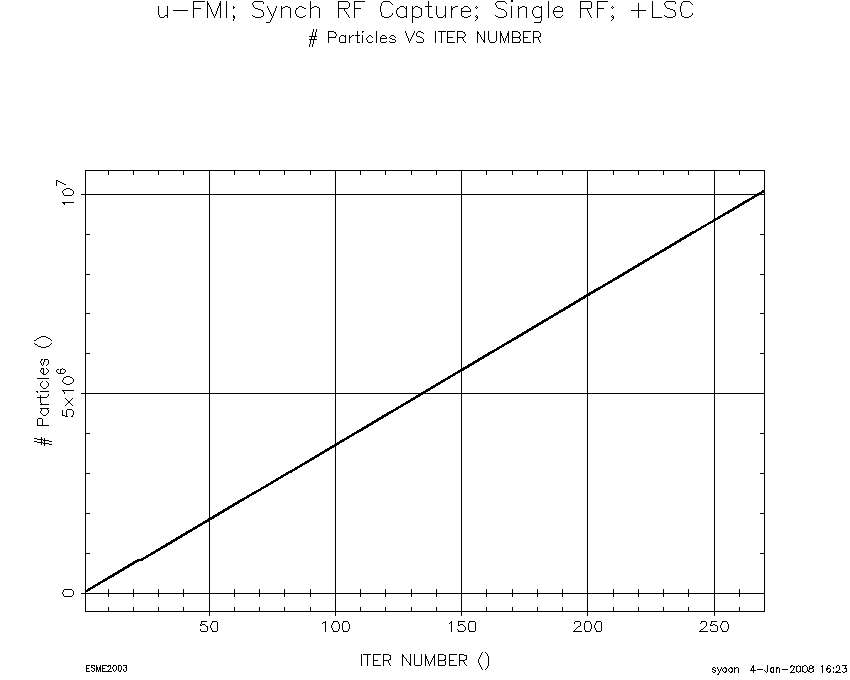}
              }
   \subfigure[\label{fig:dEturn-singlerf-noramp-270t}
             Induced $\Delta E$ per turn ($\Delta E/turn$) due to space charge  
             with a single RF harmonic at the $270^{th}$ turn] 
             {\includegraphics[scale=0.25]{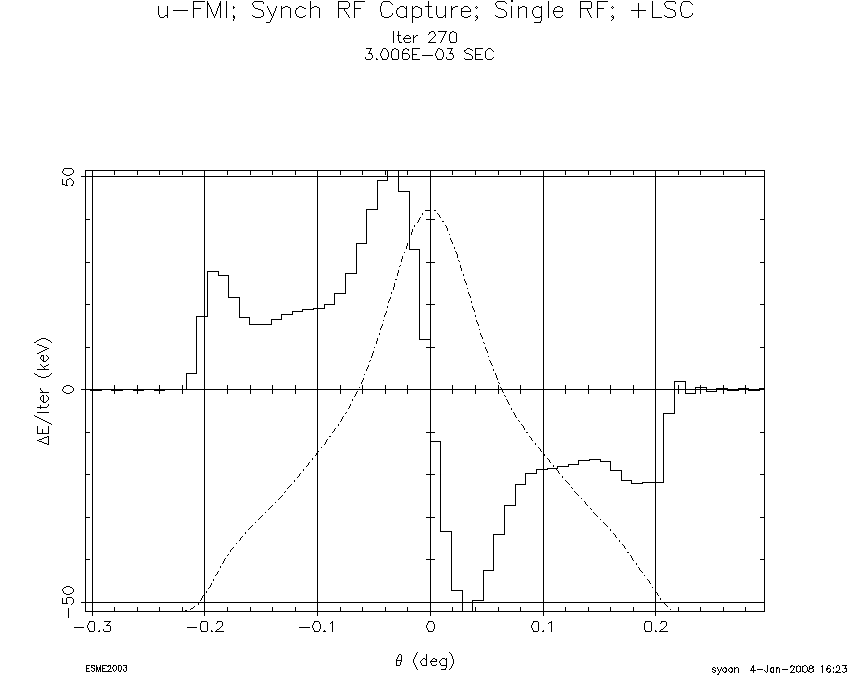}}\qquad
   \subfigure[\label{fig:vpkfd-singlerf-noramp-270t}
              Turn-by-turn evolution of collective voltage in frequency domain over 270 turns]
              {\includegraphics[scale=0.25]{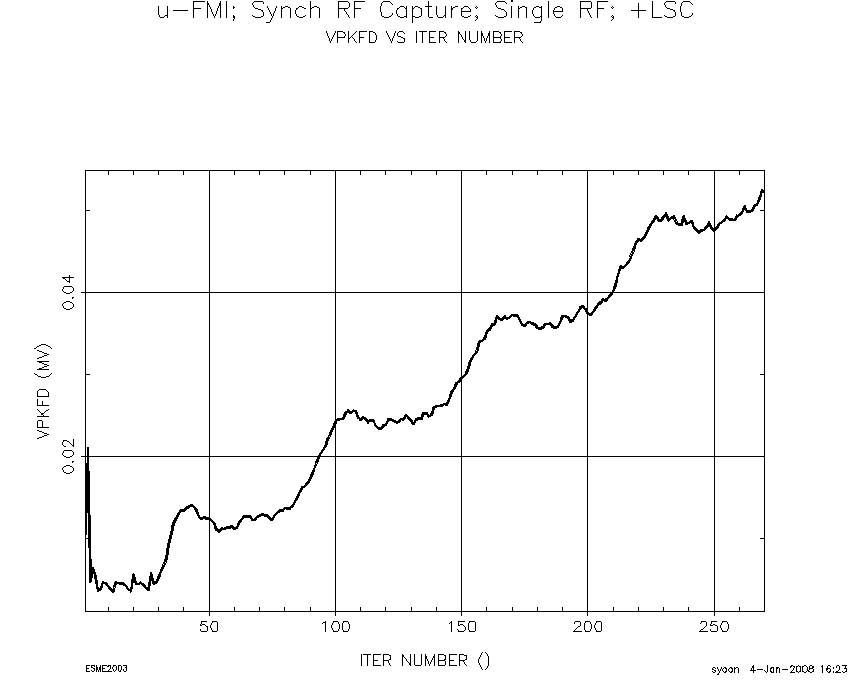}}\quad
   \subfigure[\label{fig:bf-singlerf-noramp-270t}
              Turn-by-turn evolution of bunching factor]
              {\includegraphics[scale=0.25]{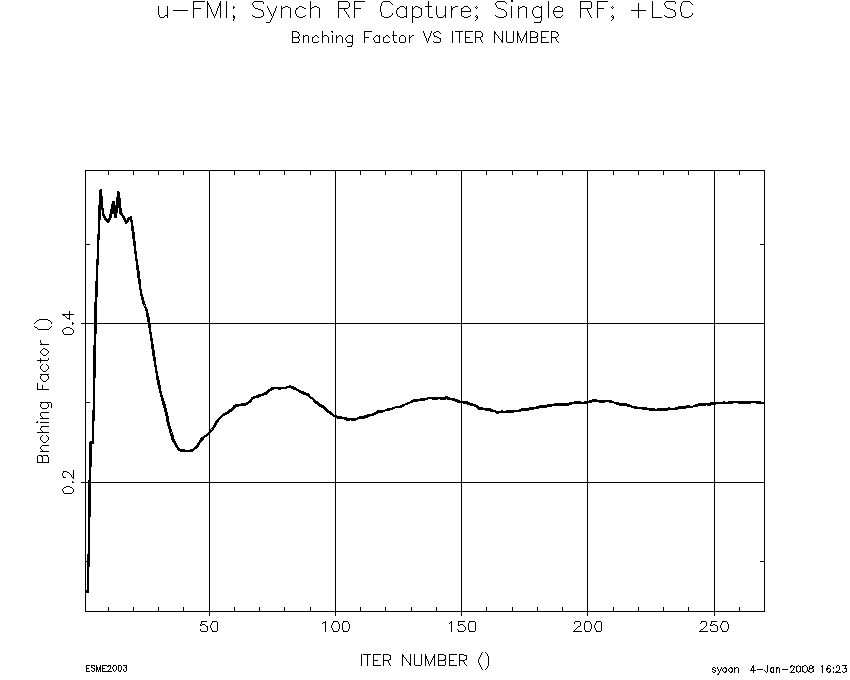}}
\caption{$\lbrack$\textbf{Scenario I}$\rbrack$~
         Additional calculations in relation to longitudinal space-charge effects}
\end{figure}

The single turn in the MI in terms of 
the average machine circumference and 
the revolution period can be expressed 
with the following relationship for use 
as conversion factors:
\begin{equation}\label{eqn:mi-deg-sec-meter}
   360~(deg) = 11.1338~(\mu s) = 3319.388~(m)
\end{equation}
As illustrated in Figure~\ref{fig:rf-bucket-fmi} 
the relationships, which can be used as conversion factors for one MI RF bucket,
are derived from Eqn.~(\ref{eqn:mi-deg-sec-meter}):
%
\begin{equation}\label{eqn:bk-deg-sec-meter}
   52.8114~(MHz) = 18.9353~(ns) = 0.6122~(deg) = 5.645~(m)
\end{equation}
%
Due to the \textit{longitudinal mismatch}, or \textit{RF mismatch} 
between the SCRF linac and the MI, inherent phase slips, or phase jitters 
can be induced in the form of \textit{uncontrolled} or \textit{parasitic} 
painting in RF-phase direction.
%
%
With $f$ and $\lambda$ being radio frequency and RF wavelength,
respectively, we computed the RF ratio ($R_{rf}$):
\newpage
\begin{equation}\label{eqn:rf-ratio}
  \begin{aligned}
   R_{rf}
   &= \frac{f_{PD}}{f_{MI}}
   = \frac{325~\text{MHz}}{52.8114~\text{MHz}} \\
   &= \frac{\lambda_{MI}}{\lambda_{PD}}
   = \frac{5.6453~(m)}{0.9173~(m)}\\
   &= 6.154,
  \end{aligned}
\end{equation}
in which PD stands for the 8-GeV SCRF Proton Driver.
\black
%
Referring to Figure~\ref{fig:rf-bucket-fmi},
Eqn.~(\ref{eqn:rf-ratio}) implies that
a total of 6 linac RF buckets can fit into 
one MI RF bucket.
Since the RF ratio is a non-integral number,
a modulus of the RF ratio is computed 
as for the case of modeling the Booster's injection\cite{yoon:thesis}:
\begin{equation}\label{eqn:mi-modulus}
   R_{rf}\Bigr\vert_{mod} = 0.154
\end{equation}
Then, the range of phase jitters within one MI RF bucket
can be determined:
\begin{equation}\label{eqn:phase-slip}
   0.154 \times 0.0995~(deg) = 0.0153~(deg)
\end{equation}
With phase slips included after the first three injection turns,
trains of micro-bunches captured in an MI RF bucket
and their \textit{discrete} charge distribution
are acquired as shown in Figure~\ref{fig:singlerf-noramp-3t}.
%
For each successive MI RF bucket,
a train of 325-MHz bunches advance by 0.0153 (deg).
Accordingly, over 6 injection turns, the trains of micro-bunches 
slip through the phase direction across one 325-MHz bucket: \\
\begin{equation}\label{eqn:phase-slip-turn}
   \vspace{0.3cm}
   \delta\theta_{rf} = \frac{0.0995~(deg)}{0.0153~(deg)} = 6.5
\end{equation}
Having this amount of uncontrolled phase slips included,
we simulated the SCRF linac-to-MI injection 
with ESME over 270 injection turns. 
During the 270 turns, longitudinal painting takes place at each turn
within the MI RF bucket under the influence of longitudinal space charge,
as illustrated by Figure~\ref{fig:lps-singlerf-noramp-270t}.
%
Shown in Figure~\ref{fig:theta-singlerf-noramp-270t-2700t}
is one-peaked \textit{continuous} distribution of
charge density\footnote{
\textbf{Charge density}, 
or \textbf{azimuthal density} is referred to as 
charge line density ($\lambda$), for convenience.} 
at the 270$^{th}$ turn.
The turn-by-turn evolution of longitudinal emittance growth 
for the case of a single RF harmonic is shown 
in Figure~\ref{fig:epsilon-l-singlerf-noramp-270t}.
As the number of macro-particles increments, 
so does the longitudinal emittance up to about 0.08 (eV-s).
Then, the emittance reaches its equilibrium and remains constant
until the end of the simulation run.
The induced voltage ($\mathcal{V}_{sc}$) 
arising from space-charge fields ($\mathcal{E}(z)$)
is proportional to the \textit{g-factor}, which is defined as
\begin{math} 1 + 2\ln(R_{w}/R_{b})\end{math}.
Thus, space-charge-induced voltage per turn can be computed
at a specific turn from the charge distribution.
In particular, we looked at the induced voltage 
arising from the symmetric charge distribution
at the end of injection (at the 270$^{th}$ turn), 
as shown in Figure~\ref{fig:dEturn-singlerf-noramp-270t}:
\begin{equation}\label{eqn:sc-voltage}
  \mathcal{V}_{sc} \propto
  \frac{1}{\beta\gamma^{2}}\Bigl[1 + 2 \ln(R_{w}/R_{b})\Bigr]\frac{d\lambda}{dz},
\end{equation}
where $R_{w}$, $R_{b}$, $\beta$, $\gamma$, $\lambda$, and $z$ denote 
effective beam pipe radius, effective beam radius, 
relativistic $\beta$ and $\gamma$, line density, and 
longitudinal path length, respectively.
In the case of the Main Injector at an injection energy of 8.0 (GeV),
the value of \textit{g-factor} is 5.64:
\black
\newpage\clearpage
\newpage\clearpage
\newpage\clearpage
\clearpage
Drawn in Figure~\ref{fig:vpkfd-singlerf-noramp-270t} 
is the peak collective voltage in frequency domain ($\hat{V}_{FD}$)
computed at each turn. The collective voltage goes
up to about 50.0 (keV) at the end of the injection process.
Since the bunching factor is usually defined as 
average current ($\langle~I~\rangle$) over 
peak current ($\hat{I}$) for convenience,
its value is upper-bounded at unity. 
Figure~\ref{fig:bf-singlerf-noramp-270t} shows 
that the bunching factor converges to the value of 
about 0.3 for the case of Scenario I.
\black
%
\section{Scenario II}
%
In Scenario II,
the injection process described in Scenario I is followed by 
RF-voltage ramping linearly up to 150$\%$ of the initial value.
Thus, the RF voltage ramping lasts for additional 27 (ms),
and the total injection time elapses about 30 (ms).
\par\bigskip
\begin{center}
\fbox{\begin{Bitemize}[t]\itemsep 0.01in
    \item $V_{rf,~i}$ = 800 (kV), $V_{rf,~f}$ = 1,200 (kV)
    \item ramping RF voltage after the injection is complete 
\end{Bitemize}}
\end{center}
\par
\begin{figure}[h!t!]\centering
     \includegraphics[scale=0.3]{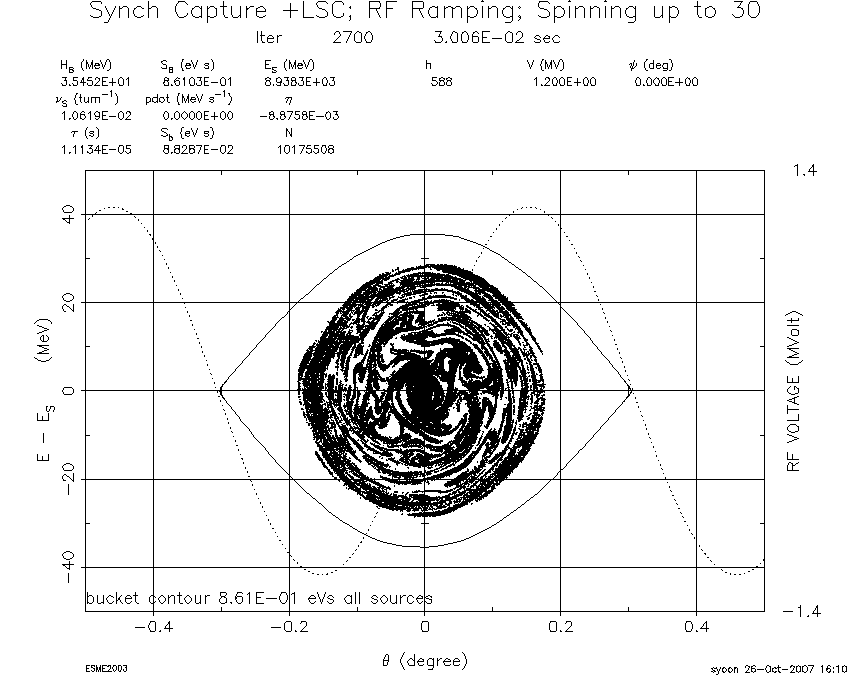}
     \caption{\label{fig:lps-singlerf-ramp-2700t}
     [\textbf{Scenario II}]~Synchronous injection of micro-bunches
     with a single RF harmonic and ramping RF voltage}
\end{figure}

\begin{figure}[h!]\centering
   \includegraphics[scale=0.3]{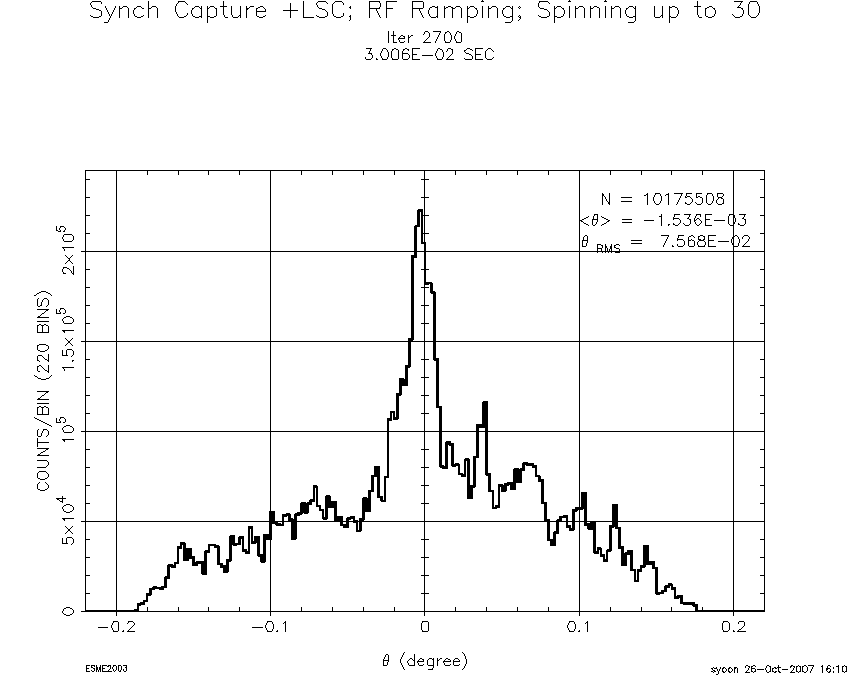}
   \caption{\label{fig:phi-singlerf-ramp-2700t}
   [\textbf{Scenario II}]~Distribution of charge density
   with a single RF harmonic and ramping RF voltage after 2,700 turns}
\end{figure}

\begin{figure}[h!t!]\centering
     \includegraphics[scale=0.3]
      {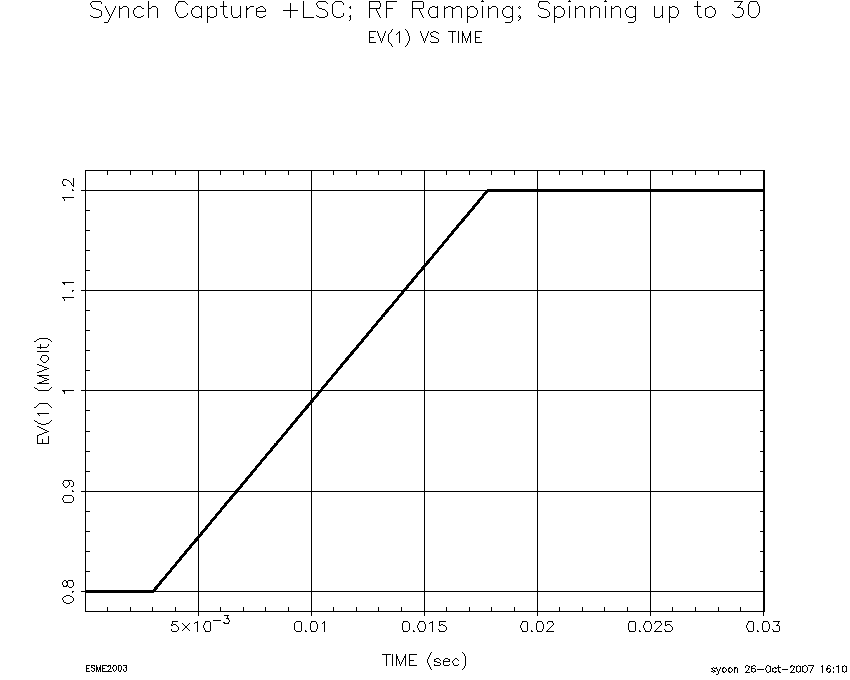}
     \caption{\label{fig:EV-1-singlerf-ramp-2700t}
     [\textbf{Scenario II}]~RF voltage curve over 2,700 turns}
\end{figure}

\begin{figure}[h!]\centering
   \includegraphics[scale=0.30]{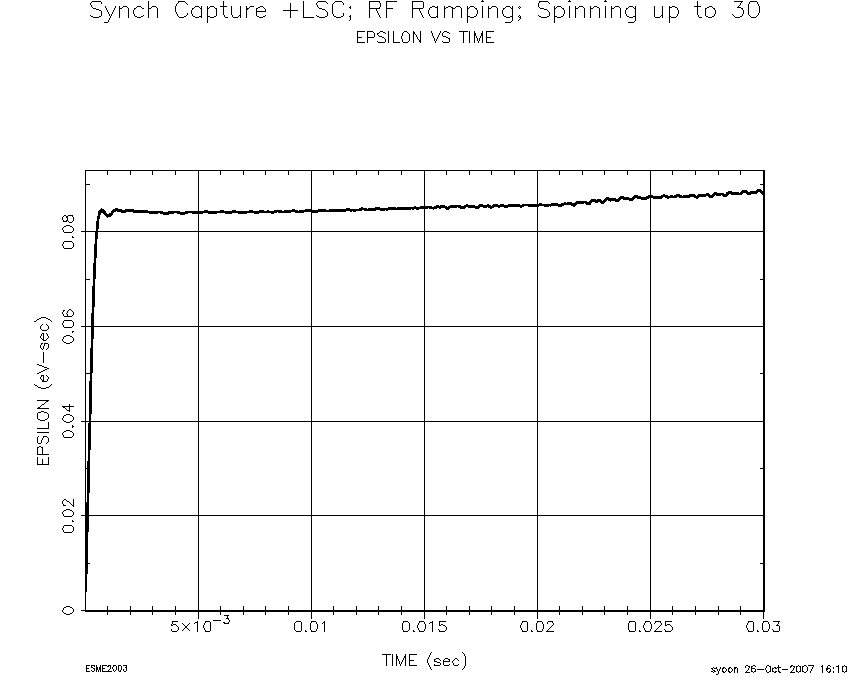}
   \caption{\label{fig:epsl-singlerf-ramp-2700t}
   [\textbf{Scenario II}]~Turn-by-turn evolution of longitudinal emittance 
   ($\epsilon_{l}$) with a single RF system}
\end{figure}

\bigskip
At the end of RF-voltage ramping, 
the entire macro-particles are well captured 
within an enlarged RF bucket as depicted in
Figure~\ref{fig:lps-singlerf-ramp-2700t}.
With RF-voltage ramping, the bucket area is enlarged,
such that beam loss arising from space-charge effect 
can be avoided.
%
%
Figure~\ref{fig:EV-1-singlerf-ramp-2700t} shows 
the variation of RF voltage from 800 (kV) to 1,200 (kV) 
during the interval from the 270$^{th}$ turn through the 2,700$^{th}$ turn.
In comparison to Figure~\ref{fig:theta-singlerf-noramp-270t-2700t} 
from Scenario I, the gradient of charge distribution 
($d\lambda/dz$) are reduced and spread more out 
with ramping RF voltage over extended 2,700 turns 
(cf.~Figure~\ref{fig:phi-singlerf-ramp-2700t}).
Yet, a peaked distribution centered around the origin is observed. 
%
The longitudinal emittance ($\varepsilon_{l}$) 
grows up to 0.085 (eV-s) with a fixed RF voltage at 800 (kV).
Once the RF voltage starts ramping linearly,
the emittance continues to grow gradually up to 0.09 (eV-s)
as plotted in Figure~\ref{fig:epsl-singlerf-ramp-2700t}.
\par In an attempt to further reduce  
the space-charge-induced voltage and 
to attain a more uniform charge distribution, 
a dual RF harmonic system is explored in Scenarios III and IV
in the following subsections.
\black
%
\newpage\clearpage
\section{Scenario III}
\par\bigskip
\fbox{\begin{Bitemize}[t]\itemsep 0.01in
    \item Dual RF Harmonics:\\
          $f_{rf,~1}$~=~53~MHz~and~$f_{rf,~2}$~=~106~MHz
    \item $H_{1}~=~588$ and $H_{2}~=~1176$\\
          $R_{H}$ = $H_{2}/H_{1}$ = 2.0
    \item $V_{rf,~1}~=~400~(kV)$ and $V_{rf,~2}~=~300~(kV)$ 
          (fixed RF voltages)\\
          $R_{H}$ = $V_{rf,~2}/V_{rf,~1}$ = 0.75
\end{Bitemize}}
\par\bigskip
\par
\begin{figure}[h!]\centering
   \subfigure[after 270 turns]
   {\includegraphics[scale=0.35]{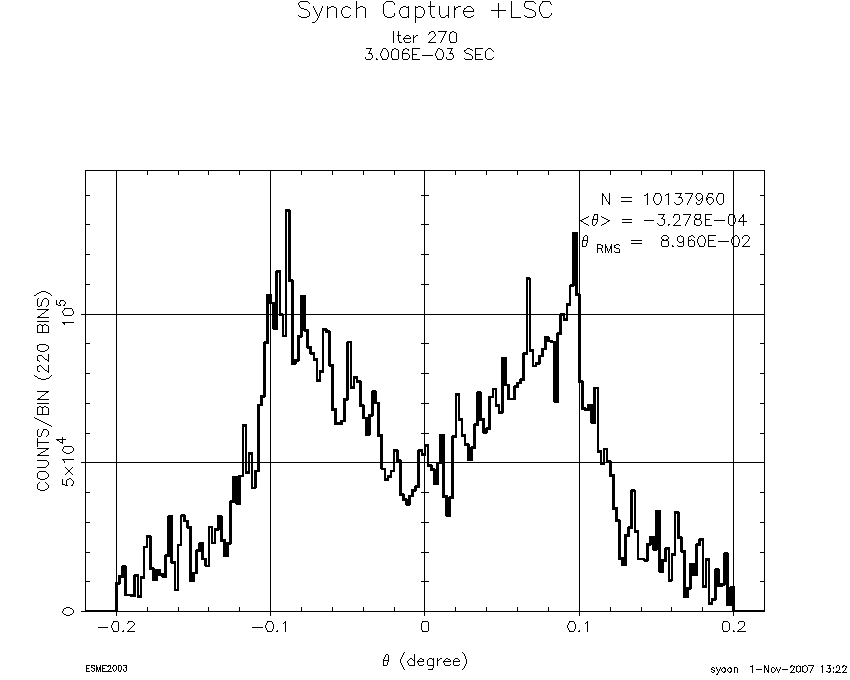}}\vspace{-0.05in}
   \subfigure[after 2,700 turns]
   {\includegraphics[scale=0.35]{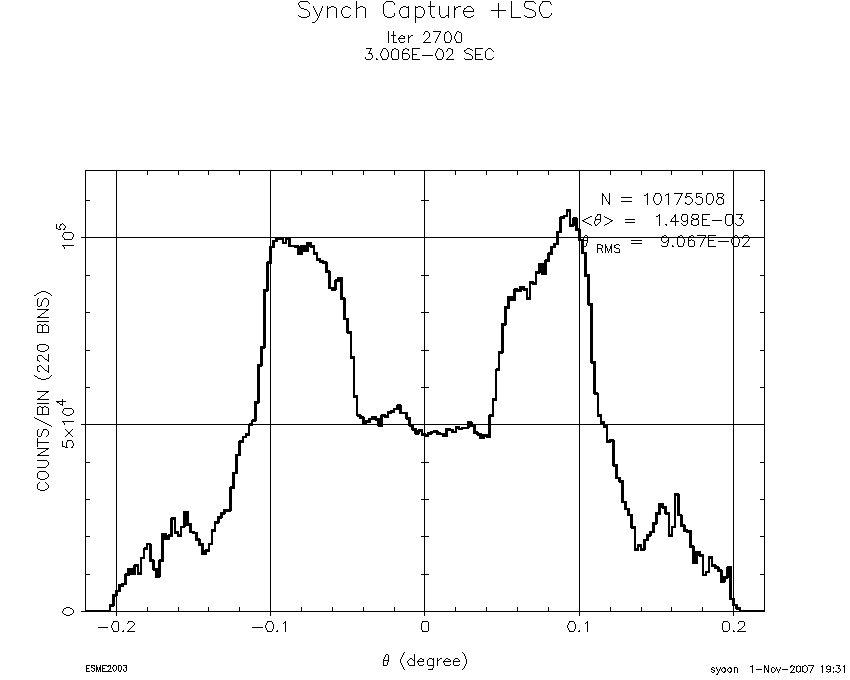}}
   \caption{\label{fig:theta-dualrf-400-300-270t-2700t}
   [\textbf{Scenario III}]~Distribution of charge density with longitudinal painting included}
\end{figure}

\begin{figure}[h!]\centering
      \subfigure[1$^{st}$ turn]
      {\includegraphics[scale=0.15]{lps_dualrf_noramp_400_300_1t_bw.png}}
      \subfigure[100$^{th}$ turn]
      {\includegraphics[scale=0.15]{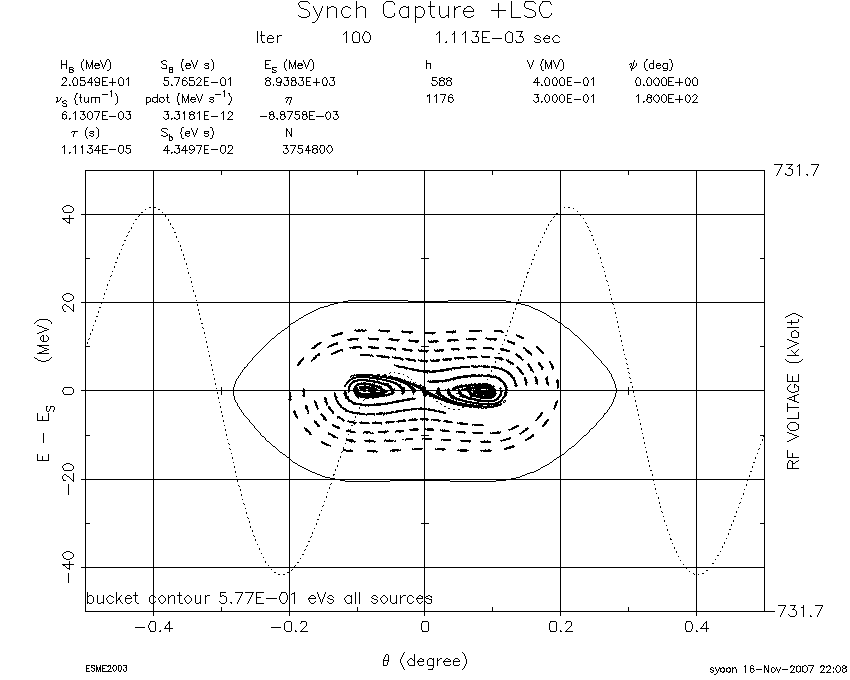}}
      \subfigure[$200^{th}$ turn]
      {\includegraphics[scale=0.15]{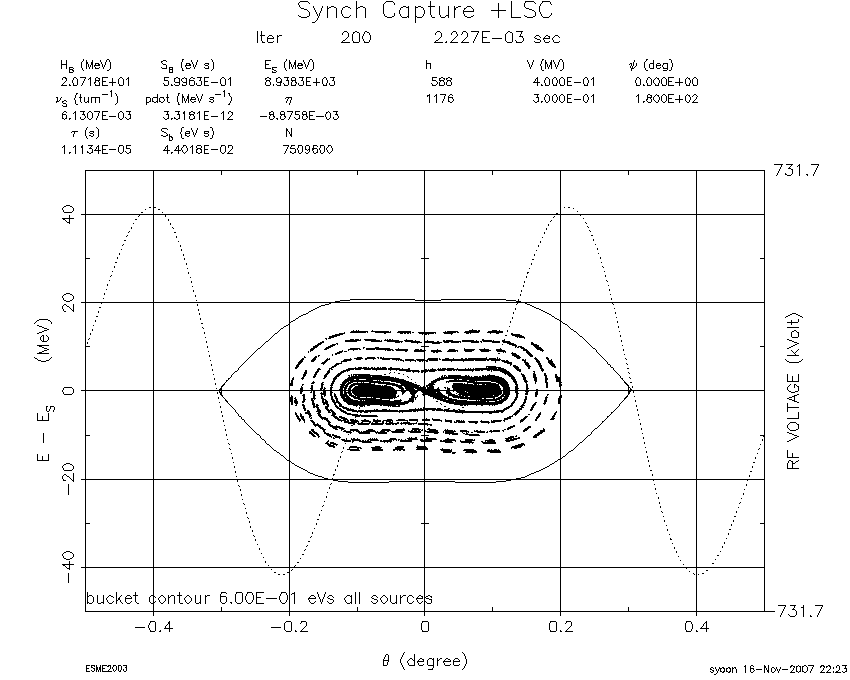}}
      \subfigure[270$^{th} turn$]
      {\includegraphics[scale=0.15]{lps_dualrf_noramp_400_300_270t_bw.png}}
      \subfigure[600$^{th}$ turn]
      {\includegraphics[scale=0.15]{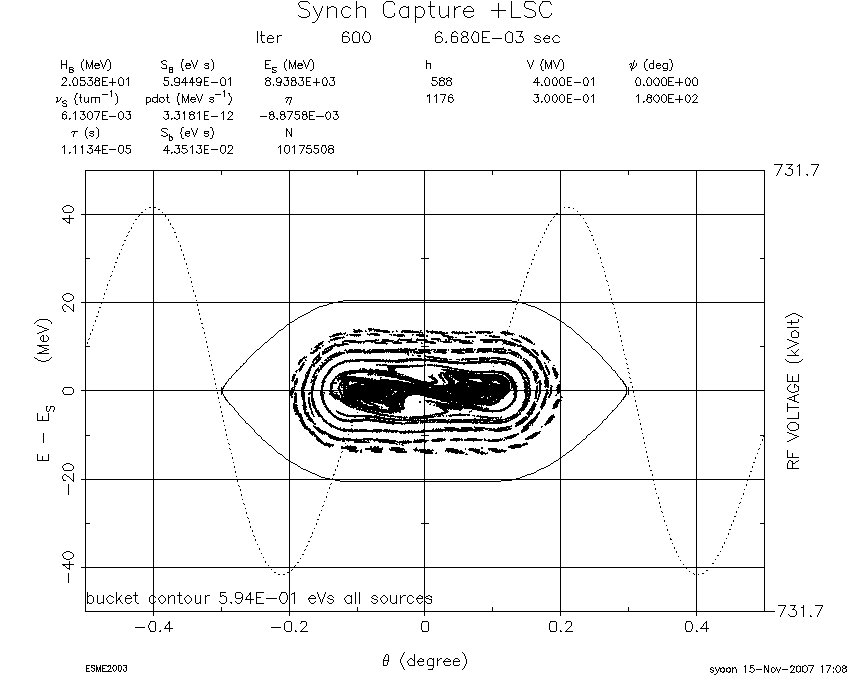}}
      \subfigure[900$^{th}$ turn]
      {\includegraphics[scale=0.15]{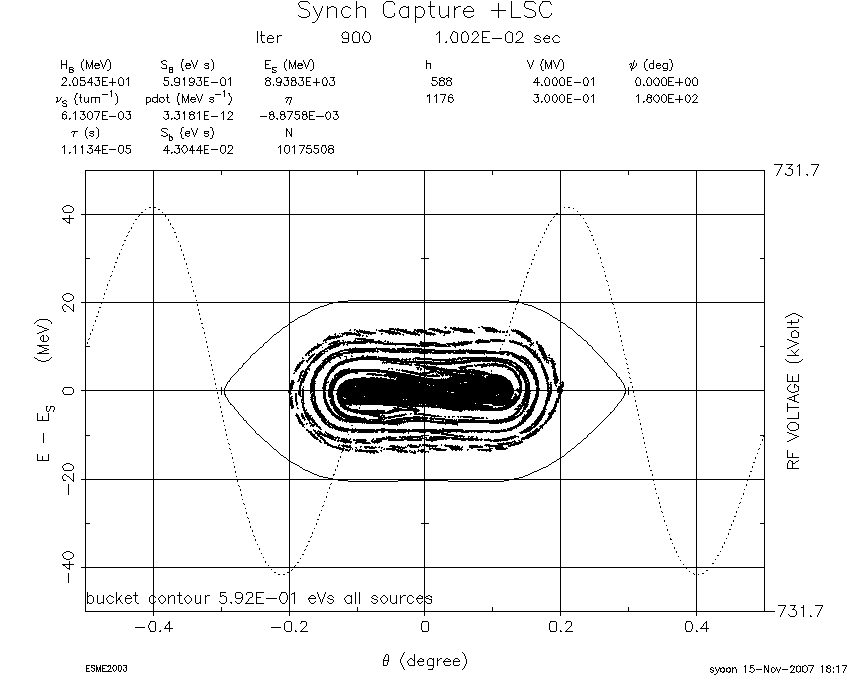}} 
      \subfigure[1,200$^{th}$ turn]
      {\includegraphics[scale=0.15]{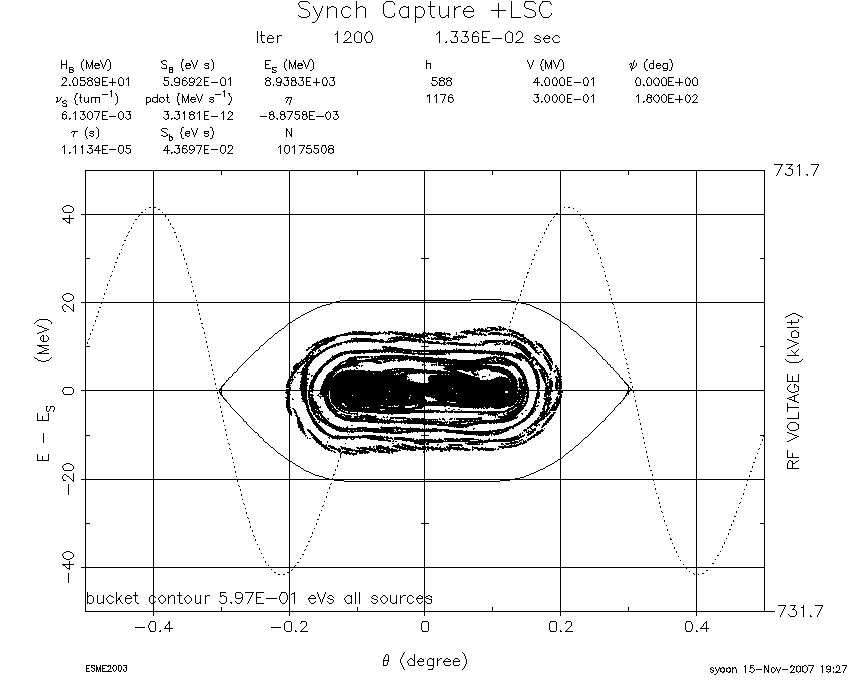}}
      \subfigure[1,500$^{th}$ turn]
      {\includegraphics[scale=0.15]{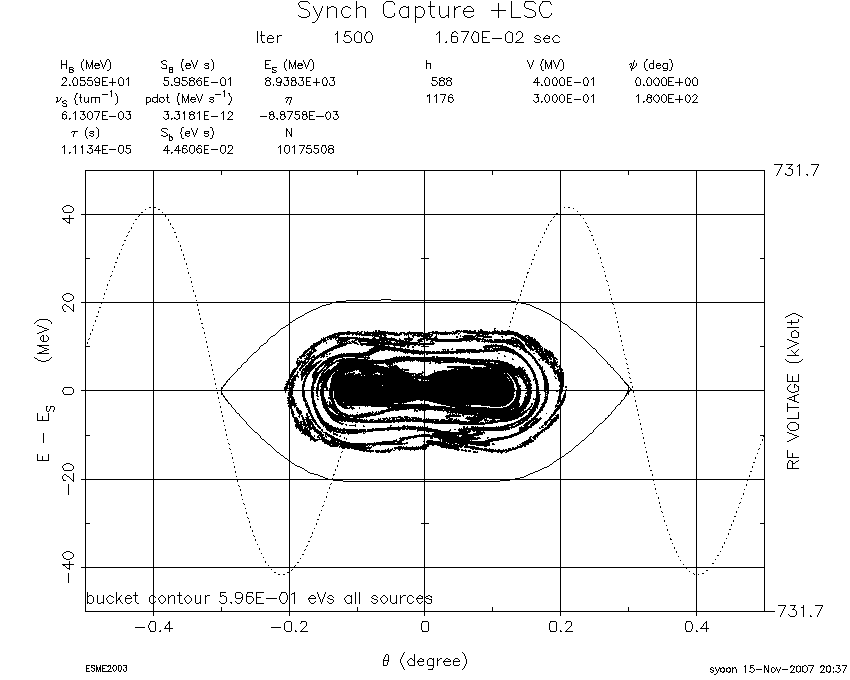}}
      \subfigure[1,800$^{th}$ turn]
      {\includegraphics[scale=0.15]{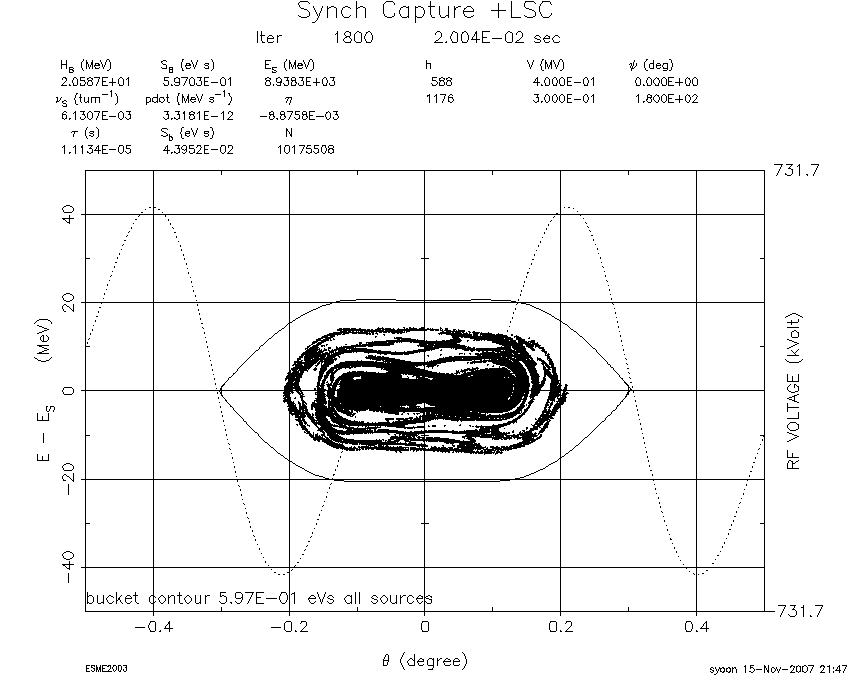}} 
      \subfigure[2,100$^{th}$ turn]
      {\includegraphics[scale=0.15]{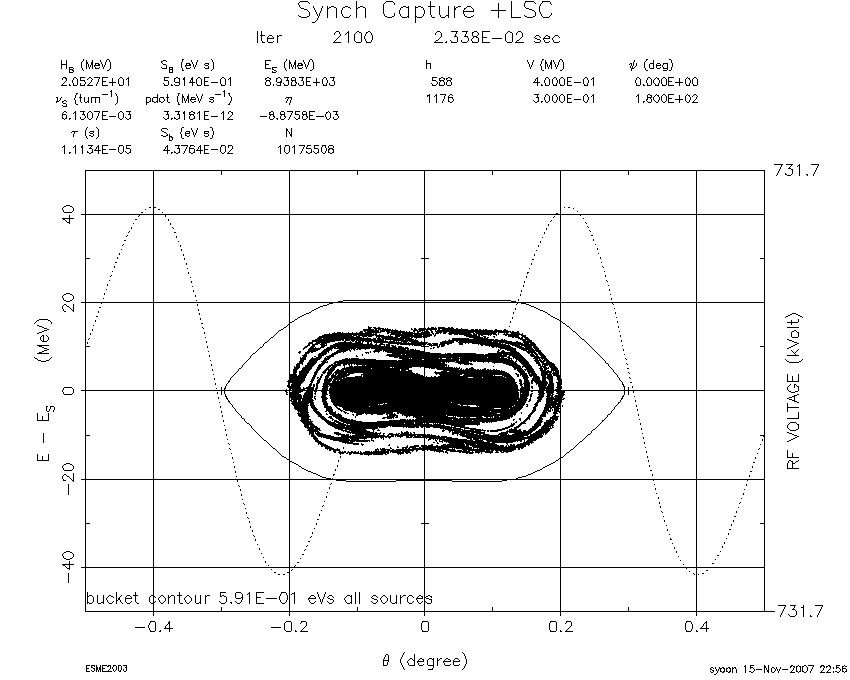}}
      \subfigure[2,400$^{th}$ turn]
      {\includegraphics[scale=0.15]{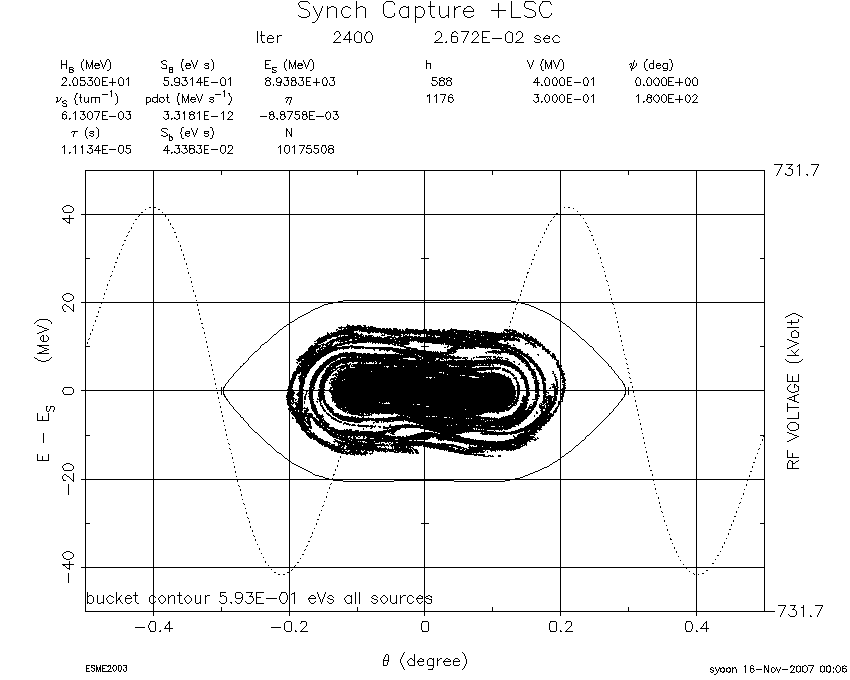}}
      \subfigure[2,700$^{th}$ turn]
      {\includegraphics[scale=0.15]{lps_dualrf_noramp_400_300_2700t_bw.png}} 
   \caption{\label{fig:s3-lps}[\textbf{Scenario III}] 
           Time evolution of phase space with
           longitudinal painting included, starting from 
           the $1^{st}$ injection turn through 2,700 turns}
\end{figure}

\begin{figure}\centering
      \subfigure[100$^{th}$ turn]
      {\includegraphics[scale=0.15]{lps_dualrf_noramp_400_300_100t_bw.png}}
      \subfigure[charge density]
      {\includegraphics[scale=0.15]{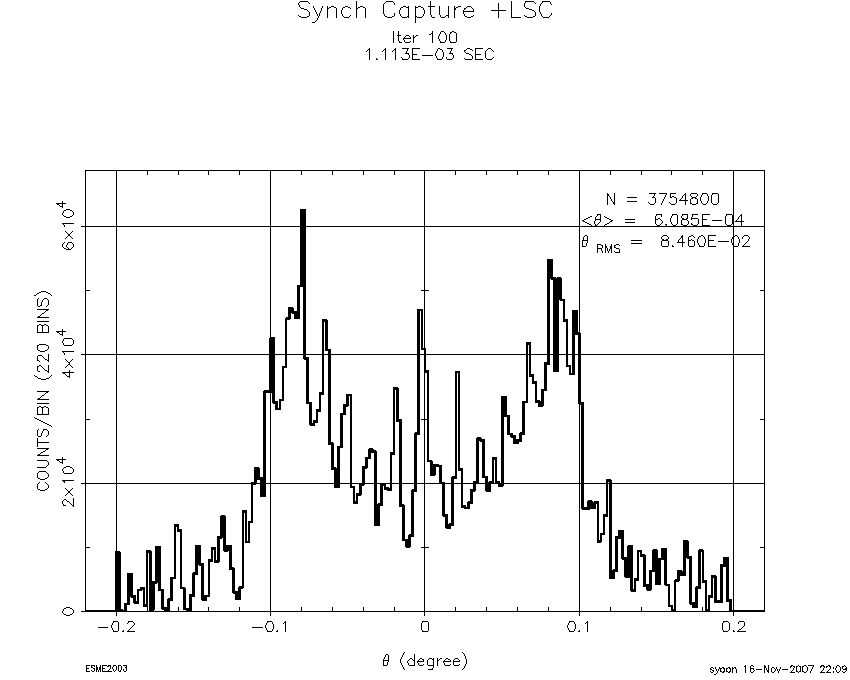}}
      \subfigure[energy density]
      {\includegraphics[scale=0.15]{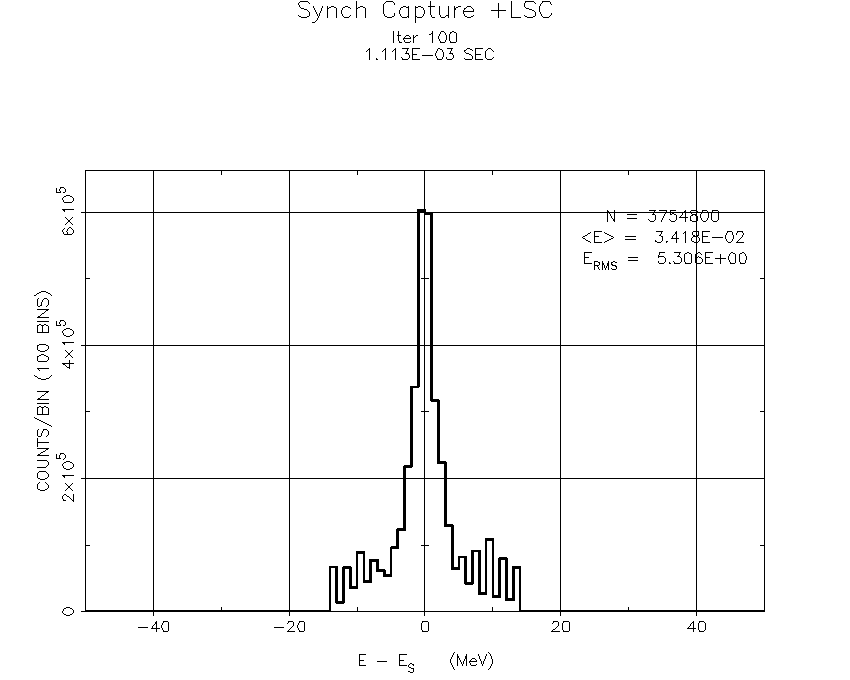}}
      \subfigure[200$^{th}$ turn]
      {\includegraphics[scale=0.15]{lps_dualrf_noramp_400_300_200t_bw.png}}
      \subfigure[charge density]
      {\includegraphics[scale=0.15]{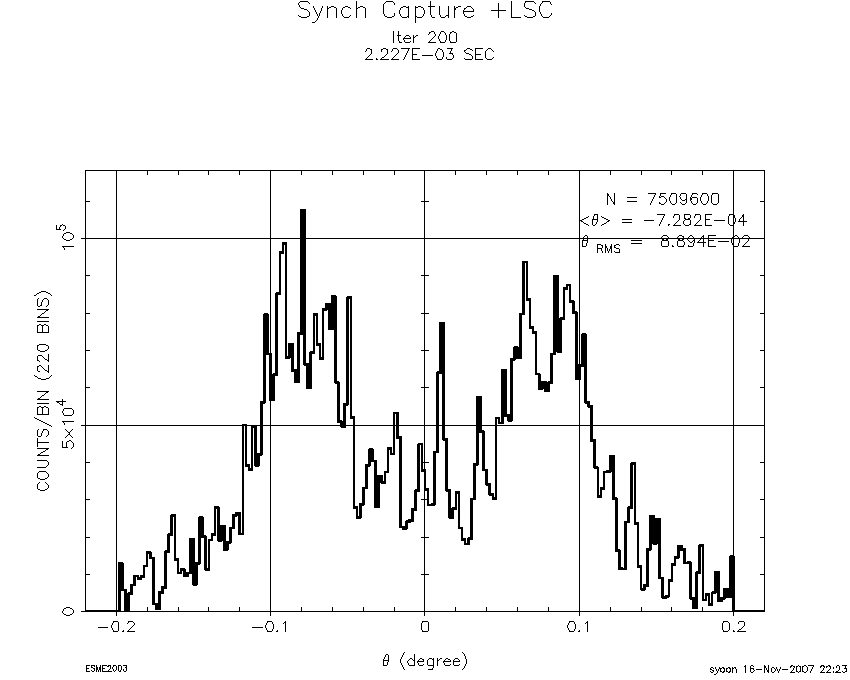}}
      \subfigure[energy density]
      {\includegraphics[scale=0.15]{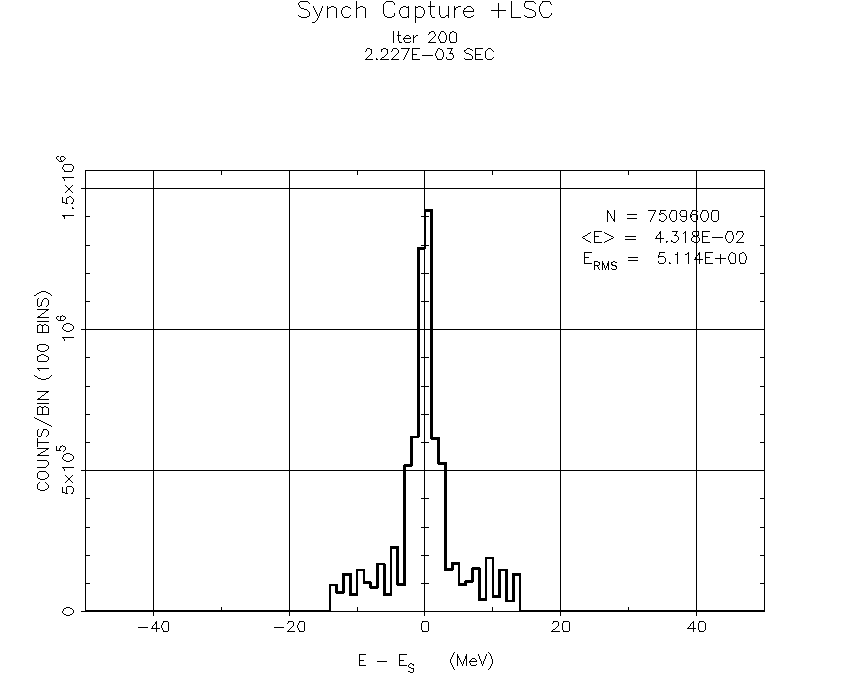}}  
      \subfigure[270$^{th}$ turn]
      {\includegraphics[scale=0.15]{lps_dualrf_noramp_400_300_270t_bw.png}}
      \subfigure[charge density]
      {\includegraphics[scale=0.15]{theta_dualrf_noramp_400_300_270t_bw.png}}
      \subfigure[energy density]
      {\includegraphics[scale=0.15]{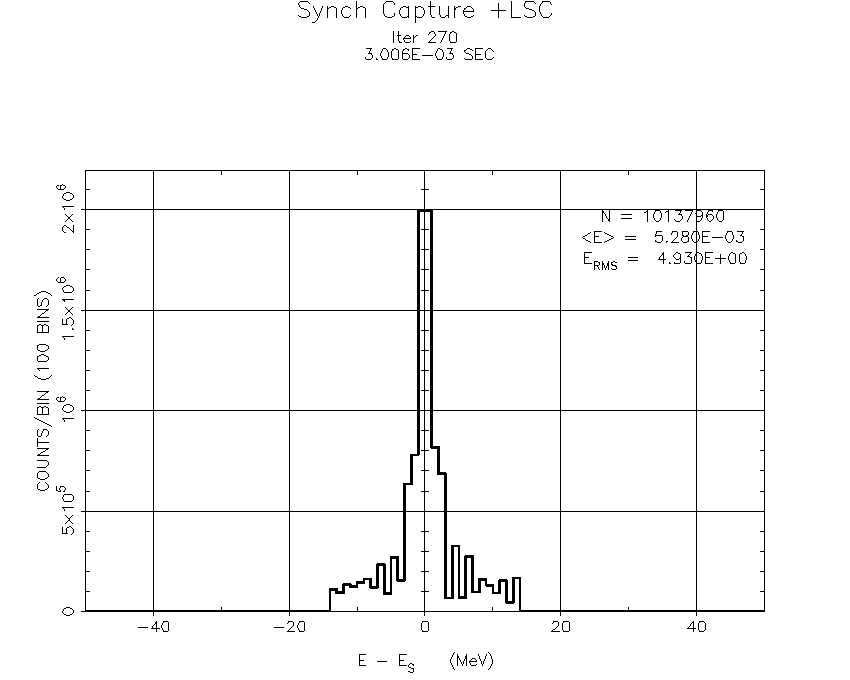}}  
      \subfigure[300$^{th}$ turn]
      {\includegraphics[scale=0.15]{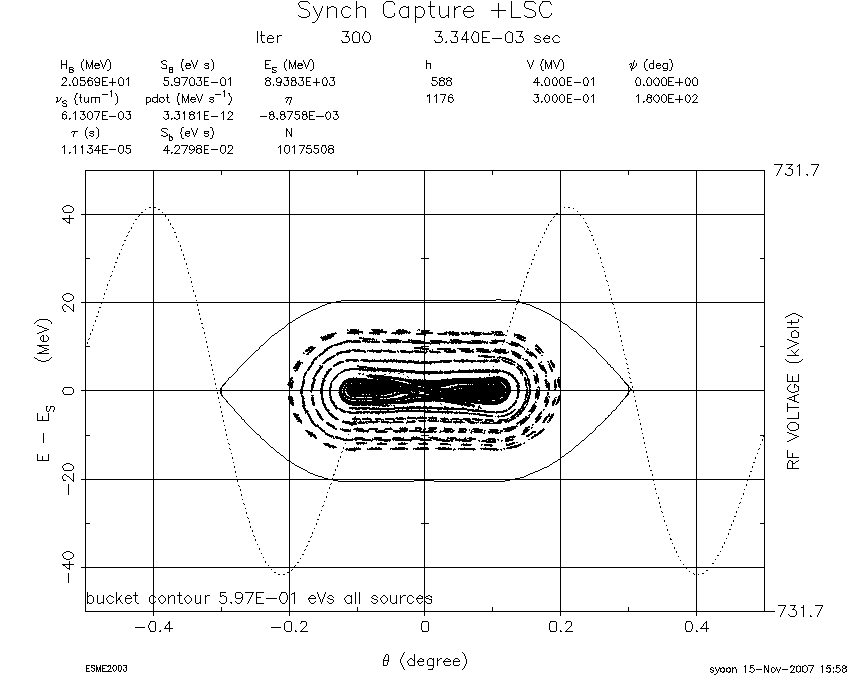}}
      \subfigure[charge density]
      {\includegraphics[scale=0.15]{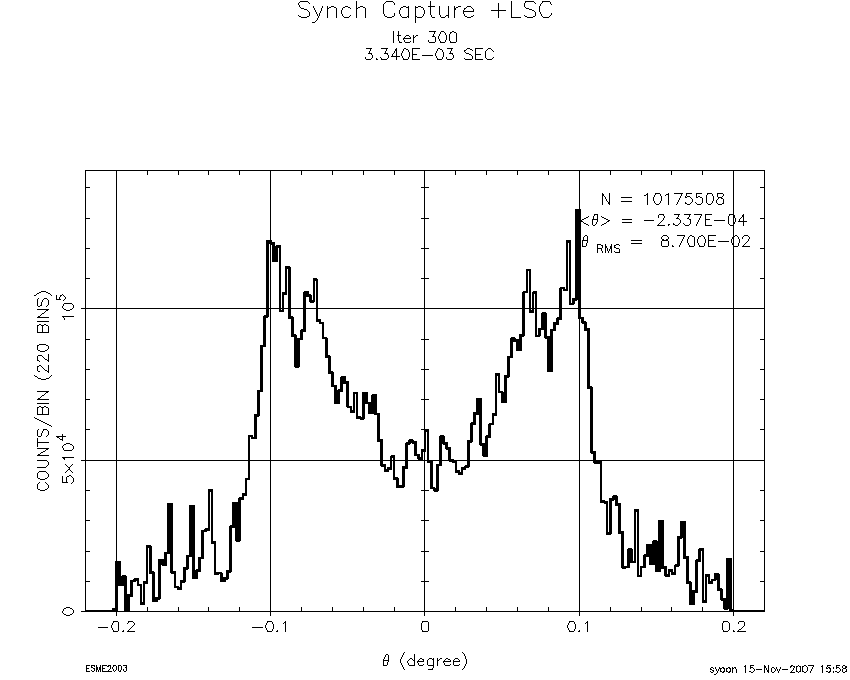}}
      \subfigure[energy density]
      {\includegraphics[scale=0.15]{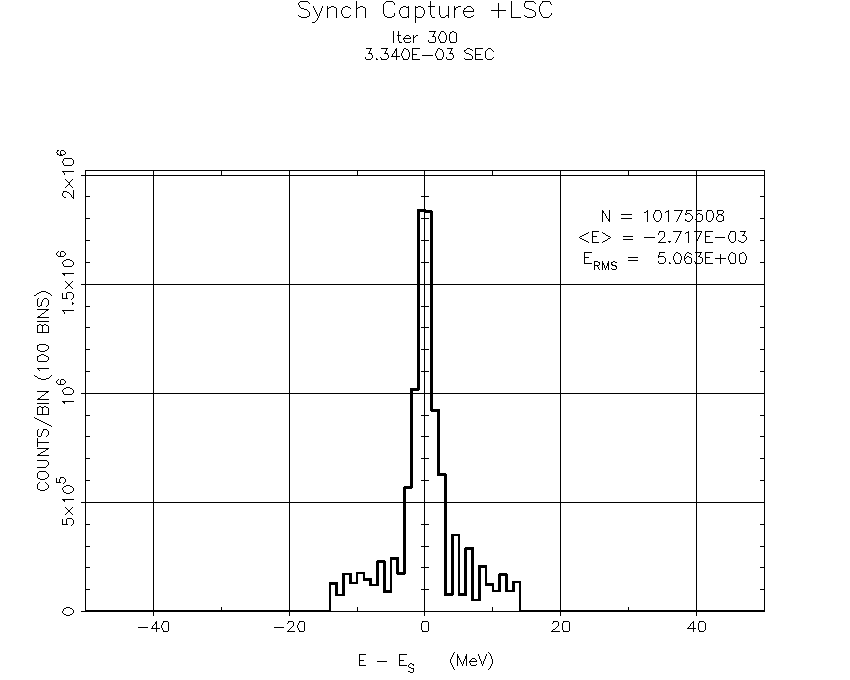}}
   \caption{\label{fig:s3-lps-theta-dE-300}[\textbf{Scenario III}] 
           Time evolution of phase space with
           longitudinal painting starting from the 
           $100^{th}$ turn through the $300^{th}$ turns}
\end{figure}

\begin{figure}
   \begin{center}
      \subfigure[600$^{th}$ turn]
      {\includegraphics[scale=0.15]{lps_dualrf_noramp_400_300_600t_bw.png}}
      \subfigure[charge density]
      {\includegraphics[scale=0.15]{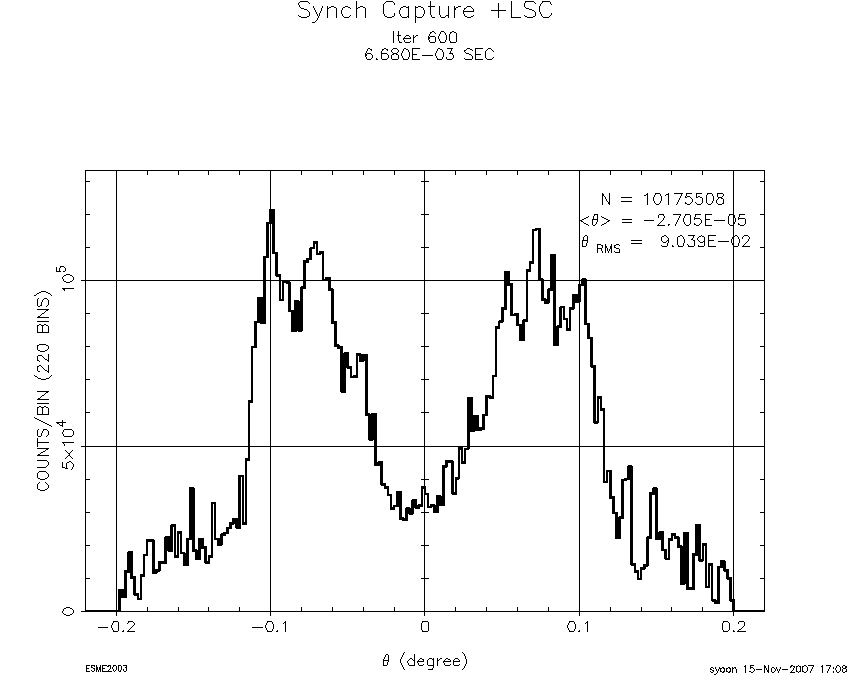}}
      \subfigure[energy density]
      {\includegraphics[scale=0.15]{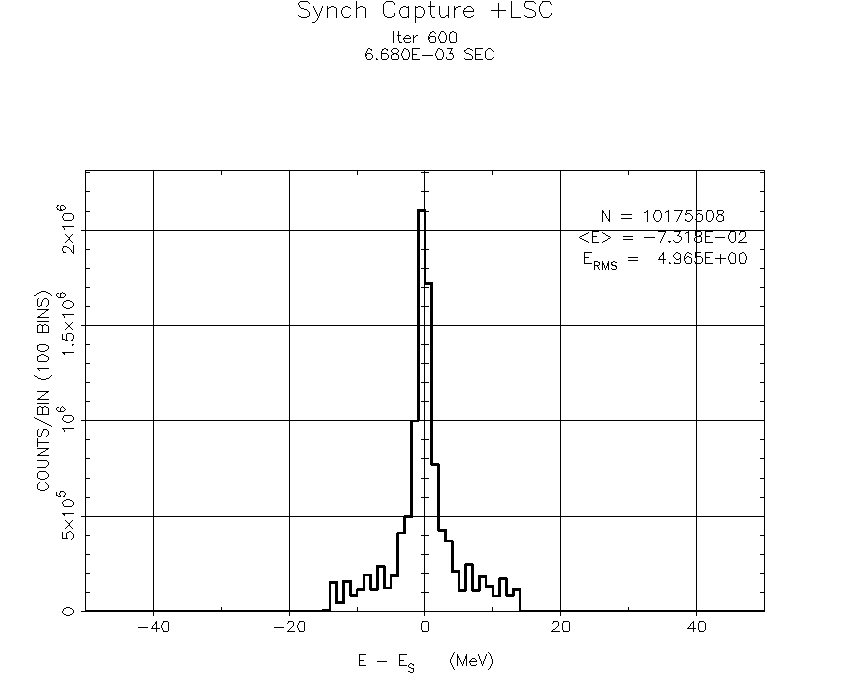}}
      \subfigure[900$^{th}$ turn]
      {\includegraphics[scale=0.15]{lps_dualrf_noramp_400_300_900t_bw.png}}
      \subfigure[charge density]
      {\includegraphics[scale=0.15]{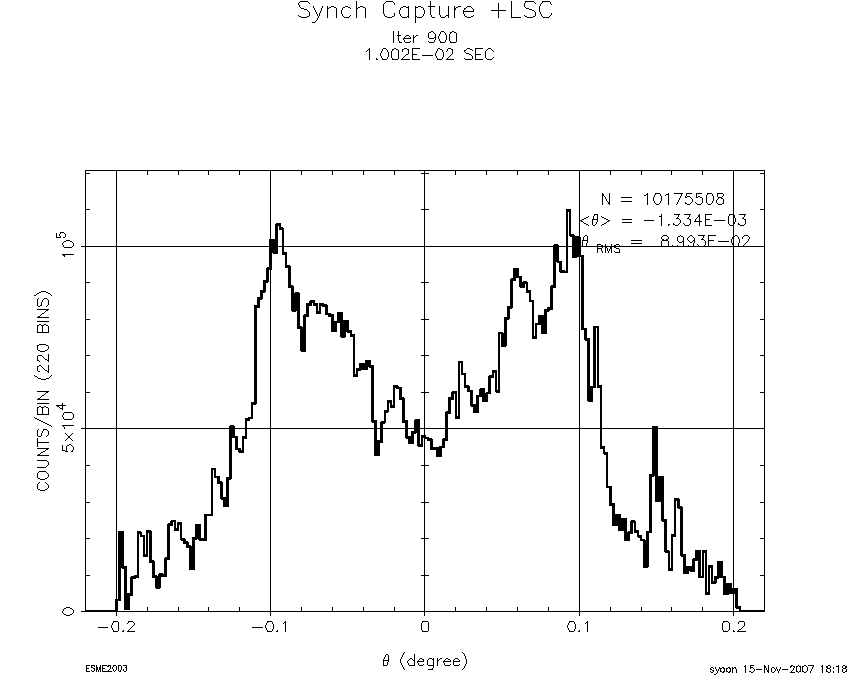}}
      \subfigure[energy density]
      {\includegraphics[scale=0.15]{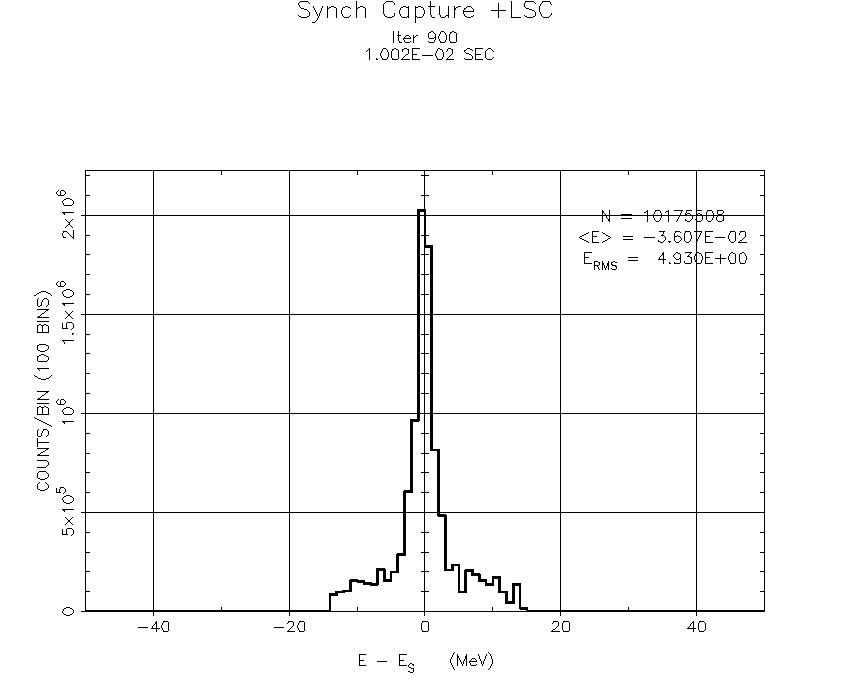}}  
      \subfigure[1,200$^{th}$ turn]
      {\includegraphics[scale=0.15]{lps_dualrf_noramp_400_300_1200t_bw.png}}
      \subfigure[charge density]
      {\includegraphics[scale=0.15]{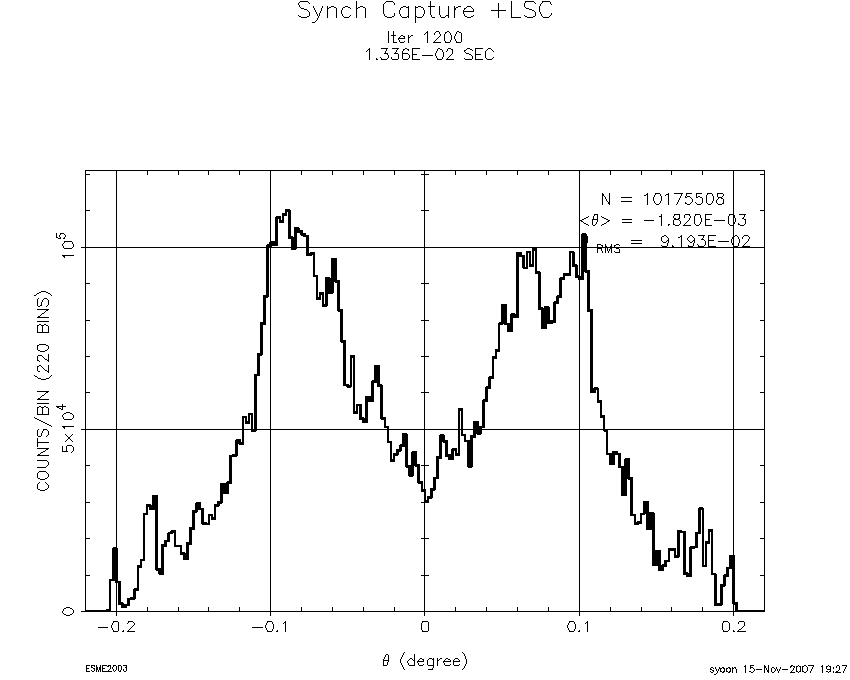}}
      \subfigure[energy density]
      {\includegraphics[scale=0.15]{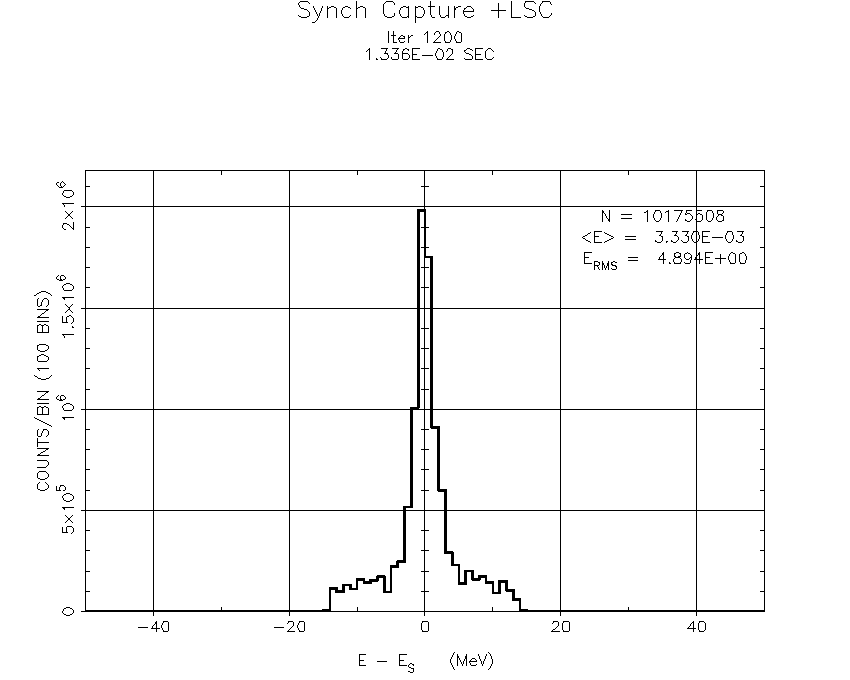}}  
      \subfigure[1,500$^{th}$ turn]
      {\includegraphics[scale=0.15]{lps_dualrf_noramp_400_300_1500t_bw.png}}
      \subfigure[charge density]
      {\includegraphics[scale=0.15]{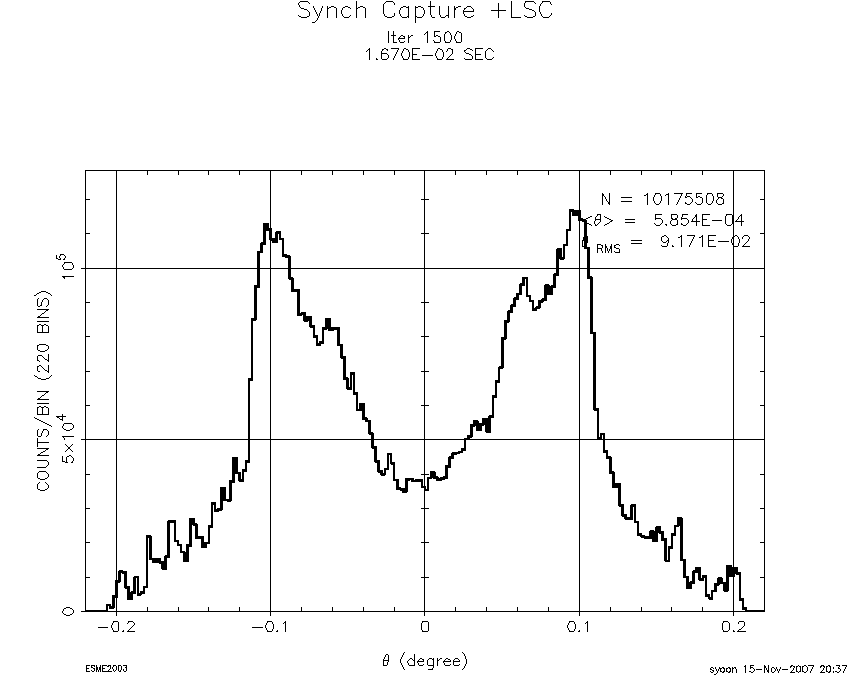}}
      \subfigure[energy density]
      {\includegraphics[scale=0.15]{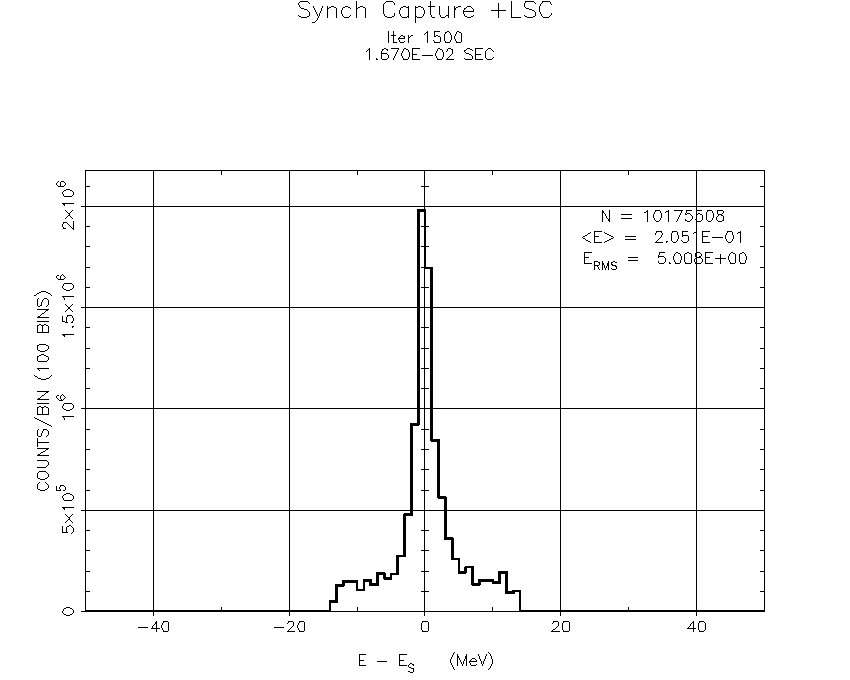}}
   \end{center}  
   \caption{\label{fig:s3-lps-theta-dE-1500}[\textbf{Scenario III}] 
           Time evolution of phase space with
           longitudinal painting starting from 
           the 600$^{th}$ turn through the 1,500$^{th}$ turns}
\end{figure}

\begin{figure}
   \begin{center}
      \subfigure[1,800$^{th}$ turn]
      {\includegraphics[scale=0.15]{lps_dualrf_noramp_400_300_1800t_bw.png}}
      \subfigure[charge density]
      {\includegraphics[scale=0.15]{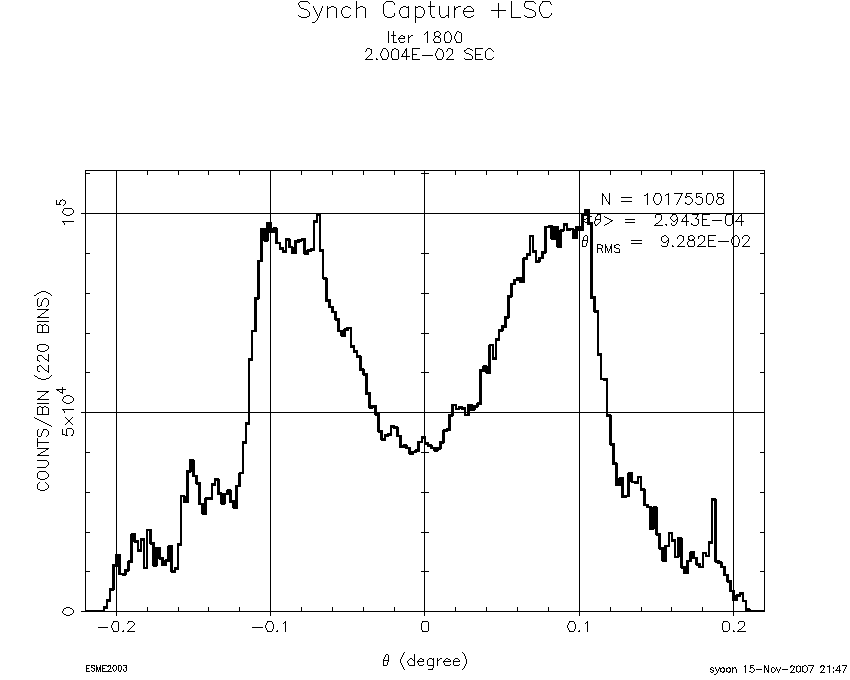}}
      \subfigure[energy density]
      {\includegraphics[scale=0.15]{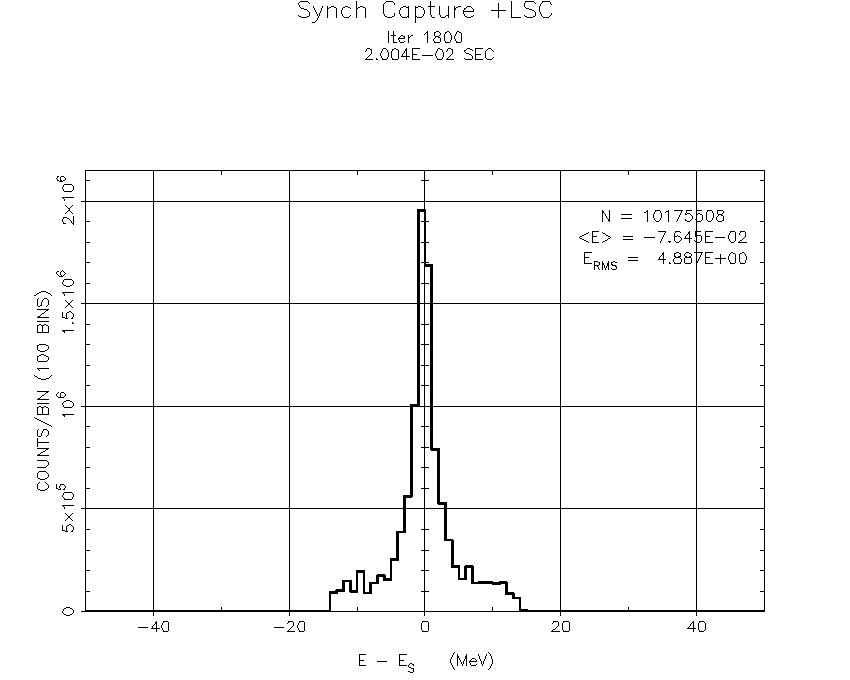}}
      \subfigure[2,100$^{th}$ turn]
      {\includegraphics[scale=0.15]{lps_dualrf_noramp_400_300_2100t_bw.png}}
      \subfigure[charge density]
      {\includegraphics[scale=0.15]{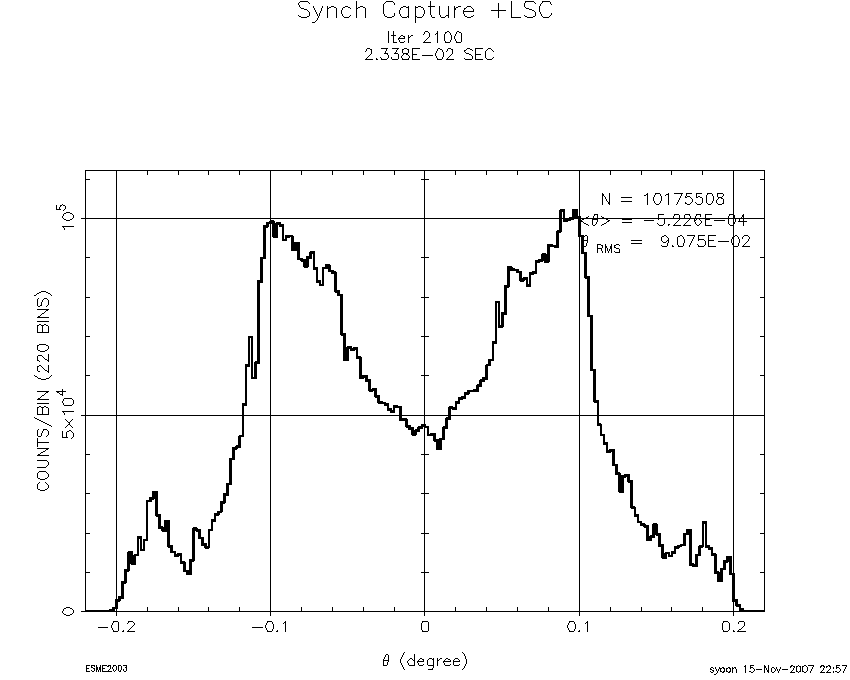}}
      \subfigure[energy density]
      {\includegraphics[scale=0.15]{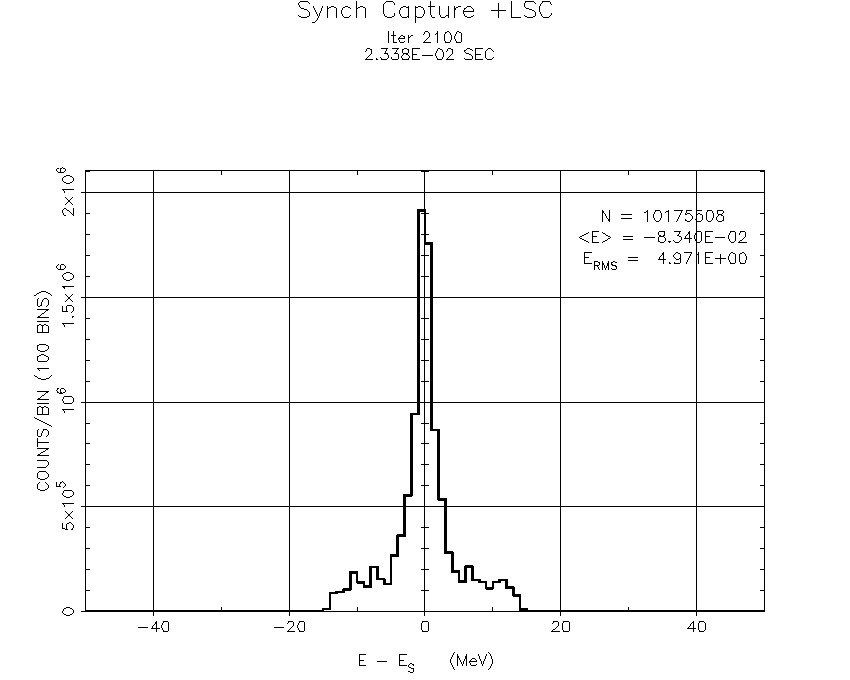}}  
      \subfigure[2,400$^{th}$ turn]
      {\includegraphics[scale=0.15]{lps_dualrf_noramp_400_300_2400t_bw.png}}
      \subfigure[charge density]
      {\includegraphics[scale=0.15]{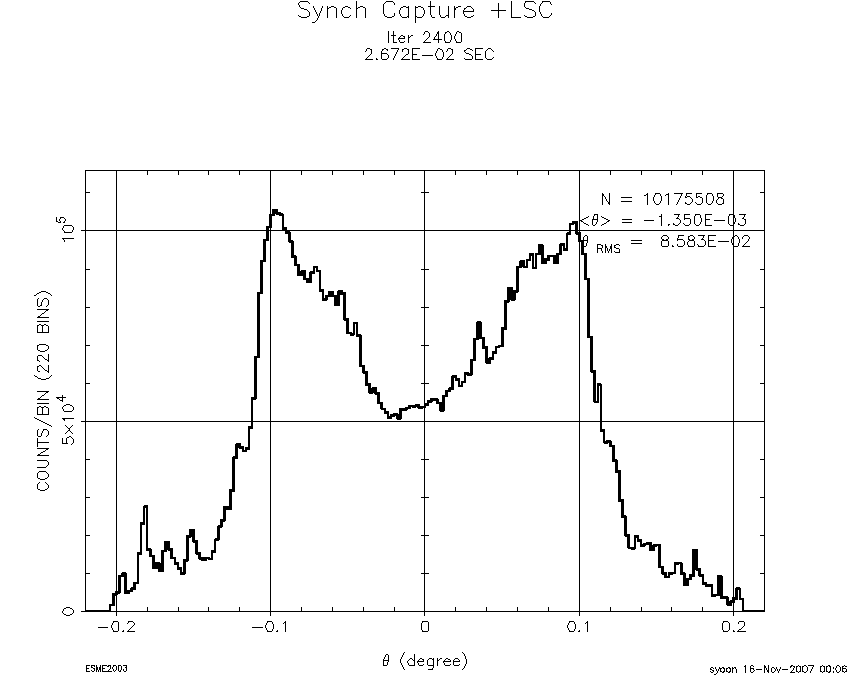}}
      \subfigure[energy density]
      {\includegraphics[scale=0.15]{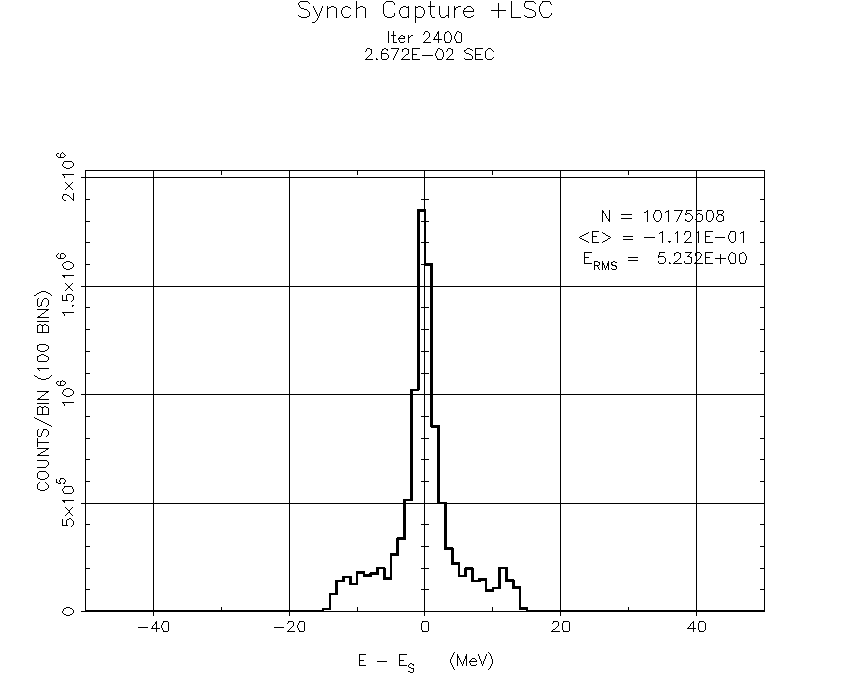}}  
      \subfigure[2,700$^{th}$ turn]
      {\includegraphics[scale=0.15]{lps_dualrf_noramp_400_300_2700t_bw.png}}
      \subfigure[charge density]
      {\includegraphics[scale=0.15]{theta_dualrf_noramp_400_300_2700t_bw.png}}
      \subfigure[energy density]
      {\includegraphics[scale=0.15]{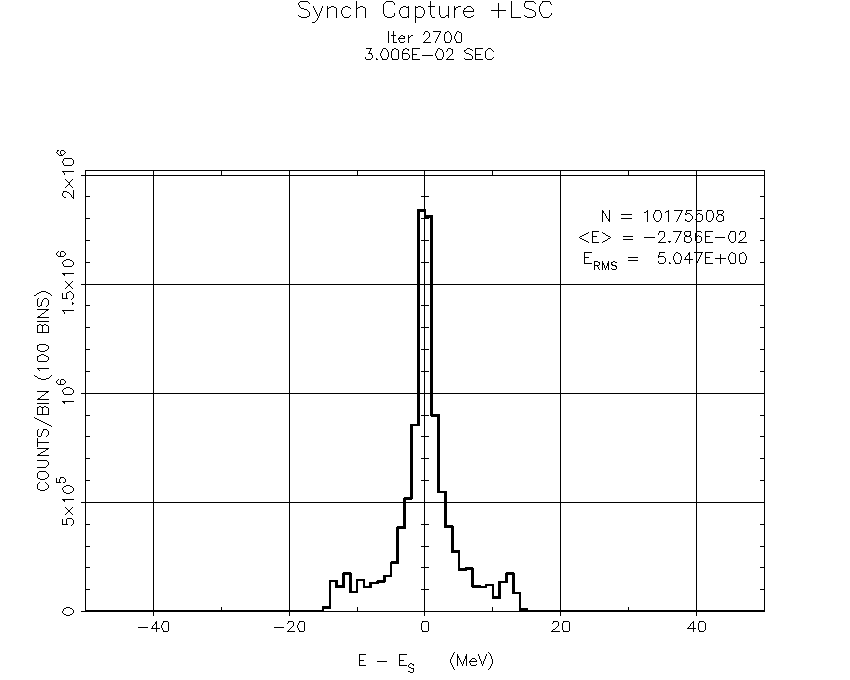}}
   \end{center}  
   \caption{\label{fig:s3-lps-theta-dE-2700}[\textbf{Scenario III}] 
           Time evolution of phase space with
           longitudinal painting starting from the 
           1,800$^{th}$ turn through the 2,700$^{th}$ turns}
\end{figure}

 
In Scenario III, we explored a scheme of injecting micro-bunches 
into an MI RF bucket with dual RF harmonics
of 400 (kV) on 53 (MHz) and 300 (kV) on 106 (MHz).
The higher-order RF harmonic ($H_{2} = 1176$)
is as double as the principal RF harmonic ($H_{1} = 588$).
The secondary RF voltage ($V_{rf,~2}$) is 75$\%$ of
the principal RF voltage ($V_{rf,~1}$).
The RF voltage waveform of the higher harmonic 
for dual RF harmonics can be obtained 
on the basis of Eqn.~(\ref{eqn:dualrf-voltage}):
\begin{equation}\label{eqn:dualrf-voltage}
   \mathcal{V}
   = \mathcal{V}_{rf,~1}\sin(H_{1}\phi_{1} + \psi_{1})
   + \mathcal{V}_{rf,~2}\sin(H_{2}\phi_{2} + \psi_{2}),
\end{equation}
where $\mathcal{V}_{rf}$, $H$, $\phi$, and $\psi$
are RF voltage, harmonic number, phase angle of each macro-particle,
and phase of RF cavity, respectively.
As can be seen in Figure~\ref{fig:dualrf-waveform-400-300},
the waveform has a negative slope
around the stable phase of 0 (deg).
By adding a higher secondary harmonic RF voltage
to a principal harmonic RF voltage, 
a flat-bottom potential energy can be produced.
With the dual RF system, 
the RF-bucket contour is also flattened
at the top and bottom on a phase-space plot,
resulting in a flattened charge distribution.
Besides, the principal RF voltage
can be half as high as the RF voltage
used for the case of single RF harmonic.
%
%
Since the flat-bottom potential energy
can iron out the peaked charge distribution,
the utilization of a dual RF system offers
a great advantage over the single RF harmonic.
Eventually, the dual RF system helps to lower 
the beam-current limits set by space-charge effects.
After 2,700 turns, the formation of localized macro-particle 
distributions is observed around three local bumps of the 
dual-harmonic voltage waveform.
The time evolution of injected macro-particles inside 
a dual RF bucket is shown in Figure~\ref{fig:theta-dualrf-400-300-270t-2700t}.
From turn 1 through turn 270, trains of four micro-bunches are injected with 
parasitic phase offsets. After the injection process is complete, 
in order to reach an equilibrium state, injected macro-particles are circulated 
up to 2,700 turns with no further injection. 
In Figure~\ref{fig:s3-lps}, trains of injected micro-bunches captured 
in a dual RF bucket drawn on a phase-space plot at every 100 turns 
starting from the 1$^{st}$ turn through the 2,700$^{th}$ turns.
Note that as time elapses after the completion of injection process,
sporadic dips in between lumps of macro-particles are gradually disappeared.
In Figures from \ref{fig:s3-lps-theta-dE-300} through \ref{fig:s3-lps-theta-dE-2700} 
three plots of longitudinal phase space, charge density, 
and energy density comprise each row corresponding to 
turn number indicated below each phase-space plot.
With the careful choice of RF voltages ($R_{V} = 0.75$), 
a bi-modal charge distribution is created due to 
a pair of potential wells around the stable phase of 0 (deg).
As the turn number increments, 
longitudinal painting progresses in a parasitic fashion,
thus leading up to a continuous bi-modal charge distribution.
\black
\begin{figure}[h!]\centering
      \subfigure[Dual RF bucket]{\includegraphics[scale=0.27]{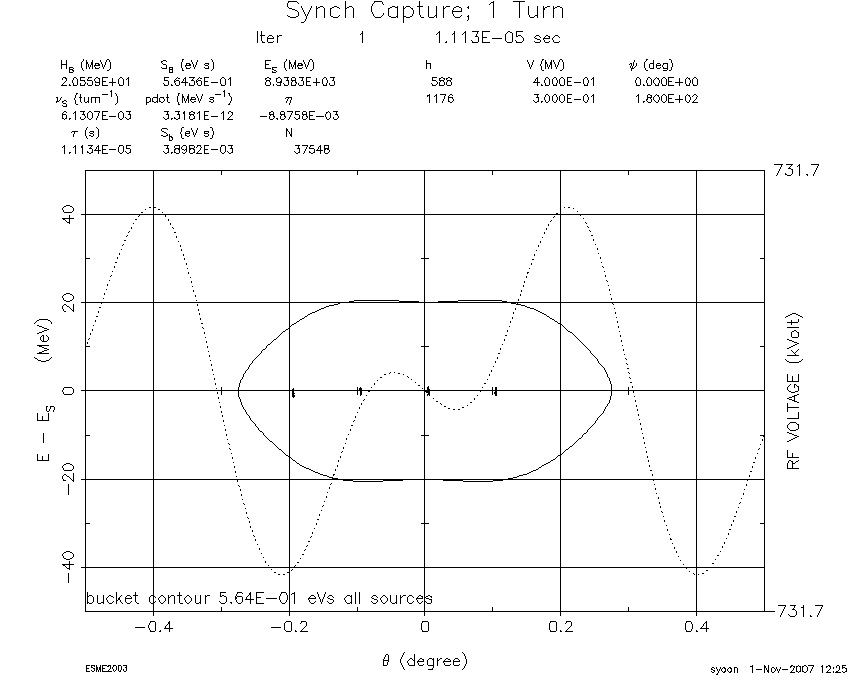}}
      \subfigure[charge distribution]{\includegraphics[scale=0.27]{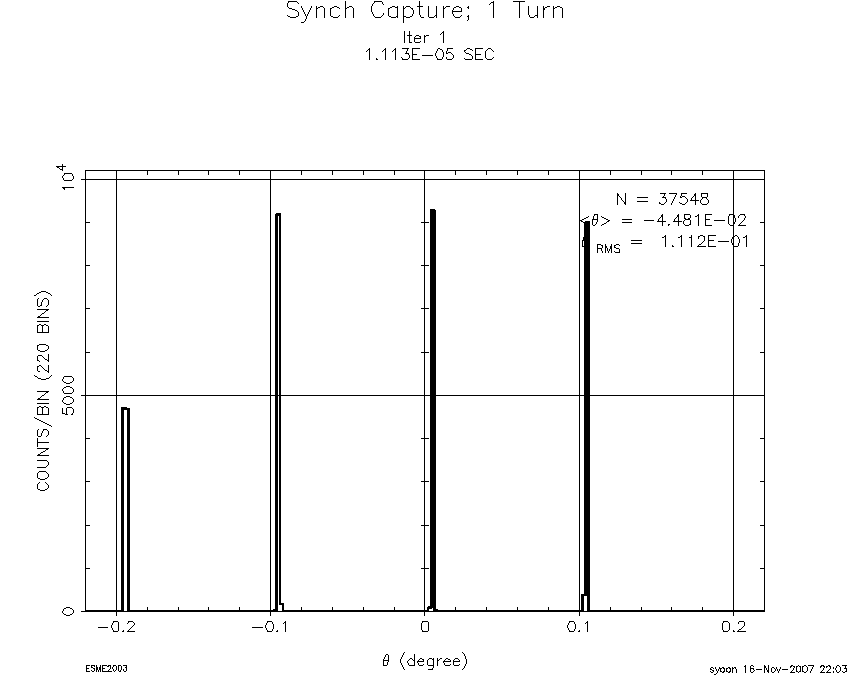}} 
      \caption{\textbf{[Scenario III]}~(a) Dual RF bucket capturing the first train of micro-bunches 
               and its RF waveform in the background (b) Its discrete charge distribution\label{fig:dualrf-waveform-400-300}}
\end{figure}
\newpage\clearpage
\begin{figure}[h!]\centering
   \subfigure[after 270 turns]{\includegraphics[scale=0.35]{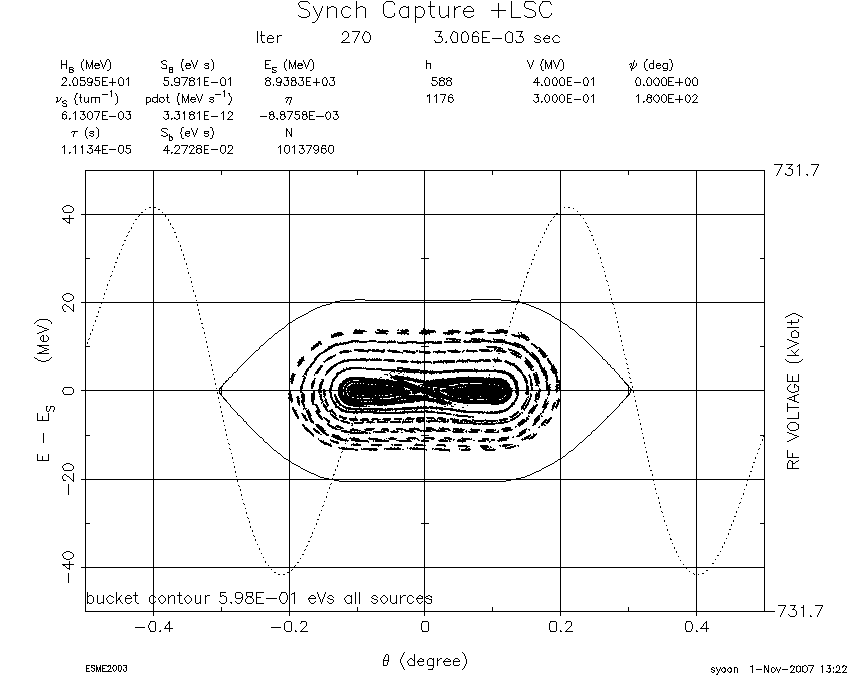}}\vspace{-0.05in}
   \subfigure[after 2,700 turns]{\includegraphics[scale=0.35]{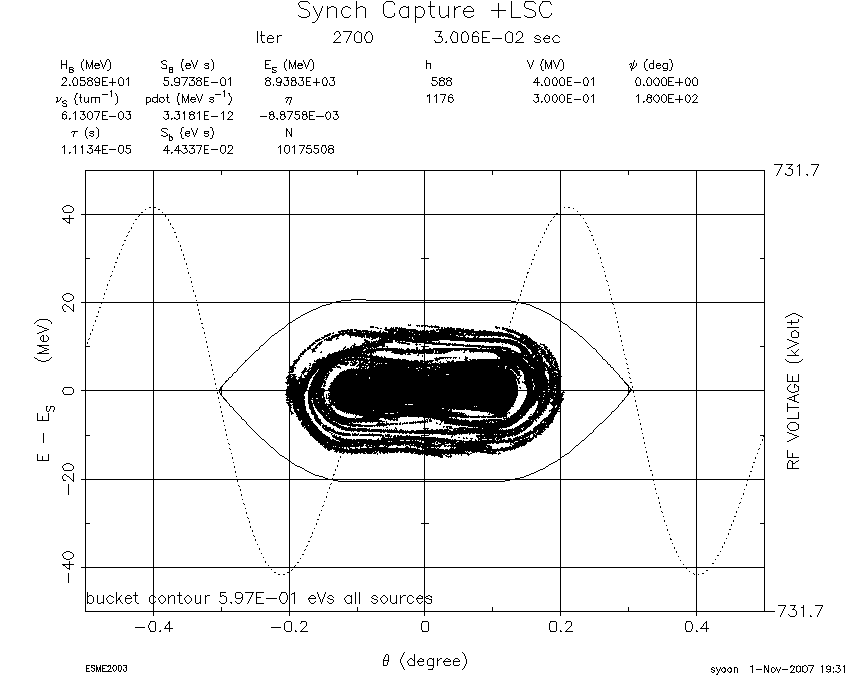}}
   \caption{\label{fig:lps-dualrf-400-300-270t-2700t}
   [\textbf{Scenario III}]~Micro-bunches captured within the dual-RF MI bucket
   and progressing longitudinal painting after 270 turns and 2700 turns
   \label{fig:dualrf-waveform-400-300}}
\end{figure}
\newpage\clearpage
\newpage\clearpage
%
\newpage\clearpage
\clearpage
\newpage\clearpage
\section{Scenario IV: \emph{Optimized MI RF}}
\smallskip
\fbox{\begin{Bitemize}[t]\itemsep 0.01in
    \item Dual RF Harmonics:\\
        $f_{rf,~1}~=~53~MHz$ and $f_{rf,~2} = 106~MHz$\\
        $H_{2} = 1176$ and $H_{1} = 588$\\
        $R_{H} = f_{rf,~2}/f_{rf,~1} = 2.0$
    \item $V_{rf,~1}$ = 400 (kV) and $V_{rf,~2}$ = 200 (kV)
          (fixed RF voltages)
\end{Bitemize}}
\bigskip
\par
\begin{figure}[h!]\centering
     \subfigure[at the first turn\label{fig:s4_lps_first}]
               {\includegraphics[scale=0.28]{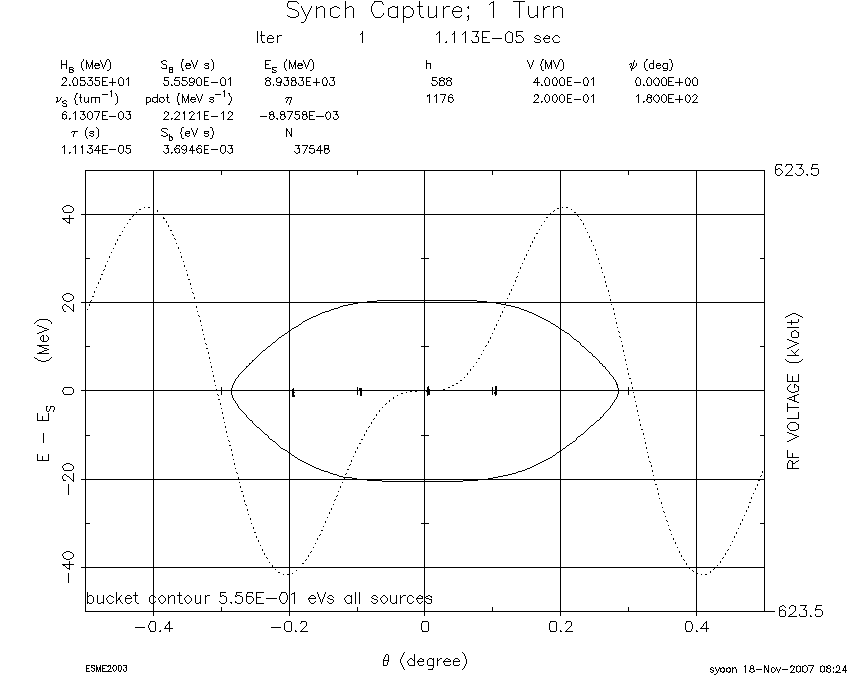}}
     \subfigure[at the 270$^{th}$ turn\label{fig:s4_lps_270}]
               {\includegraphics[scale=0.28]{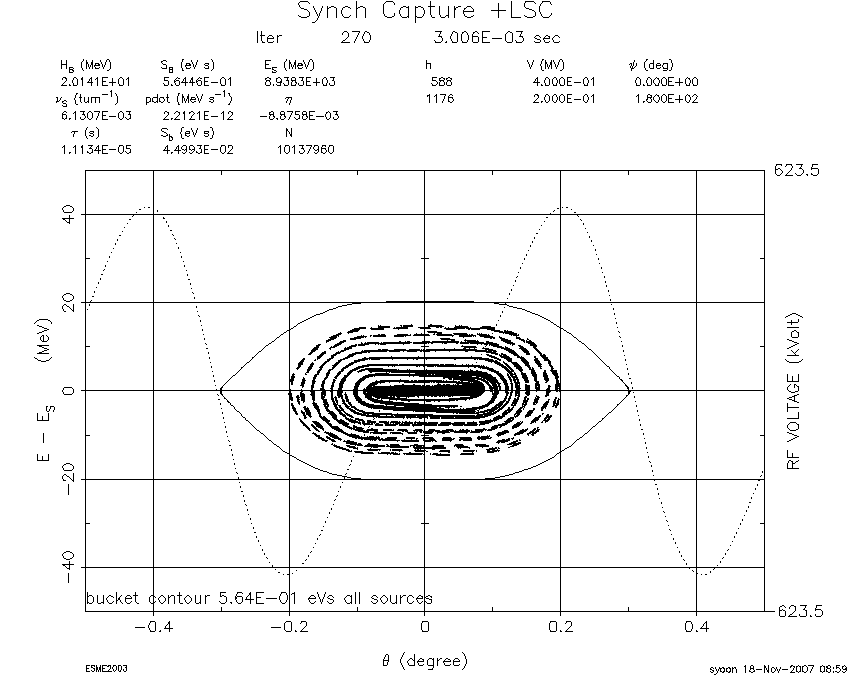}}
     \subfigure[at the 2,700$^{th}$ turn\label{fig:s4_lps_2700}]
               {\includegraphics[scale=0.28]{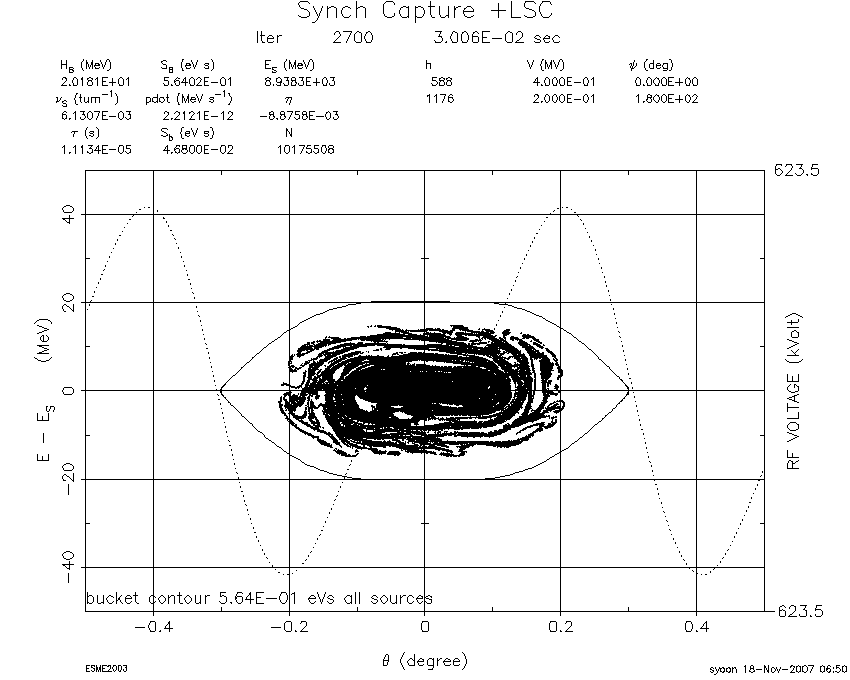}}
     \caption{\label{fig:dualrf-400-200-270t-2700t}
     [\textbf{Scenario IV}]~Optimized Dual RF System: 
     Synchronous injection of micro-bunches
     (a) at the 1$^{st}$ turn (b) at the 270$^{th}$ turn, 
     and (c) at the 2,700$^{th}$ turn\black}
\end{figure}

\begin{figure}[h!]\centering
   \subfigure[at the 270$^{th}$ turn\label{fig:s4_theta_270}]
             {\includegraphics[scale=0.35]{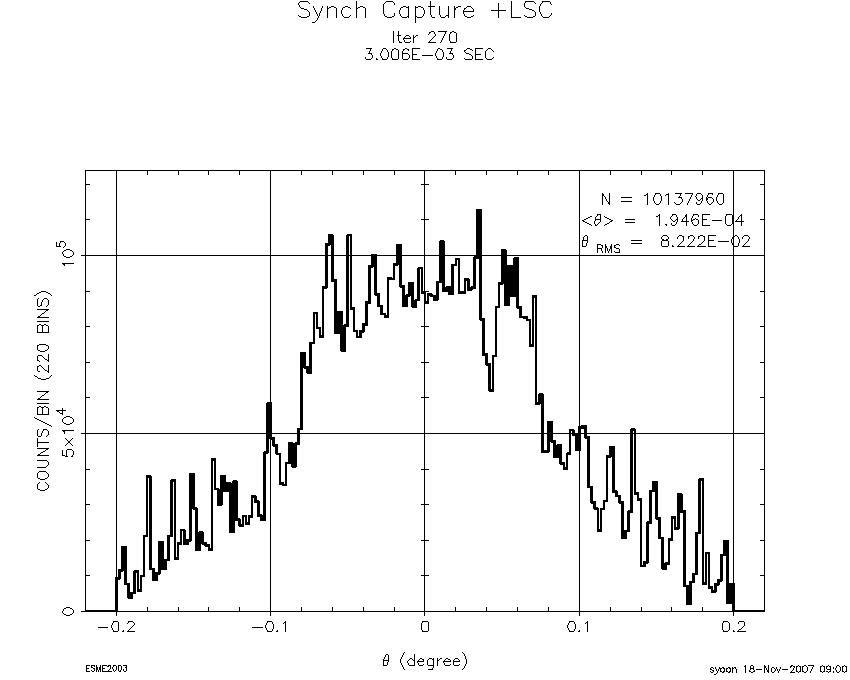}}
   \subfigure[at the 2,700$^{th}$ turn\label{fig:s4_theta_2700}]
             {\includegraphics[scale=0.35]{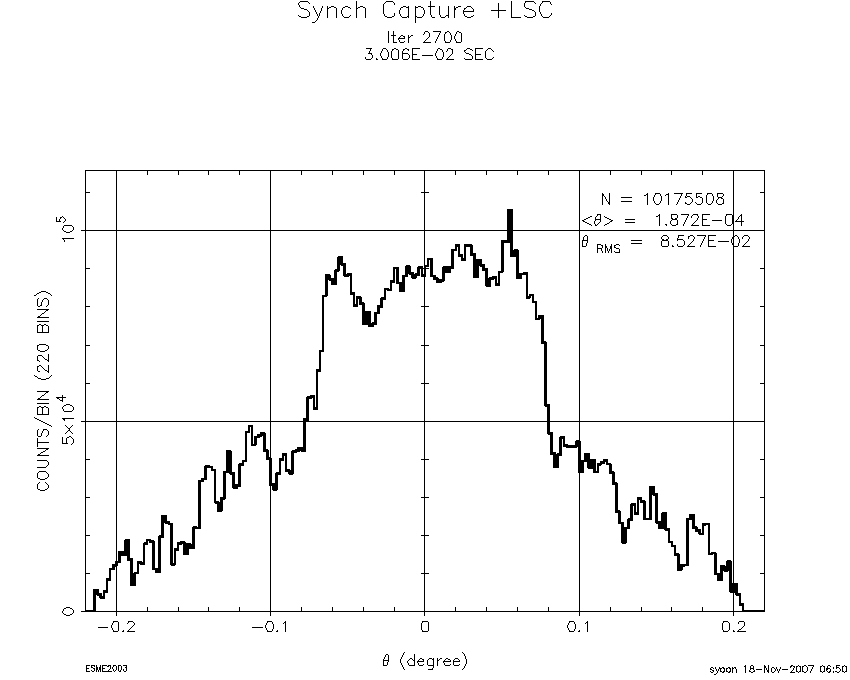}}
   \caption{\label{fig:theta-dualrf-400-200-270t-2700t}
   [\textbf{Scenario IV}]~
   Distribution of charge density with optimized dual RF harmonics
   (a) at the 270$^{th}$ turn and (b) at the 2,700$^{th}$ turn}
\end{figure}

\begin{figure}[h!]\centering
      \subfigure[1$^{st}$ turn]
                {\includegraphics[scale=0.17]{lps_dualrf_noramp_400_200_1t_bw.png}}
      \subfigure[100$^{th}$ turn]
                {\includegraphics[scale=0.17]{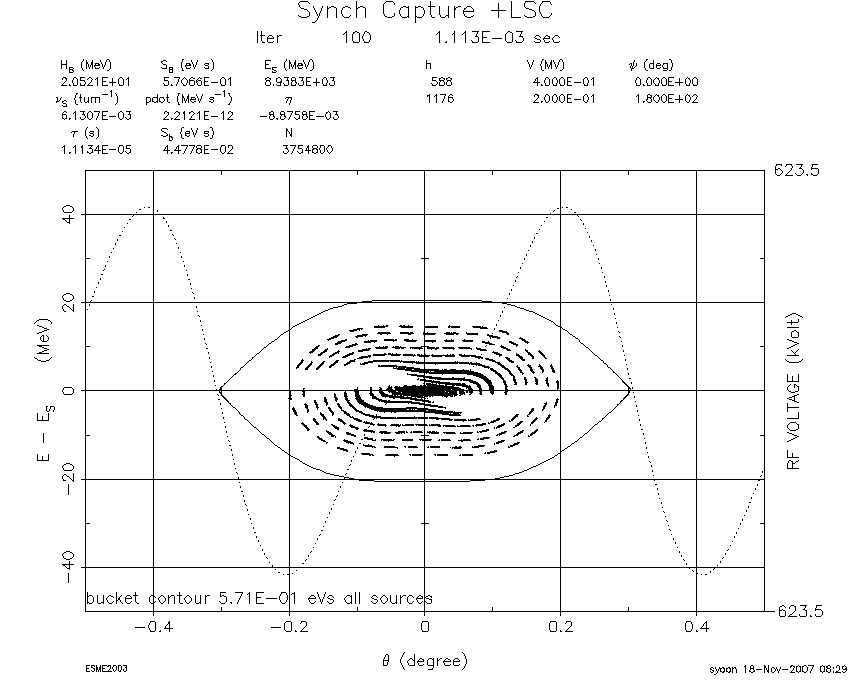}}
      \subfigure[200$^{th}$ turn]
                {\includegraphics[scale=0.17]{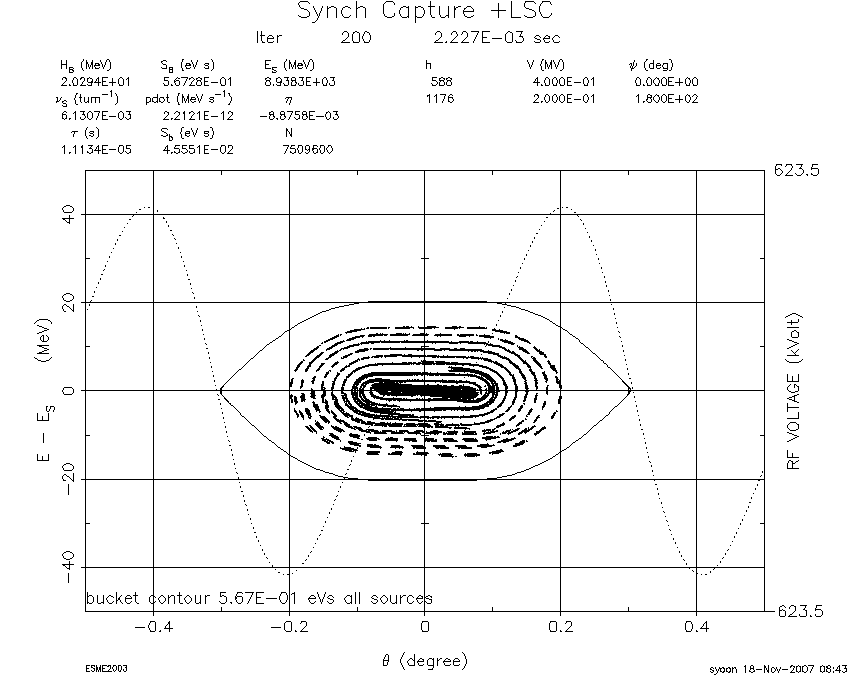}}
      \subfigure[270$^{th} turn$]
                {\includegraphics[scale=0.17]{lps_dualrf_noramp_400_200_270t_bw.png}}
      \subfigure[600$^{th}$ turn]
                {\includegraphics[scale=0.17]{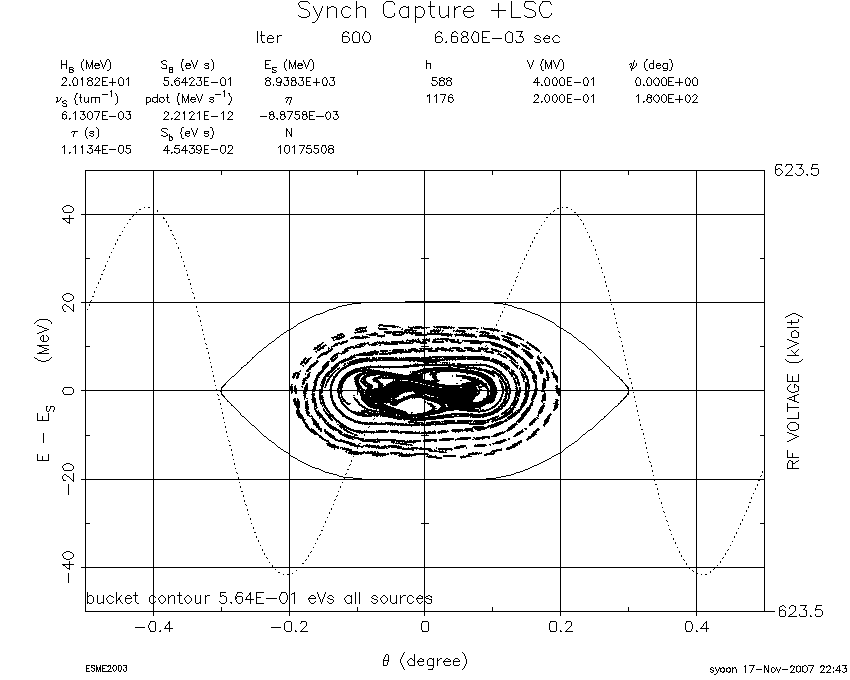}}
      \subfigure[900$^{th}$ turn]
                {\includegraphics[scale=0.17]{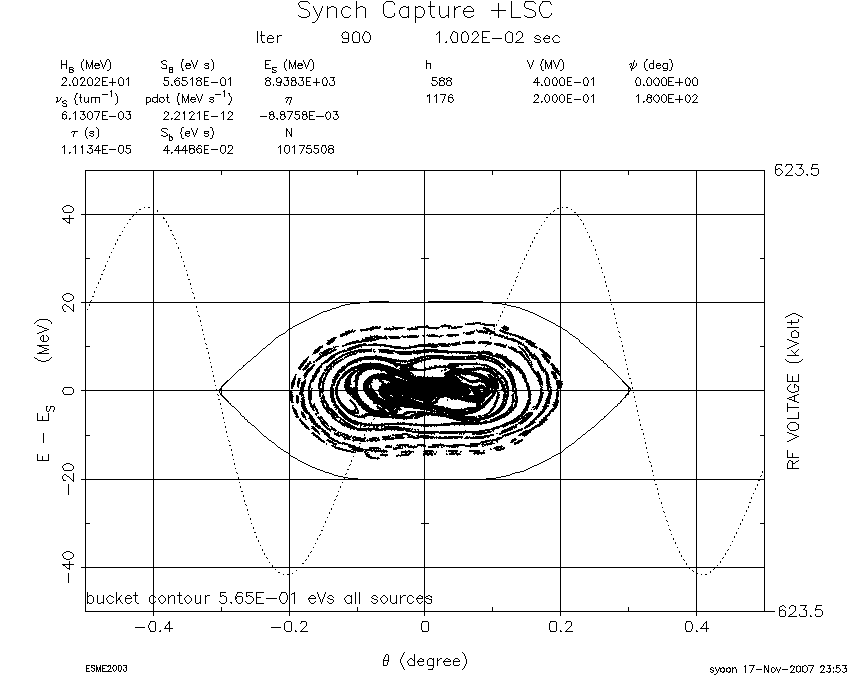}} 
      \subfigure[1,200$^{th} turn$]
                {\includegraphics[scale=0.17]{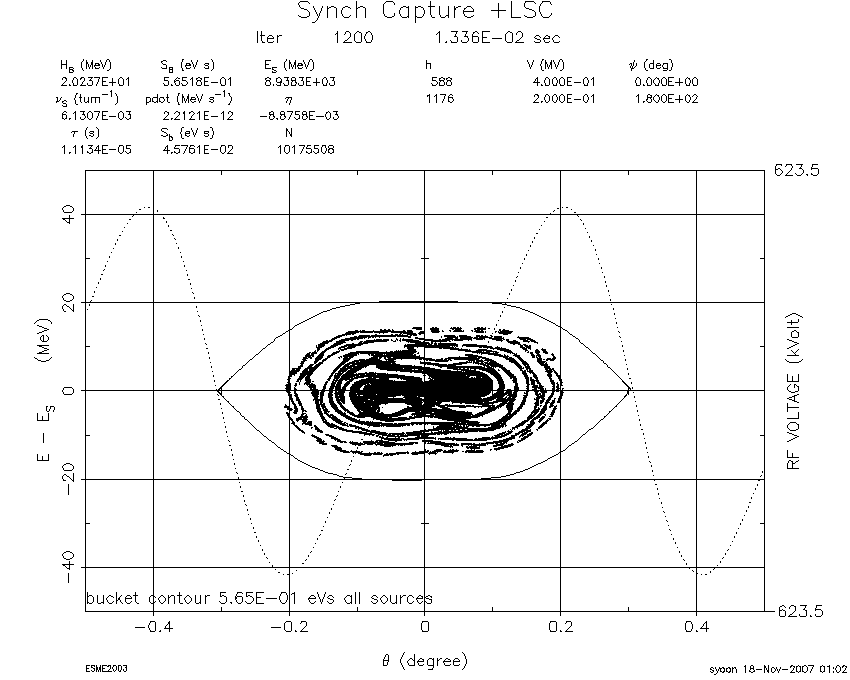}}
      \subfigure[1,500$^{th}$ turn]
                {\includegraphics[scale=0.17]{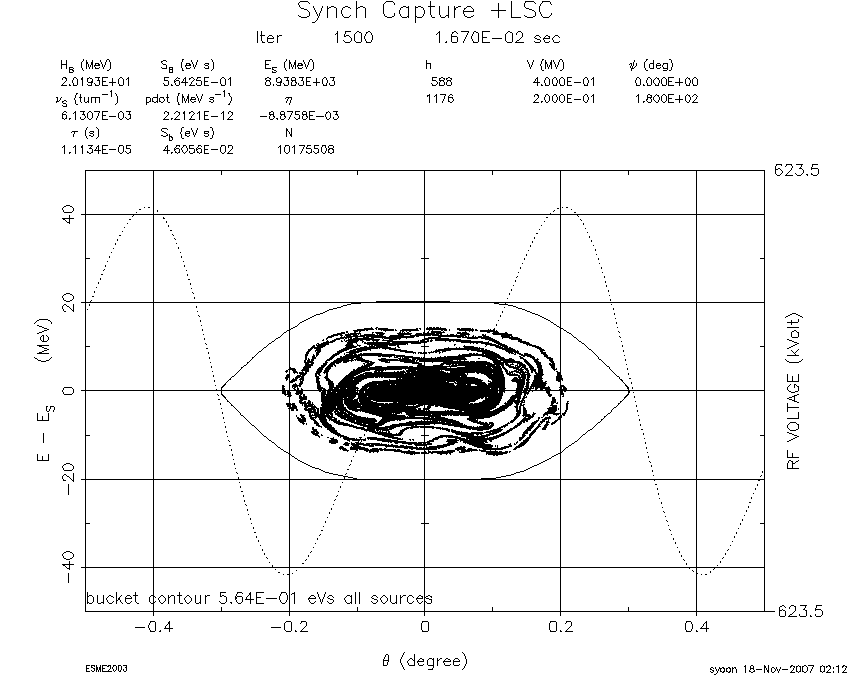}}
      \subfigure[1,800$^{th}$ turn]
                {\includegraphics[scale=0.17]{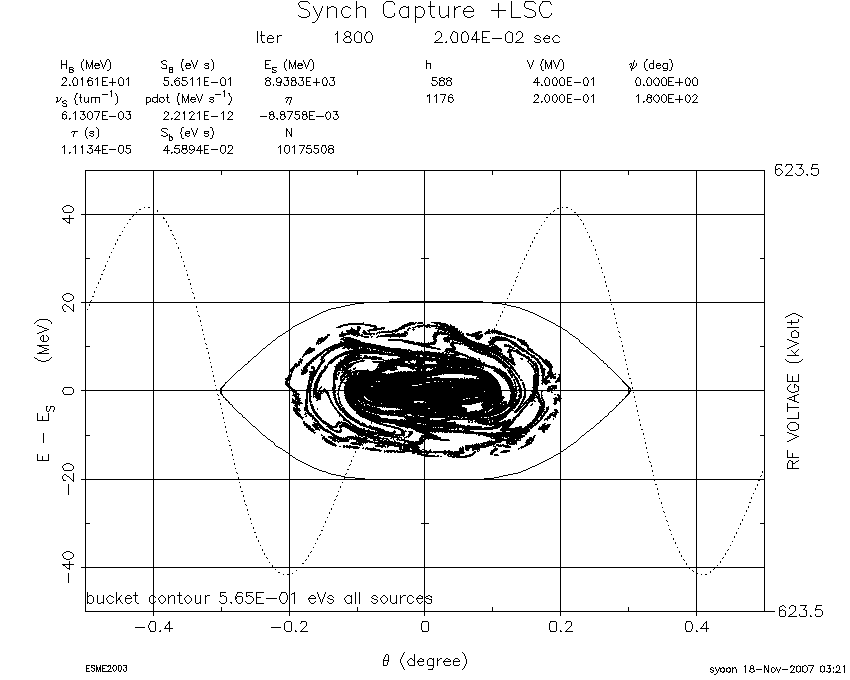}} 
      \subfigure[2,100$^{th} turn$]
                {\includegraphics[scale=0.17]{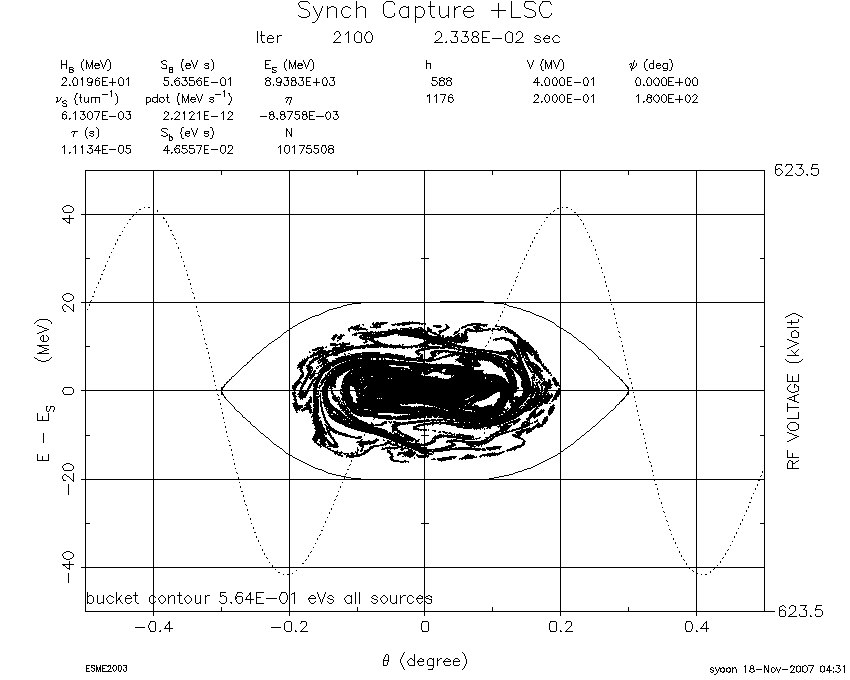}}
      \subfigure[2,400$^{th}$ turn]
                {\includegraphics[scale=0.17]{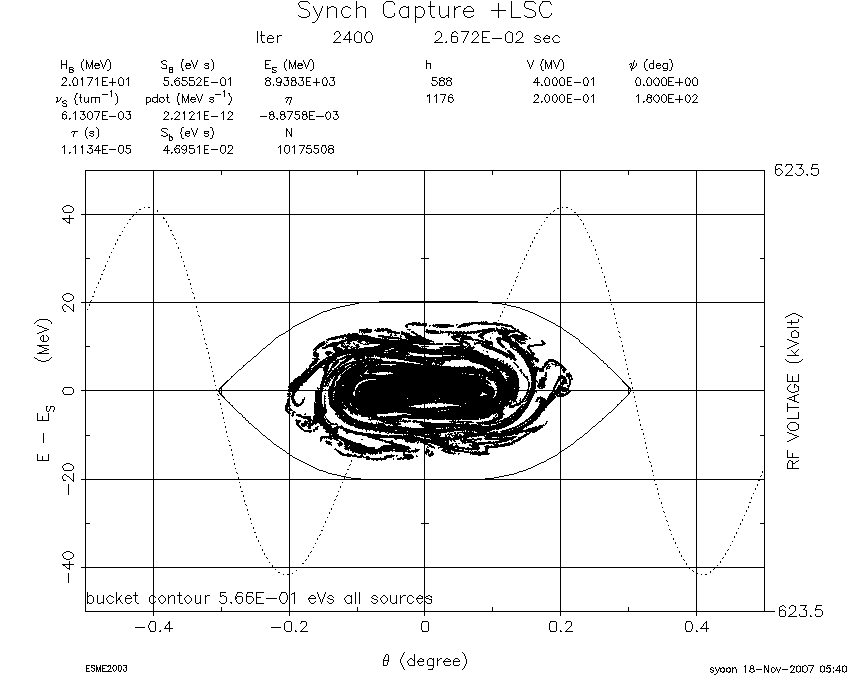}}
      \subfigure[2,700$^{th}$ turn]
                {\includegraphics[scale=0.17]{lps_dualrf_noramp_400_200_2700t_bw.png}} 
   \caption{\label{fig:s4_lps}[\textbf{Scenario IV}]
           Time evolution of injected micro-bunches captured within an optimized MI RF bucket
           with dual RF harmonics; starting from the $1^{st}$ injection turn through the $2,700^{th}$ turn\black}
\end{figure}

\begin{figure}\centering
      \subfigure[$100^{th}$ turn]
                {\includegraphics[scale=0.17]{lps_dualrf_noramp_400_200_100t_bw.png}}
      \subfigure[charge density]
                {\includegraphics[scale=0.17]{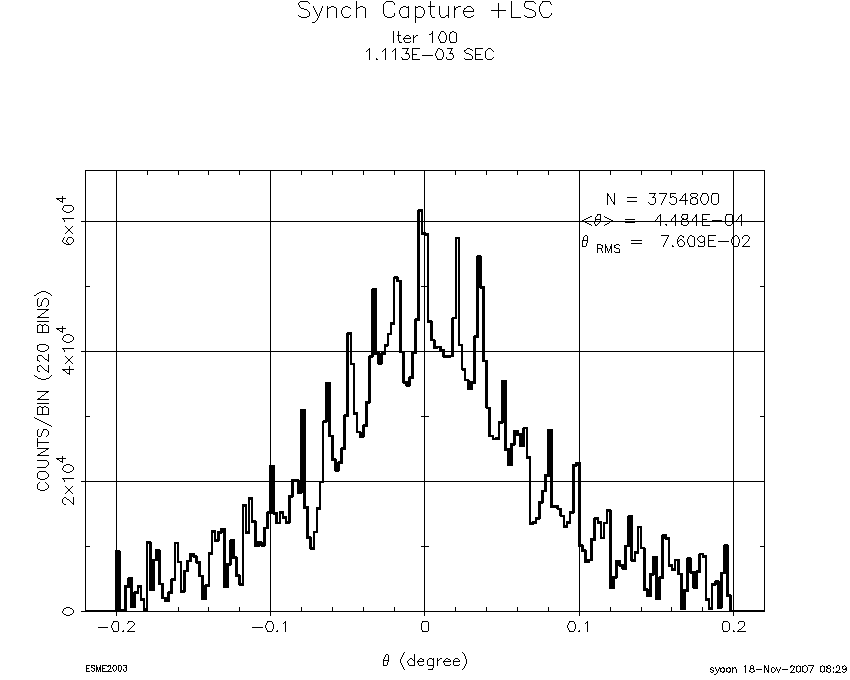}}
      \subfigure[energy density]
                {\includegraphics[scale=0.17]{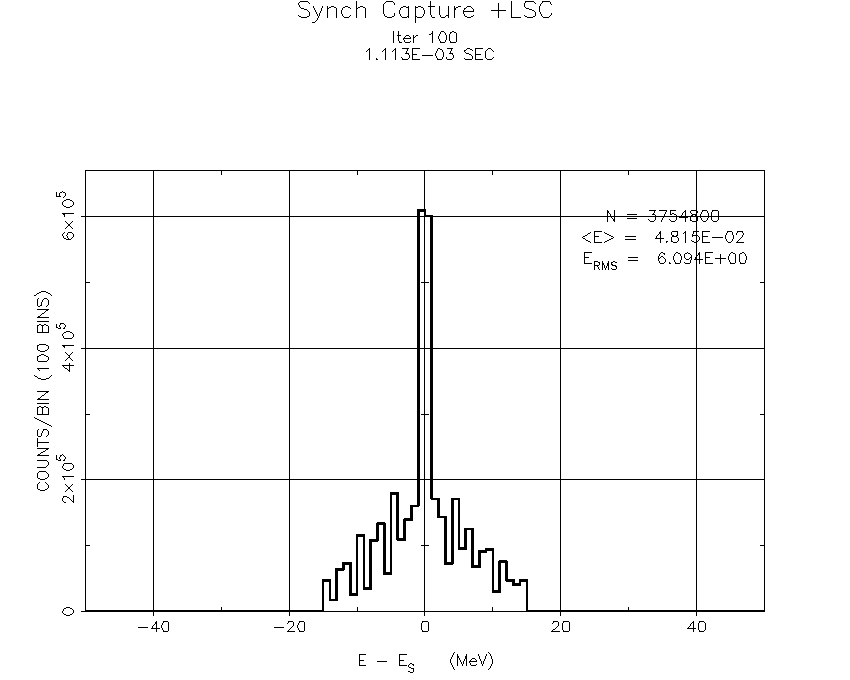}}
      \subfigure[$200^{th}$ turn]
                {\includegraphics[scale=0.17]{lps_dualrf_noramp_400_200_200t_bw.png}}
      \subfigure[charge density]
                {\includegraphics[scale=0.17]{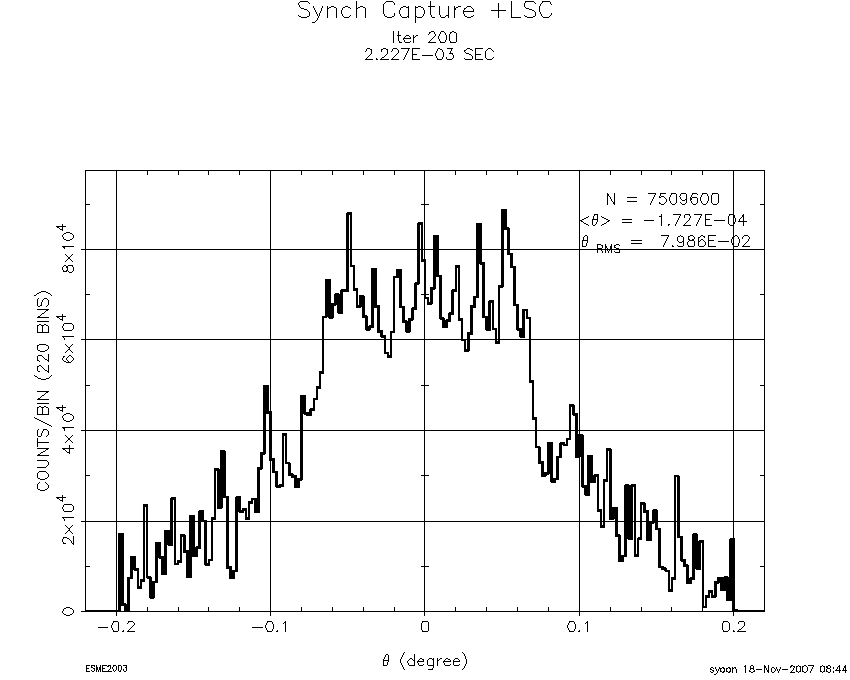}}
      \subfigure[energy density]
                {\includegraphics[scale=0.17]{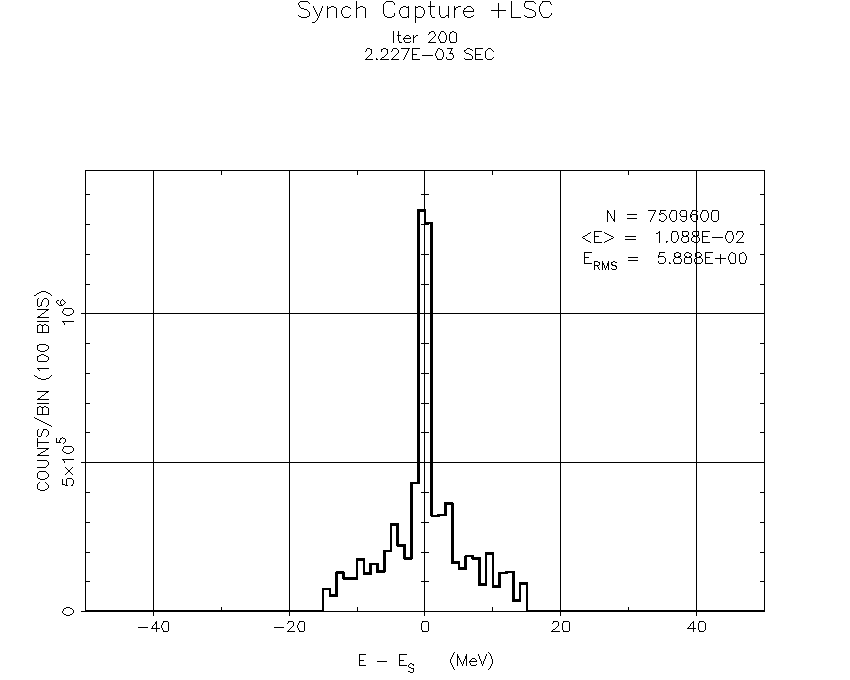}}  
      \subfigure[$270^{th}$ turn]
                {\includegraphics[scale=0.17]{lps_dualrf_noramp_400_200_270t_bw.png}}
      \subfigure[charge density]
                {\includegraphics[scale=0.17]{theta_dualrf_noramp_400_200_270t_bw.png}}
      \subfigure[energy density]
                {\includegraphics[scale=0.17]{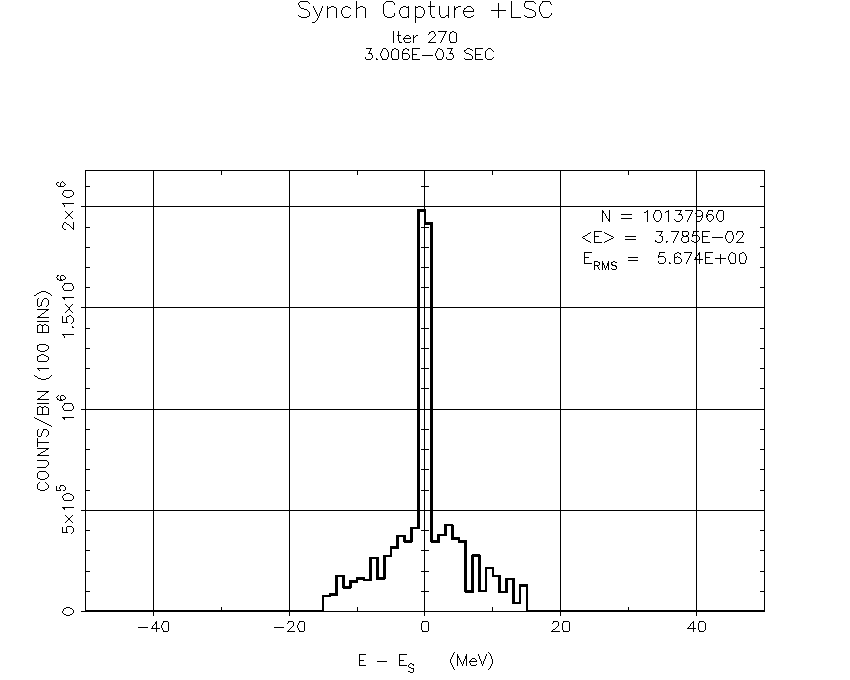}}  
      \subfigure[$300^{th}$ turn]
                {\includegraphics[scale=0.17]{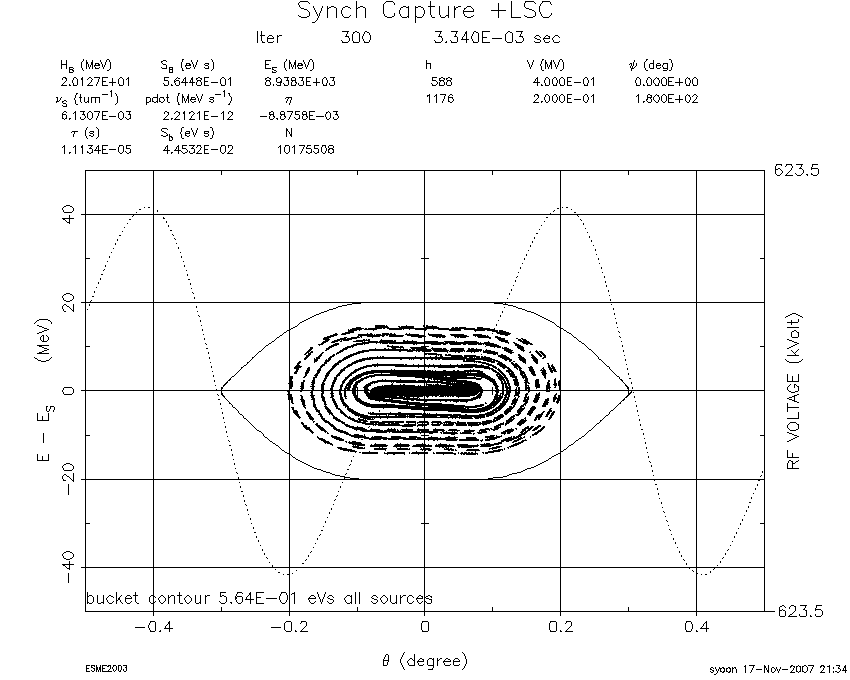}}
      \subfigure[charge density]
                {\includegraphics[scale=0.17]{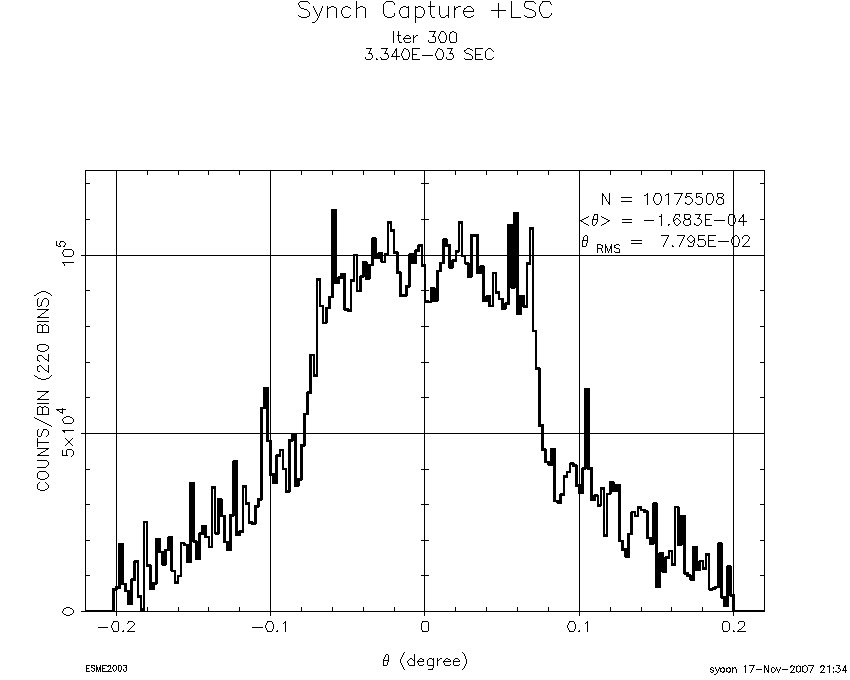}}
      \subfigure[energy density]
                {\includegraphics[scale=0.17]{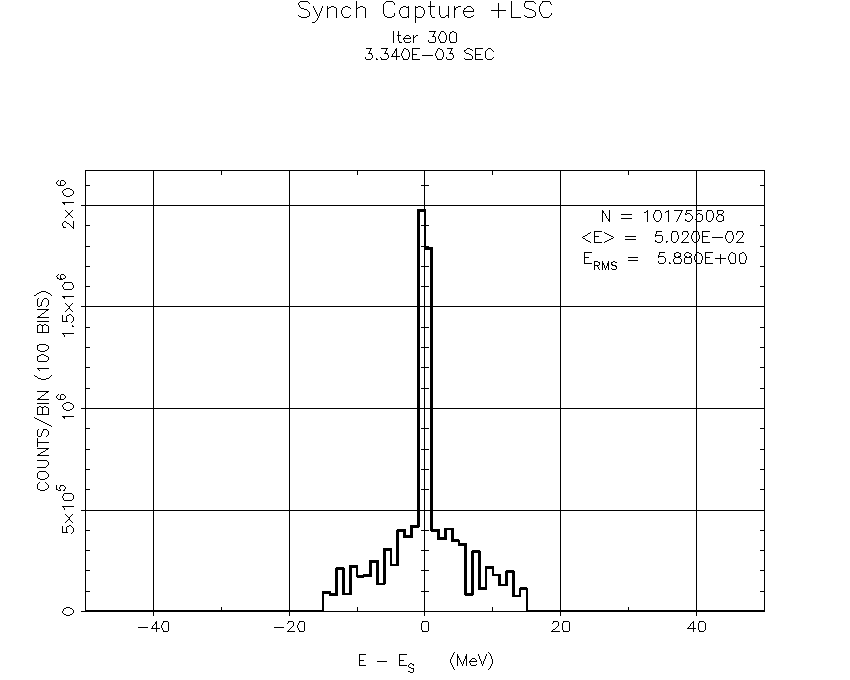}}
   \caption{\label{fig:s4_lps_theta_dE_300}[\textbf{Scenario IV}] 
           Time evolution of injected micro-bunches inclusive of
           longitudinal space-charge effect; starting from 
           the $100^{th}$ turn through the $300^{th}$ turn\black}
\end{figure}

\begin{figure}\centering
      \subfigure[600$^{th}$ turn]
      {\includegraphics[scale=0.17]{lps_dualrf_noramp_400_200_600t_bw.png}}
      \subfigure[charge density]
      {\includegraphics[scale=0.17]{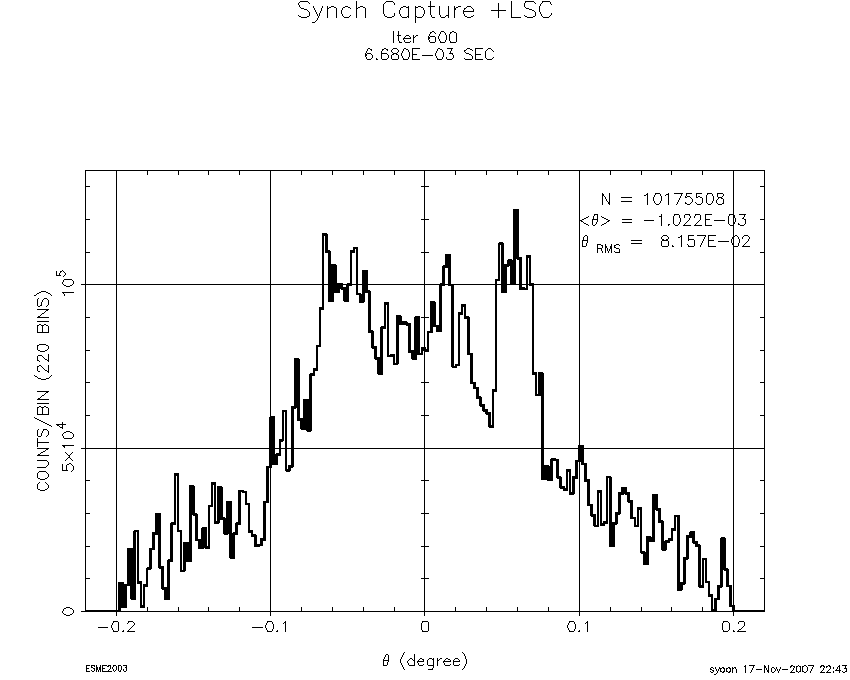}}
      \subfigure[energy density]
      {\includegraphics[scale=0.17]{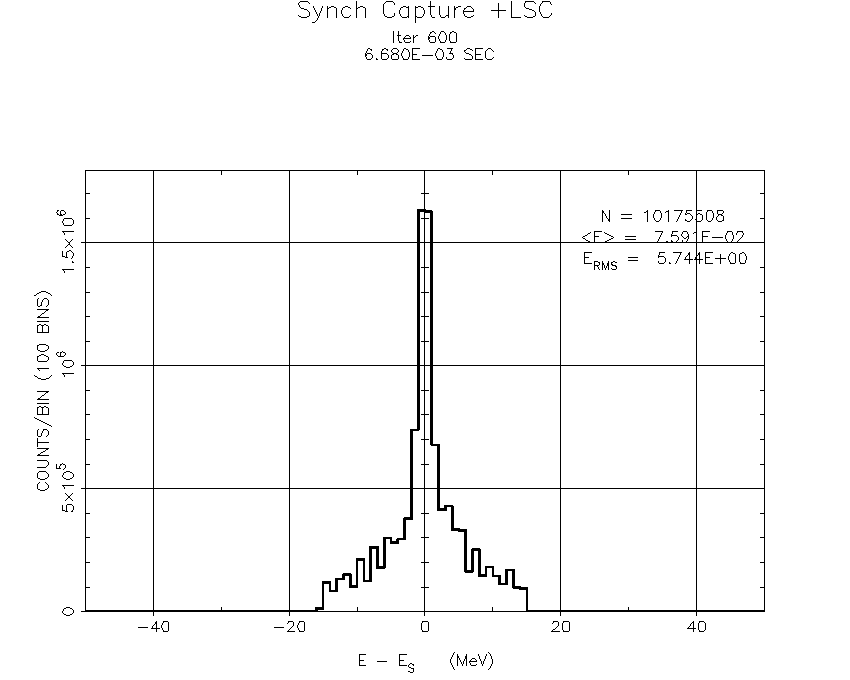}}
      \subfigure[900$^{th}$ turn]
      {\includegraphics[scale=0.17]{lps_dualrf_noramp_400_200_900t_bw.png}}
      \subfigure[charge density]
      {\includegraphics[scale=0.17]{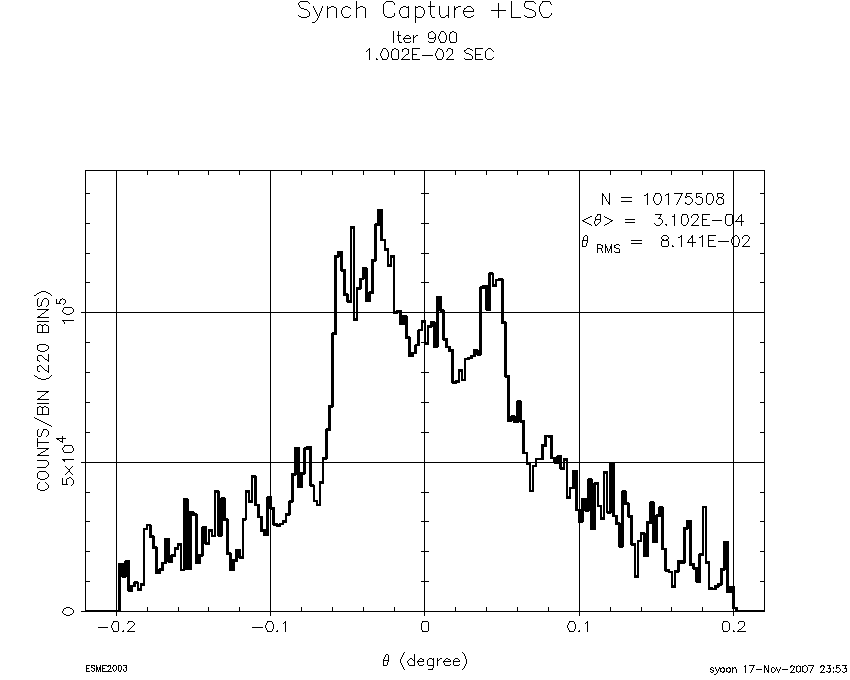}}
      \subfigure[energy density]
      {\includegraphics[scale=0.17]{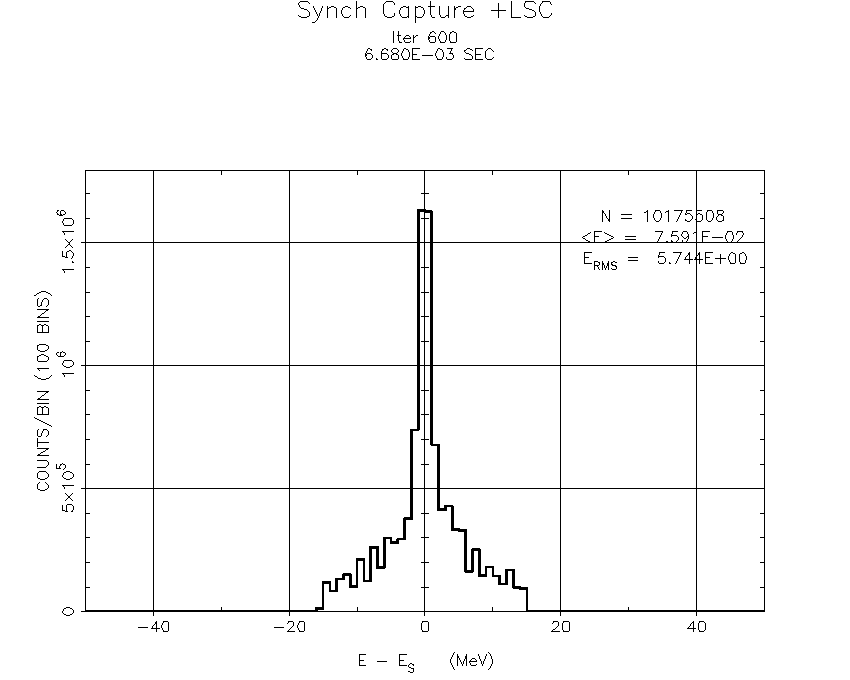}}  
      \subfigure[1,200$^{th}$ turn]
      {\includegraphics[scale=0.17]{lps_dualrf_noramp_400_200_1200t_bw.png}}
      \subfigure[charge density]
      {\includegraphics[scale=0.17]{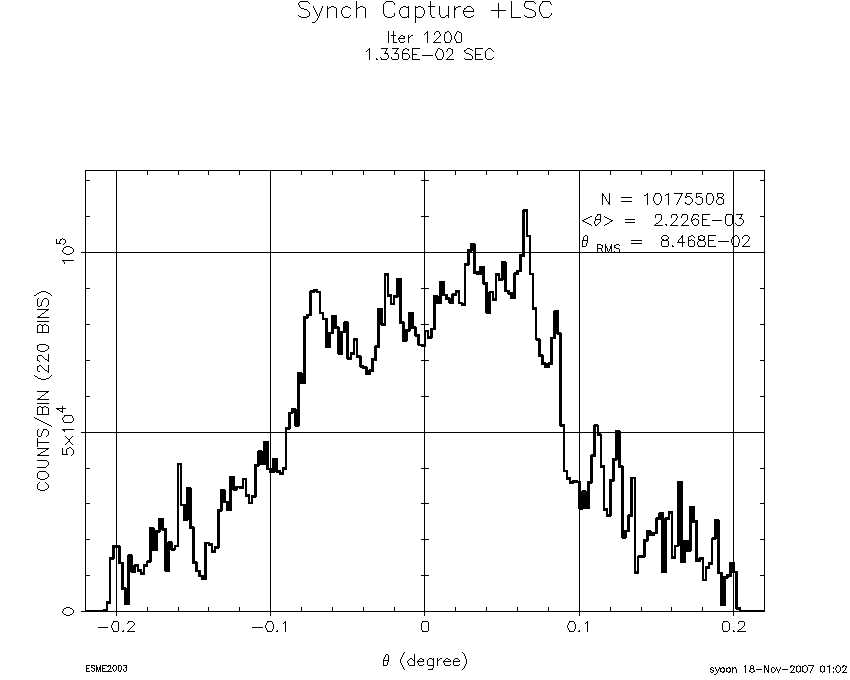}}
      \subfigure[energy density]
      {\includegraphics[scale=0.17]{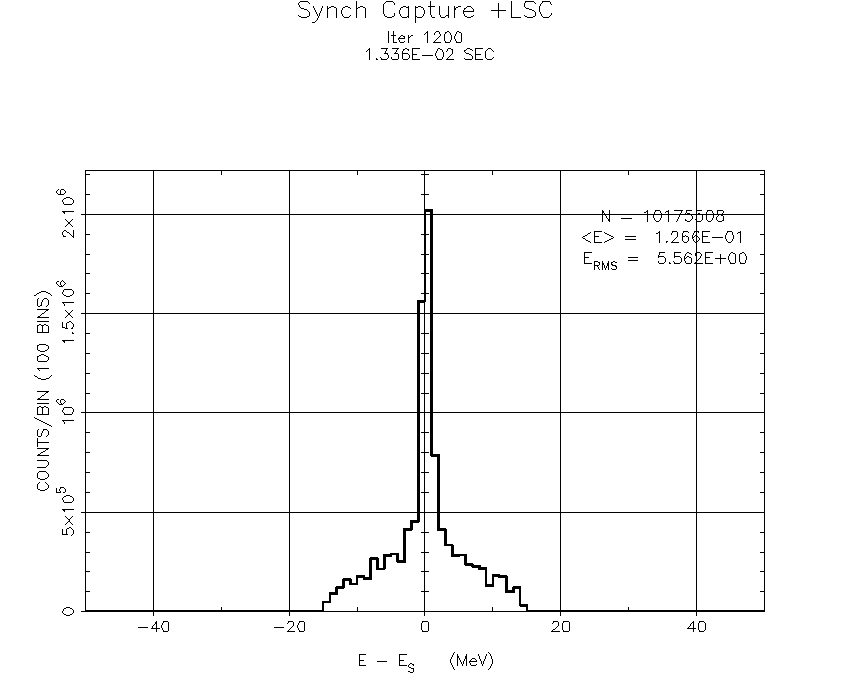}}  
      \subfigure[1,500$^{th}$ turn]
      {\includegraphics[scale=0.17]{lps_dualrf_noramp_400_200_1500t_bw.png}}
      \subfigure[charge density]
      {\includegraphics[scale=0.17]{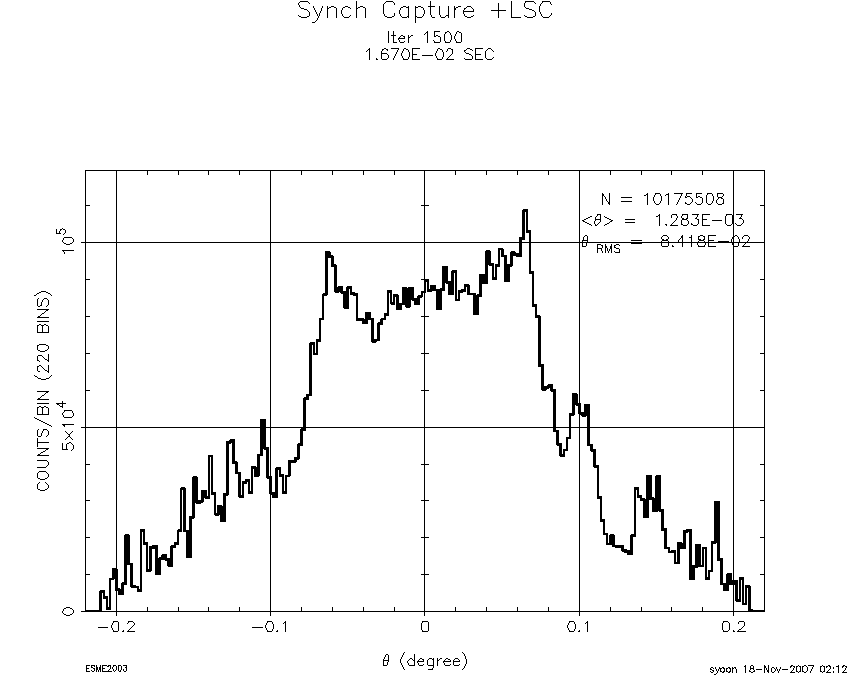}}
      \subfigure[energy density]
      {\includegraphics[scale=0.17]{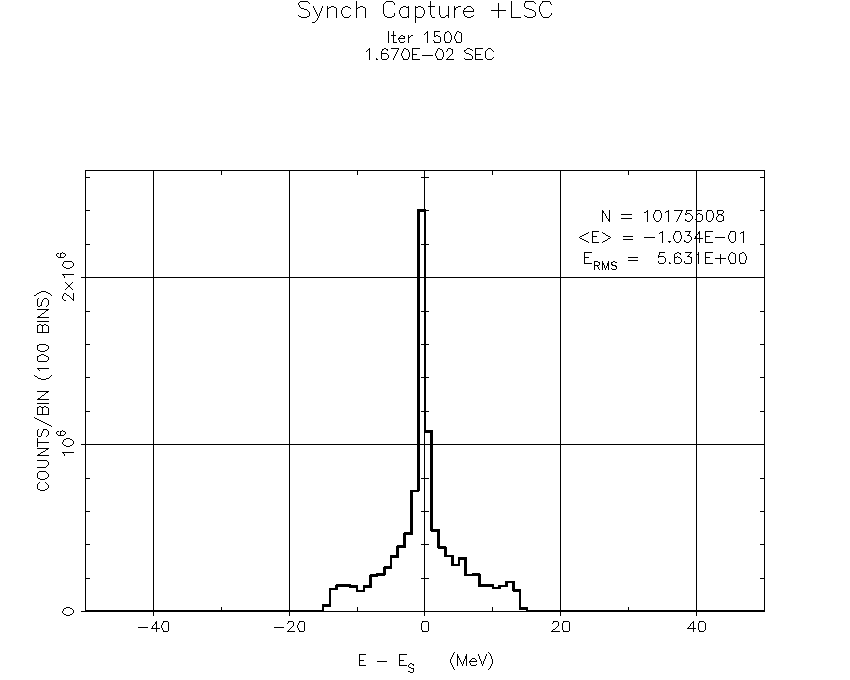}}
   \caption{\label{fig:s4_lps_theta_dE_1500}[\textbf{Scenario IV}] 
           Time evolution of injected micro-bunches inclusive of 
           longitudinal space-charge effect; starting from 
           the $600^{th}$ turn through the $1500^{th}$ turn\black}
\end{figure}

\begin{figure}\centering
      \subfigure[$1800^{th}$ turn]
      {\includegraphics[scale=0.17]{lps_dualrf_noramp_400_200_1800t_bw.png}}
      \subfigure[charge density]
      {\includegraphics[scale=0.17]{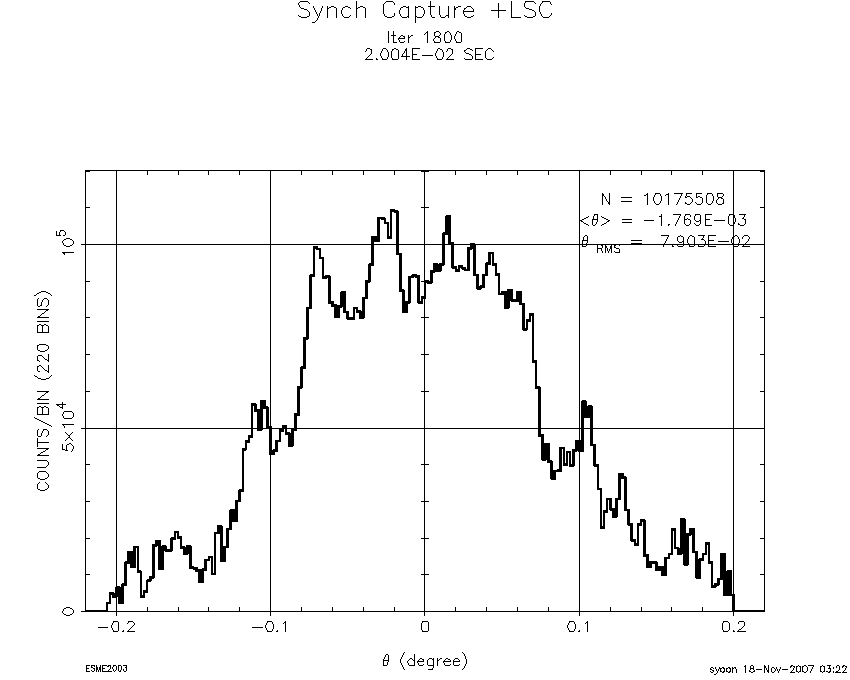}}
      \subfigure[energy density]
      {\includegraphics[scale=0.17]{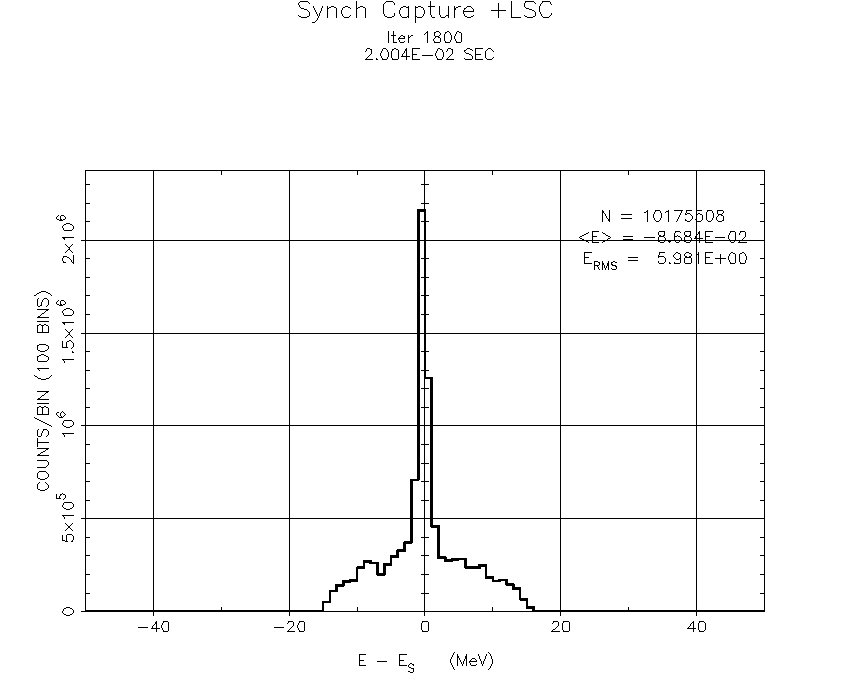}}
      \subfigure[$2100^{th}$ turn]
      {\includegraphics[scale=0.17]{lps_dualrf_noramp_400_200_2100t_bw.png}}
      \subfigure[charge density]
      {\includegraphics[scale=0.17]{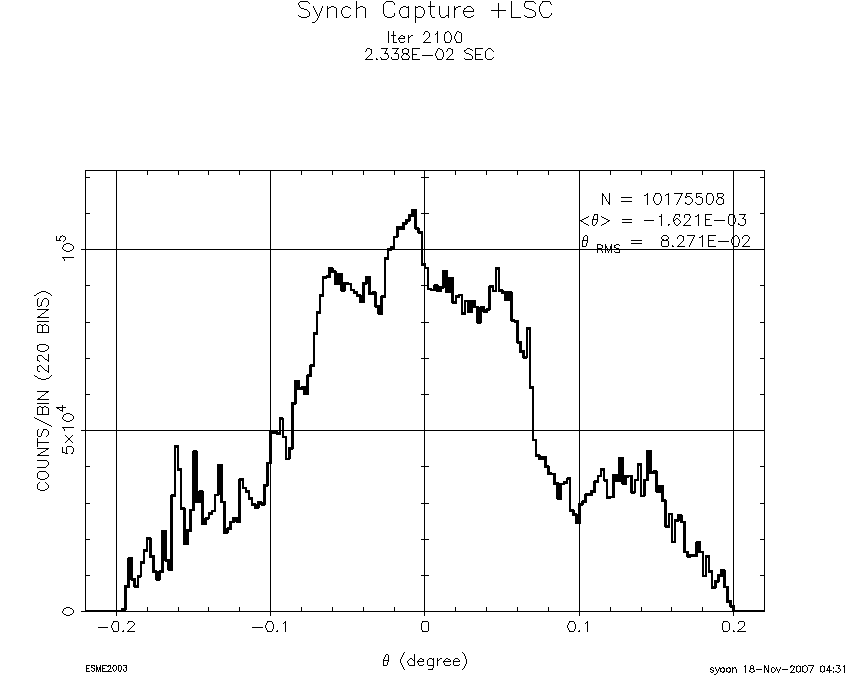}}
      \subfigure[energy density]
      {\includegraphics[scale=0.17]{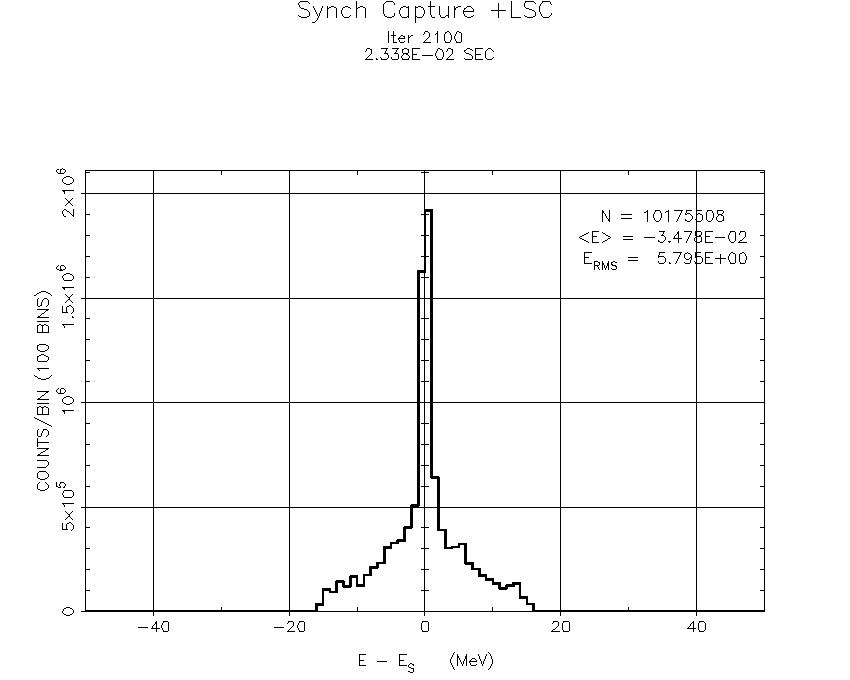}}  
      \subfigure[$2400^{th}$ turn]
      {\includegraphics[scale=0.17]{lps_dualrf_noramp_400_200_2400t_bw.png}}
      \subfigure[charge density]
      {\includegraphics[scale=0.17]{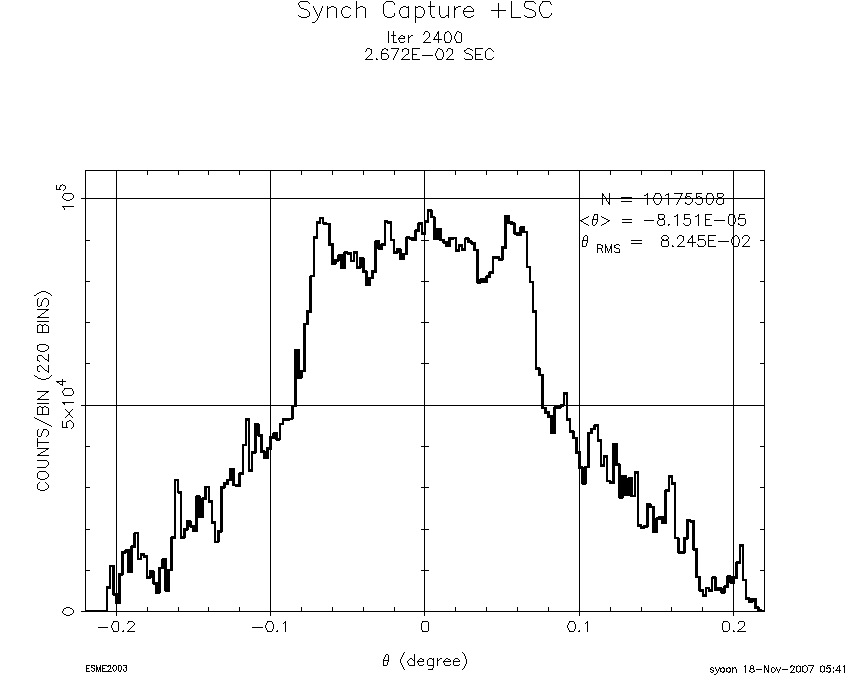}}
      \subfigure[energy density]
      {\includegraphics[scale=0.17]{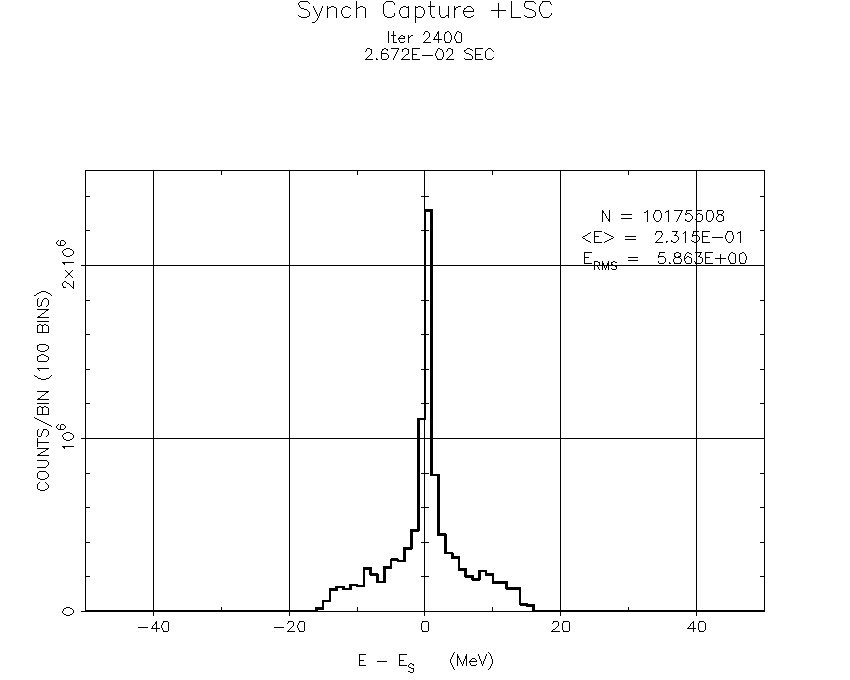}}  
      \subfigure[2,700$^{th}$ turn]
      {\includegraphics[scale=0.17]{lps_dualrf_noramp_400_200_2700t_bw.png}}
      \subfigure[charge density]
      {\includegraphics[scale=0.17]{theta_dualrf_noramp_400_200_2700t_bw.png}}
      \subfigure[energy density]
      {\includegraphics[scale=0.17]{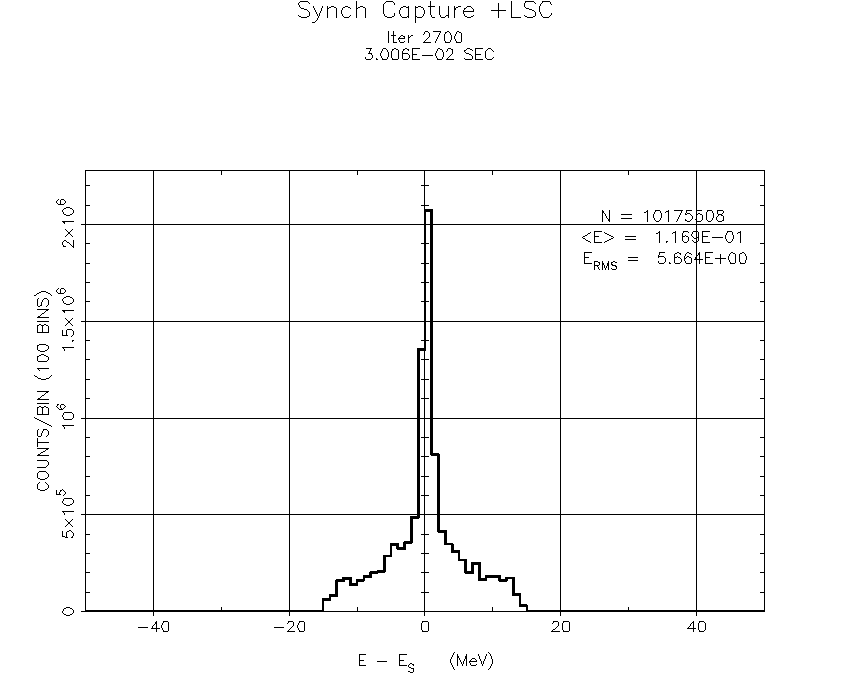}}
   \caption{\label{fig:s4_lps_theta_dE_2700}[\textbf{Scenario IV}] 
           Time evolution of injected micro-bunches inclusive of 
           longitudinal space-charge effect; starting from 
           the $1,800^{th}$ turn through the $2,700^{th}$ turn}
\end{figure}

\begin{figure}[h!]\centering  
\subfigure[Growth of longitudinal emittance over 2,700 turns\label{fig:s4_eps}]
          {\includegraphics[scale=0.26]{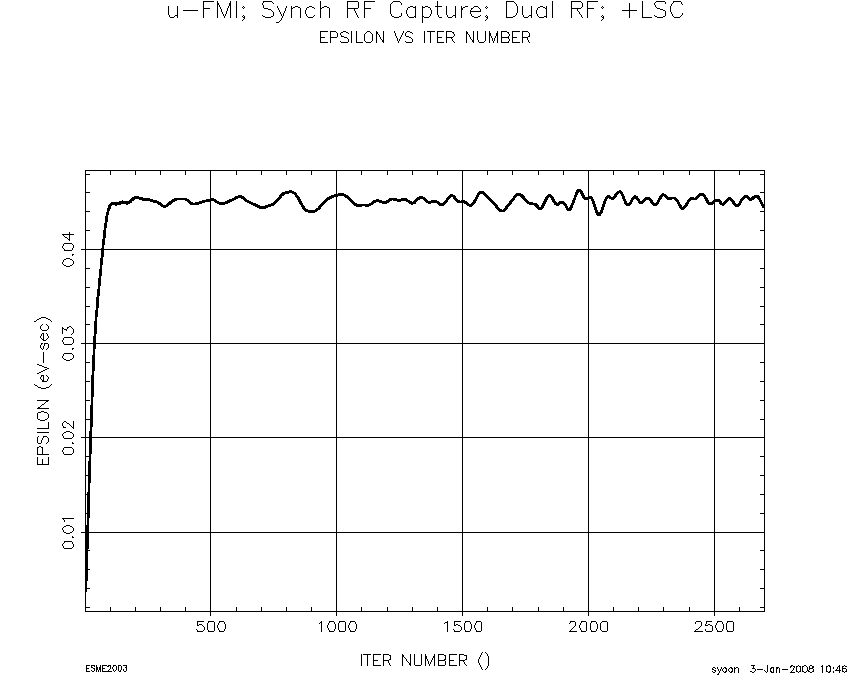}}
\subfigure[Additional $\Delta E$ induced by longitudinal space charge after 2,700 turns\label{fig:s4_vsc}]
          {\includegraphics[scale=0.26]{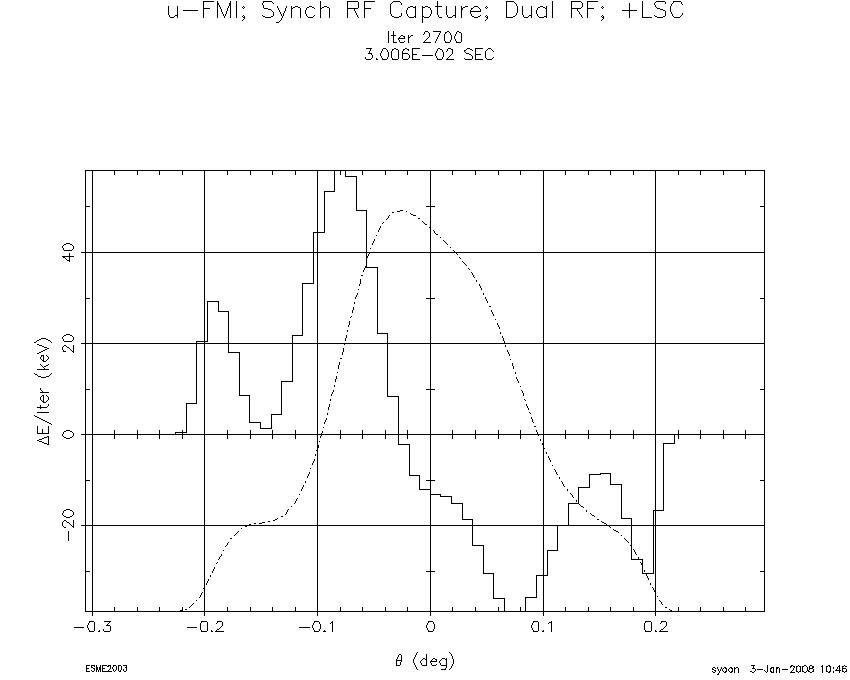}}
\subfigure[Time evolution of peak voltage in frequency domain over 2,700 turns\label{fig:s4_vpkfd}]
          {\includegraphics[scale=0.26]{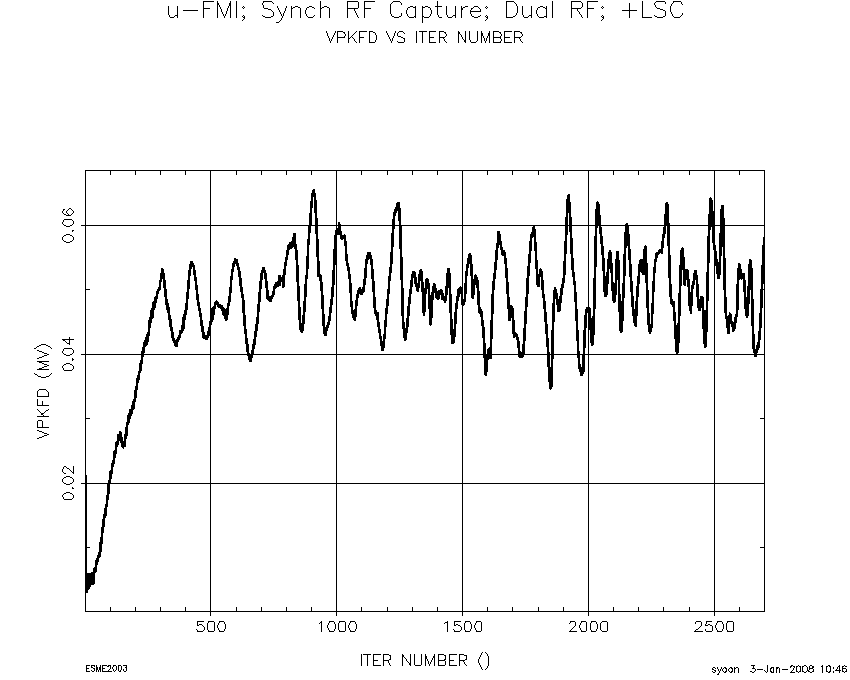}}
\subfigure[Time evolution of bunching factor\label{fig:s4_bf}]
          {\includegraphics[scale=0.26]{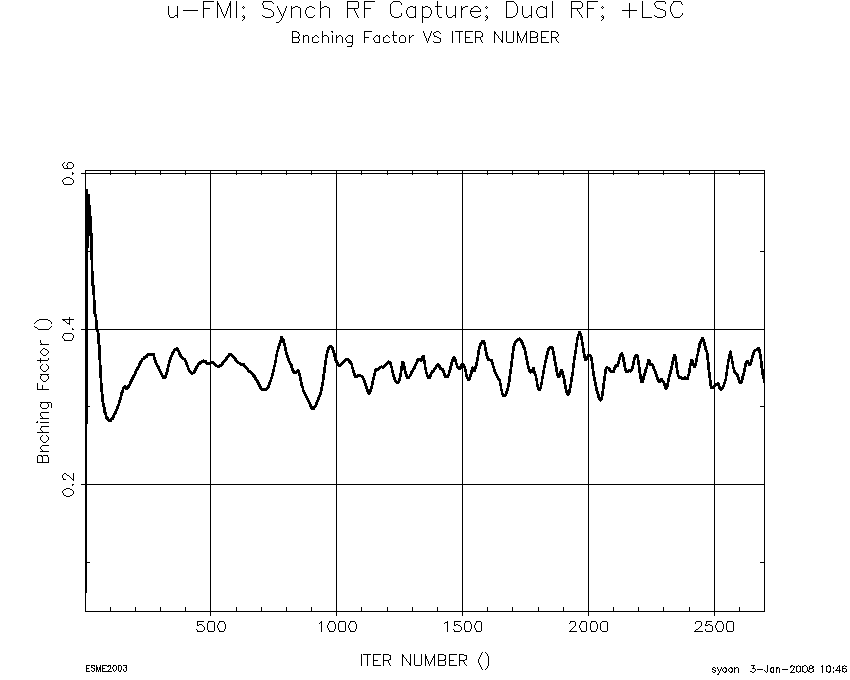}}
\caption{\textbf{[Scenario IV]} 
         Time evolution of longitudinal emittance and energy gain/loss arising from 
         longitudinal space-charge effect\label{fig:s4_eps_vsc}}
\end{figure}

 
Scenario IV is under the same conditions as Scenario III,
except that the secondary RF voltage setting is 50$\%$ of 
the principal RF voltage setting.
Scenario IV is more advantageous than Scenario III in that 
its \textit{optimized} RF waveform 
has a nice plateau around the stable phase of 0 (deg) 
as seen in Figure~\ref{fig:s4_lps_first}.
Through the parasitic longitudinal painting 
under the influence of space charge at each turn, 
a total of 1,080 (4 $\times$ 270) micro-bunches 
are injected over 270 turns. 
Afterwards, those injected micro-bunches are circulated 
for as long as 30 (ms), thereby transforming gradually into 
a continuous \textit{macro-bunch}, or \textit{long bunch} 
spanning over the region between -0.2 (deg) and +0.2 (deg)
within the MI RF bucket. 
Unlike in Figure~\ref{fig:theta-dualrf-400-300-270t-2700t} (a) for Scenario III, 
localization of macro-particles at two positions is not observed 
in Figure~\ref{fig:theta-dualrf-400-200-270t-2700t} (a)  
when the injection over 270 turns is complete.
After extended circulation of beams up to 2,700 turns 
with no further injection, the beam charge distribution becomes
continuous and smoother in between -0.2 (deg) and +0.2 (deg). 
(see Figure~\ref{fig:theta-dualrf-400-200-270t-2700t}).
As a consequence, the bi-modal distribution of charge density 
observed from Scenario III (Figures~\ref{fig:s3-lps-theta-dE-300} 
through \ref{fig:s3-lps-theta-dE-2700}) is not observed any longer 
(Figures~\ref{fig:dualrf-400-200-270t-2700t} 
through \ref{fig:s4_lps_theta_dE_2700}). 
As the number of injected macro-particles increments, 
the process of longitudinal painting continues. 
Subsequently, the fine structure of charge distribution gradually vanishes,
and the contour of charge distribution becomes smoother after 2,700 turns.
With phase offsets alone without energy jitter, 
this painting effect stands out in charge distribution, 
rather than energy distribution as observed in Scenario III.
\black
As shown in Figures~\ref{fig:s4_eps} and \ref{fig:s4_vsc},
an encouraging result is obtained with the dual RF harmonic system.
The longitudinal emittance grows and reaches its equilibrium value,
which is about a half of the emittance in the cases of the single RF harmonic system.
%
The peak induced voltage ($\hat{V}_{sc}$) 
due to space charge was computed using asymmetric charge distributions seen in Figure~\ref{fig:s4_vsc}.
Its peak space-charge voltage rises up to 50 (kV) per turn according to Figure~\ref{fig:s4_vsc}.
%
Plotted in Figure~\ref{fig:s4_vpkfd} is
the evolution of peak voltage induced by space charge.
Using the phase modulation in a controlled fashion 
enables us to maneuver charge distribution, thus 
resulting in further lowering space-charge voltage.
%
\newpage\clearpage
\newpage\clearpage
%
\newpage\clearpage
\newpage\clearpage
\newpage\clearpage
%
\newpage\clearpage
As presented in Figure~\ref{fig:s4_bf},
the bunching factor calculated at the completion of multi-turn injection 
in Scenario IV turns out to be close to that of Scenario I.
It is worthwhile to note that the bunching factor (BF) calculations 
are important in that the BF gives us an idea of how large tune spreads 
will be prior to calculating 3-D space charge effects while tracking beams.
\vspace{1cm}
%
\newpage\clearpage
\section{Concluding Remarks}\label{sec:concluding-remarks}
%
      We have investigated different scenarios of injecting
      micro-bunches from a SCRF linac into the MI ring
      under the influence of longitudinal space charge:
      {\em from the 8-GeV linac Proton Driver
           to the Main Injector}
      \par The RF mismatch between a linac and a ring
      can induce phase slips with trains of micro-bunches,
      leading to the longitudinal painting phenomenon in an uncontrolled fashion.
      Thus, it is considered rather advantageous to use harmonics
      of \textit{non-integral} ratio between a linac and the ring 
      in order to increase proton intensity by benefiting from 
      parasitic longitudinal painting.
      Besides, subsequent charge \textit{redistribution} 
      in longitudinal space can take place by circulating 
      injected beams with no further injection.  
      In other words, charge roaming within the RF bucket 
      in longitudinal phase space can help lower 
      the gradient of charge distribution,
      thereby reducing its induced voltage arising from 
      the space-charge effect.
      In addition to longitudinal painting,
      future simulations are planned to include both phase 
      and energy jitters due to errors in the SCRF linac.
      Because of the short bunch length of the linac beam,
      it is anticipated that the impact of the broad-band
      impedance may play an important role 
      in the context of longitudinal dynamics\cite{gjackson}.
      An optimized dual RF system and longitudinal painting
      can overcome beam-intensity limits set by the space-charge 
      effect in high-intensity and low-energy machines.
      Four different scenarios of injecting multi-bunches demonstrate 
      that a double RF system with the harmonic ratio 
      ($R_{H} = 1176/588$) of 2.0 and the voltage ratio 
      ($R_{V} = 200 kV/400 kV$) of 0.5 are most favored for minimizing 
      the longitudinal space-charge effect.
      All of the scenarios for the time-structured multi-turn injection
      including animations are available on a Fermilab website~\cite{pd2:web}.
\black
%
%
\section{Acknowledgment}
%
Authors wish to thank J. Maclachlan of the Proton Source Department 
and J-P. Carneiro of Accelerator Physics Center (APC) of Fermilab.
Authors had useful discussions with J. Maclachlan modeling multi-turn 
injection with \textit{macro-bunches} at the outset.  
J-P. Carneiro provided us with an input file of the 8-GeV SCRF linac 
distribution that were generated by the TRACK code\cite{ostroumov}.
\black
%

%
\end{document}